\documentclass[rmp,aps,reprint,amsmath,amssymb,graphicx,longbibliography]{revtex4-2}
\usepackage{bm}
\usepackage{amssymb}
\usepackage{amsmath}
\usepackage{graphicx}
\usepackage{here}
\usepackage{color}
\usepackage{booktabs}

\newcommand{\red}{\textcolor{black}}

\graphicspath{{Figures/}}

\begin{document}

\title{Kitaev Quantum Spin Liquids}
\author{Yuji Matsuda}
\affiliation{Department of Physics, Kyoto University, Kyoto 606-8502, Japan} 
\affiliation{and Los Alamos National Laboratory, Los Alamos, NM 87545, USA}
\author{Takasada Shibauchi}
\affiliation{Department of Advanced Materials Science, University of Tokyo, Kashiwa, Chiba 277-8561, Japan}
\author{Hae-Young Kee}
\affiliation{Department of Physics, University of Toronto, Toronto, Ontario, Canada M5S 1A7}
\affiliation{and Canadian Institute for Advanced Research, Toronto, Ontario, Canada M5G 1M1}
\date{\today{}}

\begin{abstract}
Quantum spin liquids (QSLs) represent exotic states of matter where quantum spins interact strongly yet evade long-range magnetic order down to absolute zero. Characterized by non-local quantum entanglement and resultant fractionalized excitations, QSLs have emerged as a frontier in condensed matter physics, bolstered by the recent identification of several candidate materials. This field holds profound implications for understanding strong correlations, topological order, and emergent phenomena in quantum materials. Among them, the spin-1/2 Kitaev honeycomb model, featuring bond-directional Ising interactions, provides a rare exactly solvable QSL example. Its ground state is a topological QSL, with spin degrees of freedom fractionalized into emergent Majorana fermions. Under an applied magnetic field, the Kitaev QSL transitions to a topologically non-trivial chiral spin liquid state with non-Abelian anyons, offering potential resources for topological quantum computation. The non-Abelian character of these anyons in the Kitaev QSL demonstrates a profound connection to certain topological superconductors and even-denominator fractional quantum Hall states. Since the theoretical prediction that the Kitaev model could manifest in spin-orbit-coupled materials such as honeycomb iridates and ruthenates, research has focused on identifying candidate compounds. In particular, experimental evidence suggests spin fractionalization and topological phenomena akin to the Kitaev model in the spin-orbit Mott insulator $\alpha$-RuCl$_3$. However, results and interpretations remain actively debated. This review {begins with a brief review on QSLs in other systems, followed by} a comprehensive survey of existing studies on Kitaev candidate materials, with a particular focus on $\alpha$-RuCl$_3$. Rather than offering conclusive remarks, our aim is to inspire future research by examining several key aspects of the current literature and perspectives.
\end{abstract}

\maketitle

\tableofcontents{}

\section{Introduction}
\label{intro}
At elevated temperatures, materials exhibit structural symmetry.  However,  upon cooling through phase transitions, spontaneous symmetry breaking occurs, giving rise to emergent ordered states. For example, in the liquid-to-solid transition, translational symmetry is broken as atoms condense into a periodic lattice arrangement. In ferroelectrics, the inversion symmetry of the crystal structure is broken.  The high-temperature paraelectric phase exhibits centrosymmetric ion configurations, while below the Curie point, a structural distortion reduces the point group symmetry, yielding a polar state with aligned electric dipoles.  
Symmetry breaking is a ubiquitous phenomenon accompanying phase transitions, { where the system adopts a lower-symmetry ground state such as in a liquid-to-solid transition.}

On the other hand, in systems with significant quantum fluctuations, the ground state can maintain liquid-like properties without undergoing spontaneous {translational} symmetry breaking, even at absolute zero temperature, despite strong interparticle interactions.  Such systems are classified as quantum liquids and notable examples { including $^3$He and $^4$He (which become superfluids). They resist solidification due to large zero-point energy, and their behavior is governed by quantum-mechanical effects.}

In insulating magnets, where the electrons cannot move freely due to strong electron interactions, magnetism is mainly governed by the spin {(or pseudospin via spin-orbit coupling) degrees of freedom of electrons.
Quantum spins defined at lattice points in a crystal interact with each other via exchange processes, and in most cases, the ground state is an ordered state characterized by a local order parameter. For example, antiferromagnetic (AFM) or ferromagnetic (FM) orders are represented by a non-zero local uniform or sublattice magnetization, even in the absence of an external magnetic field. These states with a local order parameter can be viewed as a solid state of spins.}

Quantum spin liquids (QSLs) represent a novel class of quantum states of matter in certain magnetic systems, predominantly in insulating materials. QSLs can be conceptualized as spin-based analogs of quantum liquids \cite{balents2010spin,savary2016quantum,zhou2017quantum,knolle2019field,broholm2020quantum,clark2021quantum}.  Despite of strong interactions between spins, QSL does not exhibit {local} ordering and {its classification goes beyond the conventional notion of the symmetry breaking. Note, however, that some QSLs may still exhibit symmetry breaking.}
Unlike a conventional disordered paramagnetic phase, a QSL exhibits highly entangled spin configurations that cannot be described by simple product states. Strong spin-spin interactions constrain the low-energy excitations, leading to the emergence of fractionalized degrees of freedom fundamentally distinct from the original spin variables. These excitations give rise to exotic phenomena, such as collective behaviors described by emergent gauge fields and particles with fractional statistics---neither fermionic nor bosonic. {In gapped systems, such fractionalization is closely tied to topological order \cite{misguich2010quantum}, characterized by ground state degeneracy that is robust against local perturbations. Gapless QSLs also exist, and a wide range of theoretical models has been proposed, featuring both gapped and gapless QSL phases with various types of fractionalized excitations.}


The crystal structures of archetypal QSL candidates are shown in Figs.\,\ref{fig:Frustration}(a)-(e). In the domain of QSLs, the one-dimensional (1D) spin-1/2 Heisenberg chain serves as the quintessential example (Fig.\,\ref{fig:Frustration}(a)), amenable to rigorous analytical solution.  For spin systems in higher dimensions, it is generally believed that frustrations caused by competing exchange interactions are necessary to stabilize the QSL states \cite{anderson1973resonating,fazekas1974ground}. Geometrical frustration arises when the local interactions between magnetic moments cannot be simultaneously satisfied due to the specific arrangement of magnetic ions in the lattice structure. Inspired by the resonating-valence-bond (RVB) idea, which was applied to high-temperature superconductivity \cite{anderson1987resonating,baskaran1993resonating}, pioneering research on the enigmatic nature of QSLs in two-dimensional (2D) systems has led to the proposal of several pivotal developments. The AFM Heisenberg models on the 2D triangular and kagome lattices are archetypal examples of geometrically frustrated magnets (Figs.\,\ref{fig:Frustration}(b)(c)). In three-dimensional (3D) systems, the pyrochlore lattices (Fig.\,\ref{fig:Frustration}(e)), characterized by a network of corner-sharing tetrahedra, are regarded as prime candidates for realizing QSL states due to the significant geometrical frustration experienced by the localized magnetic moments.

Most theories of the QSLs were based on certain approximations, effective theories, or numerical studies, and there have been debates on whether the QSLs could be realized in real systems.  In the past two decades, however, several candidate QSL materials have been found experimentally mainly in 2D triangular and kagome and 3D pyrochlore lattices. In these systems, no spontaneous symmetry-breaking phenomena, such as magnetic long-range order or structural phase transitions, manifest even at temperatures far below the energy scale set by the nearest neighbor (n.n.) exchange interactions. In addition, experimental observations increasingly suggest the presence of fractionalized quasiparticle excitations, {with theoretical studies providing complementary support.}  However,  the signatures of the {fractionalization and emergent gauge fields}
remain significantly elusive. These systems exhibit intricate emergent phenomena arising from the interplay of competing interactions at low energy scales. Significant theoretical and experimental challenges remain in fully elucidating the underlying physics across different parameter regimes. 

The introduction of the Kitaev model marked a pivotal breakthrough \cite{kitaev2006anyons}, focusing on spin-1/2 particles arranged on a 2D honeycomb lattice (Fig.\ref{fig:Frustration}(d)). This model is distinguished by its bond-dependent Ising couplings between n.n.\ spins, which can be either FM or AFM in nature. Remarkably, the Kitaev model possesses an exact analytical ground state solution, a property exceedingly rare for {bilinear} quantum spin models in dimensions higher than one. This exact solvability enables rigorous investigations, offering valuable insights into the model's ground state and excitation properties. 

A crucial consequence of the Kitaev model's exact solution is the fractionalization of spin degrees of freedom, wherein spin excitations fragment into emergent Majorana quasiparticles. These Majorana quasiparticles, representing a unique class of fermions characterized by their self-conjugate nature, serve as the system's fundamental excitations. Their exotic properties play a pivotal role in the emergence of composite particles known as non-Abelian anyons when time reversal symmetry is broken.   Non-Abelian anyons exhibit remarkable properties that set them apart from conventional particles, obeying anyonic statistics that fall outside the traditional categorization of bosonic or fermionic behavior.\cite{moore1991nonabelions,nayak2008nonabelians} This remarkable phenomenon provides profound insights into the nature of fractionalized excitations and strongly correlated quantum matter, which have significant implications for the study of quantum materials and the development of novel quantum technologies.

The field of Kitaev QSL gained significant attention after Jackeli and Khaliullin discovered a mechanism for generating Kitaev interactions in honeycomb systems \cite{jackeli2009mott}.
They have proposed a mechanism for realizing the characteristic spin-exchange interactions of the Kitaev model in certain strongly correlated electron systems with strong spin-orbit coupling (SOC), such as $5d^5$ iridates. Consequently, various Kitaev candidates, mainly Ir$^{4+}$ and Ru$^{3+}$ based compounds, have been intensively explored to date. However, it was noted that Kitaev-type QSL is difficult to realize experimentally, because of Heisenberg exchange interaction, which drives the spin system into long-range magnetic order. Experimental efforts in this direction have focused on the honeycomb lattice iridates, including $\alpha$-$A_2$IrO$_3$($A=$ Na and Li) \cite{singh2010antiferromagnetic,singh2012relevance,ye2012direct,gretarsson2013crystal,gegenwart2015kitaev}, as well as 3D analogs $\beta$- and $\gamma$-Li$_2$IrO$_3$, and H$_3$LiIr$_2$O$_6$\cite{kitagawa2018spin}. However, these compounds are structurally complex and the simple applicability of Kitaev physics has also been questioned \cite{mazin20122,mazin2013origin,foyevtsova2013ab,yamaji2014first}.

In 2014, the spin-orbit Mott insulator  $\alpha$-RuCl$_3$  has emerged as a promising candidate material hosting the long-sought Kitaev QSL \cite{plumb2014alpha}, garnering significant attention. The confluence of strong SOC, electron correlations, and the honeycomb lattice structure in this $4d$ transition metal compound renders it a prime platform to explore the intriguing physics predicted by the Kitaev model \cite{kim2015kitaev}.  $\alpha$-RuCl$_3$  exhibits AFM order with a zigzag spin arrangement below the N\'{e}el temperature $T_N \approx 7$ K \cite{sears2015magnetic}. However, theoretical studies suggest that the bosonic magnon excitations characteristic of AFM order at low energies transfer into long-lived fractionalized fermionic quasiparticles, specifically Majorana fermions, at higher energies \cite{rousochatzakis2019quantum,banerjee2016proximate,banerjee2017neutron,do2017majorana}. Furthermore, the application of magnetic fields parallel to the 2D honeycomb plane disrupts the zigzag AFM order, leading to the emergence of a field-induced quantum disordered (FIQD) state above a critical field strength of approximately 8\,T \cite{kubota2015successive,johnson2015monoclinic,zheng2017gapless,leahy2017anomalous,wolter2017field,sears2017phase,yadav2016kitaev,banerjee2018excitations,gass2020field}. Unveiling the physical properties of this FIQD phase could provide crucial insights and signatures of the Kitaev QSL state.
Consequently, $\alpha$-RuCl$_3$ has become a focal point of intensive experimental and theoretical investigations aimed at probing the existence of fractionalized excitations and topological order.

A rich variety of experimental techniques can probe the fractionalized quasiparticle excitations in Kitaev QSL states. Spectroscopic methods such as Raman \cite{glamazda2017relation,sandilands2015scattering,wang2020range,wulferding2019raman,wulferding2020magnon}, inelastic neutron scattering (INS) \cite{banerjee2016proximate,banerjee2017observed,banerjee2018excitations,do2017majorana,balz2021field}, inelastic X-ray scattering (IXS) \cite{li2021giant}, THz \cite{little2017antiferromagnetic,reschke2018sub,reschke2019terahertz,shi2018field,wang2017magnetic,wu2018field,ponomaryov2017unconventional,bera2021review},  nuclear magnetic resonance (NMR) \cite{baek2017evidence,janvsa2018observation,nagai2020two,zheng2017gapless}, muon spin resonance ($\mu$SR) \cite{zi2020muon}, electron spin resonance (ESR) \cite{wellm2018signatures}, ultrasound \cite{hauspurg2024fractionalized}, and scanning tunneling microscope (STM) \cite{kohsaka2024imaging,wang2022direct,zheng2023tunneling,zheng2024insulator,qiu2024evidence} spectroscopies  offer valuable insights into spin-spin correlations, revealing the microscopic magnetic properties of fractionalized quasiparticles. Complementing these approaches, thermodynamic measurements including specific heat \cite{do2017majorana,widmann2019thermodynamic,tanaka2022thermodynamic,wolter2017field,imamura2024majorana}, magnetic torque \cite{leahy2017anomalous,modic2021scale}, magnetic Gr\"{u}neisen parameter \cite{bachus2020thermodynamic}, thermal expansion \cite{gass2020field}, and {magnetocaloric effect} \cite{balz2019finite,li2024magnetocaloric} provide crucial information about very low-energy quasiparticle excitations. Thermal transport measurements further contribute to our understanding by investigating the itinerant nature of these quasiparticles \cite{hentrich2018unusual,hirobe2017magnetic,kasahara2018unusual,leahy2017anomalous,lefranccois2022evidence,czajka2021oscillations,zhang2024thermal}. Of particular significance is the thermal Hall effect in the FIQD state \cite{kasahara2018unusual,kasahara2018majorana,kasahara2022quantized,yokoi2021half,xing2024magnetothermal,czajka2023planar,suetsugu2022evidence,bruin2022robustness,yamashita2020sample,zhang2024stacking}, which can capture the non-trivial Berry phase associated with these quasiparticles, thereby offering unique insights into their topological nature \cite{kitaev2006anyons,katsura2010theory,zhang2024thermal}. This comprehensive suite of experimental techniques enables researchers to probe various aspects of fractionalized excitations in Kitaev QSL states, from their magnetic properties to their thermodynamic behavior and topological characteristics.

Several review articles explore phenomena related to general SOC and Kitaev models \cite{takagi2019concept,trebst2022kitaev,rousochatzakis2024beyond,motome2020hunting,hermanns2018physics}.  Our review aims to complement these works by highlighting diverse interpretations of experimental results and addressing discrepancies in the data obtained through different experimental techniques, discuss challenges encountered in theoretical studies, and explore sample dependencies observed across multiple investigations. By synthesizing these diverse perspectives and findings, we seek to provide a nuanced understanding of the field's current state and stimulate further investigation into these intriguing materials.
Before we survey the Kitaev models and related systems, we will first provide a brief review of well-established 1D QSL and geometrically frustrated higher-dimensional systems.
{Readers primarily interested in the Kitaev QSL may skip the next section and proceed directly to Sec. III. }

\section{Quantum spin liquid candidates in quantum magnets}

\subsection{1D quantum spin chain}

\begin{figure*}[t]
	\includegraphics[clip,width=0.9\linewidth]{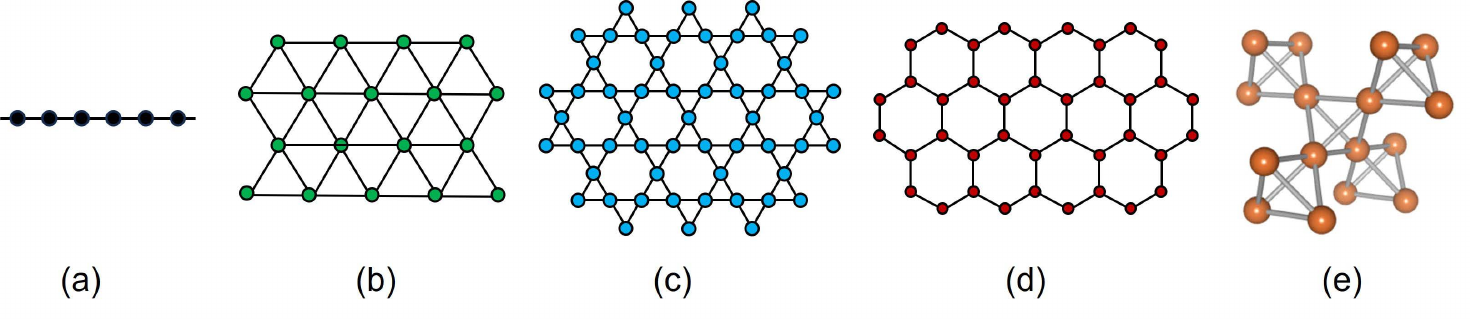}
	\caption{Crystal structures of archetypal QSL candidates. (a) 1D chain, (b) 2D triangular, (c) 2D kagome, (d) 2D honeycomb, and (e) 3D pyrochlore lattices. The triangular, kagome, and pyrochlore lattices exhibit strong geometric frustration.
	{We include the honeycomb lattice here, as Kitaev's bond-dependent interaction leads to spin frustration, even though it does not display geometrical frustration.}}
    \label{fig:Frustration}
\end{figure*}

\begin{figure}[b]
	\includegraphics[clip,width=0.9\linewidth]{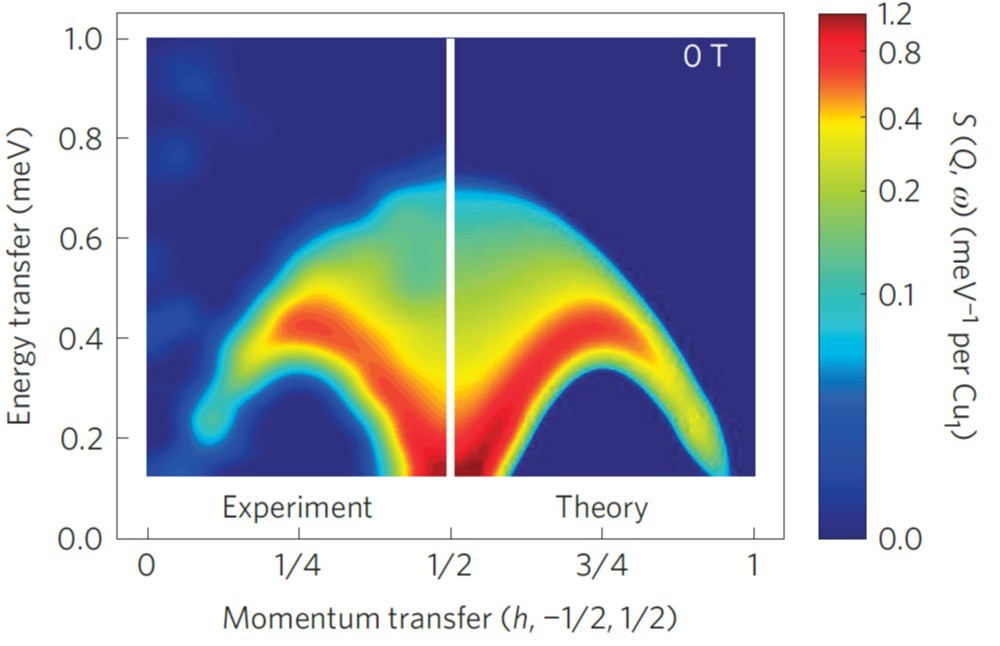}
	\caption{The excitation spectrum of the 1D Heisenberg antiferromagnet CuSO$_4$·5D$_2$O revealed by INS measurements above 3D N\'{e}el  temperature. The elementary excitations are spin-1/2 quasiparticles known as spinons, which are created exclusively in pairs. The observed excitation continuum comprises both two-spinon and four-spinon states. The experimental data are shown alongside exact theoretical calculations demonstrating the characteristic continuous spectra of these multi-spinon contributions \cite{mourigal2013fractional}.
	}
    \label{fig:1Dspinon}
\end{figure}

1D spin chains (Fig.\ref{fig:Frustration}(a)) represent a paradigmatic model system for studying quantum magnetism and the associated fractionalized excitations known as spinons \cite{faddeev1981spin,giamarchi2003quantum}.   
{The des Cloizeaux-Pearson mode represents the lower bound of the two-spinon continuum in the 1D Heisenberg AFM. It has been characterized as the most spin-wave-like of the fractionalized excited states \cite{des1962spin}.}  The spinons carry spin-1/2 but no charge and emerge from the collective behavior of the strongly correlated spins.  The spinons are distinct from the conventional spin excitations $\Delta S=\pm 1$ in ordinary magnets.  The total spin of the excitation in the INS experiment is 1. Consequently, neutrons can create or annihilate pairs of spinons, leading to a continuum spectrum in the INS experiments.   

{An important extension of 1D physics occurs in weakly coupled chain systems. Schulz showed theoretically that even in the antiferromagnetically ordered state of weakly coupled chains, the imaginary part of the susceptibility contains both the spinon continuum characteristic of isolated chains and sharp spin-wave modes at low energies \cite{schulz1996dynamics}. This coexistence has been observed experimentally in materials like KCuF$_3$, where both continuum features and conventional magnon modes appear in the excitation spectrum.}

Although a purely 1D AFM material without magnetic ordering remains theoretical due to unavoidable interchain interactions that lead to magnetic ordering, the characteristic gapless spinon continuum has been experimentally observed in several quasi-1D magnetic systems above the ordering temperature.  Figure \ref{fig:1Dspinon} depicts the INS spectrum of the Cu1 chain spins in the zero-field phase of CuSO$_4$·5D$_2$O, and theoretical two- and four-spinon dynamic structure factor at  $T\sim$100\,mK, above the N\'{e}el temperature to 3D AFM ordering \cite{mourigal2013fractional}. The continuum arises from the fact that spinons, which carry spin-1/2, are deconfined quasiparticles that can propagate independently over a finite distance.  The energy transfer, $E$, and the momentum transfer, ${\bm p}$, of the neutron are shared by two spinon excitations that are created by the neutron spin flip.  According to energy-momentum conservation, we have $E({\bm p})=\omega({\bm k})+\omega({\bm p}-{\bm k})$, where $E$ is the energy transfer,  ${\bm p}$ is the momentum transfer and $\omega({\bm k})$ is the spinon dispersion. This relation implies the presence of an excitation continuum in the INS spectrum.  

The observation of this continuum is a hallmark of the fractionalized spin dynamics in 1D quantum magnets, where the elementary excitations are not conventional bosonic spin waves but rather these deconfined spinon quasiparticles. {The coexistence of continuum and sharp modes observed in systems like $\alpha$-RuCl$_3$ may be analogous to this weakly coupled chain physics, where the interplay between different dimensionalities and coupling strengths leads to complex excitation spectra containing signatures of both fractionalized and conventional magnetic excitations.}

\subsection{Triangular lattice}
In a 2D triangular lattice antiferromagnetic (Fig.\ref{fig:Frustration}(b)), the ground state is frustrated because AFM interactions between the three magnetic moments on each triangular plaquette cannot be simultaneously satisfied.  The QSL state on a triangular lattice has been proposed by considering the RVB state, which is a superposition of different singlet spin configurations \cite{anderson1973resonating,fazekas1974ground}. In this scenario, the resonance between highly degenerate spin configurations leads to a liquid-like wavefunction. When the valence bond is short-ranged, comparable to the n.n.\ distance, the spin excitations are gapped, and a possible Ising ($Z_2$)  topologically ordered state described by the quantum dimer model can be realized \cite{moessner2001resonating}. When the valence bond is long-ranged, the spins are less tightly bound, and consequently, they can be more easily excited into a non-zero spin state, leading to a gapless QSL that consists of fermionic spinons coupled to a $U(1)$ gauge field \cite{lee2005u}.

In the n.n.\ Heisenberg model, however,  it was demonstrated that ground state ($T = 0$) exhibits long-range order characterized by a three-sublattice N\'{e}el state with 120$^{\circ}$ spin structure. This occurs even in spin-1/2 systems, where quantum effects are most pronounced \cite{capriotti1999long,white2007neel}.  Further frustration effect, such as the inclusion of ring exchange, is required to obtain QSL states \cite{misguich1999spin,liming2000neel}. 

The first QSL state in a triangular lattice was discovered in a low-density solid $^3$He film with nuclear spin-1/2, adsorbed on a graphite surface \cite{masutomi2004gapless}. In this system, the hard-core potential between $^3$He atoms facilitates higher-order multiple spin exchange processes. For bulk materials, triangular QSL candidates encompass a broad spectrum of compounds, including both organic and inorganic systems, which are characterized as Mott insulators exhibiting higher-order ring-exchange interactions. {Some of the most promising candidate materials are catalogued in Table\,\ref{tab:triangle}.}

{
\begin{table*}
\footnotesize
 \caption{\red{Spin-1/2 quantum spin liquid candidates on frustrated lattices with Heisenberg interactions: 2D triangular antiferromagnets.}}
\begin{tabular}{p{2.5cm}p{1.6cm}p{1.6cm}p{2.5cm}p{2.2cm}p{1.5cm}p{5.0cm}}
	\hline\hline
	\textbf{Material} &\textbf{Magnetic} &\textbf{Specific } & \textbf{~~Thermal } & \textbf{Susceptibility} & \textbf{Exchange} &  \textbf{~~~~~~~~Notes} \\
	&\textbf{~ordering} &\textbf{~~heat} & \textbf{conductivity} &  & ~~~~(K) &   \\
	\hline 
\multicolumn{7}{l}{\hspace{6cm}\textbf{2D triangular antiferromagnets}} \\

	\textbf{{\footnotesize EtMe$_3$Sb-[Pd(dmit)$_2$]$_2$}}  & $<$19\,mK&Finite $\gamma$ &  Finite $\kappa/T|_{T\rightarrow 0}$ & Pauli-like& $\sim$200&  Gapless QSL \cite{itou2008quantum,watanabe2012novel,yamashita2011gapless}.  The presence \cite{yamashita2010highly} or absence \cite{bourgeois2019thermal,ni2019absence} of $\kappa/T|_{T\rightarrow 0}$ had been controversial, but  its value has been later shown to be strongly dependent on the cooling rate \cite{yamashita2020presence}. \\
	\addlinespace

	\textbf{{\footnotesize $\kappa$-(BEDT-TTF)$_2$Cu$_2$(CN)$_3$}} & $<$32\,mK&Finite $\gamma$ & Zero $\kappa/T|_{T\rightarrow 0}$  &  Pauli-like & $\sim$250 &  Initially claimed as a QSL, recently suggested to be a gapped valence bond glass.\cite{shimizu2003spin,yamashita2008thermodynamic,yamashita2009thermal,miksch2021gapped}\\

\textbf{{\footnotesize $\kappa$-H$_3$(Cat-EDT-TTF)$_2$}} & $<$50\,mK&Finite $\gamma$ &  Finite $\kappa/T|_{T\rightarrow 0}$ & Pauli-like & $\sim$80-100  & Gapless QSL realized by proton fluctuations. \cite{isono2014gapless,shimozawa2017quantum}\\

\textbf{1T-TaS$_2$} & $<$20\,mK&Finite $\gamma$& Finite $\kappa/T|_{T\rightarrow 0}$ & Pauli-like &  $>1000$  &  Gapless QSL. Magnetic continuum by INS and NMR. \cite{law20171t,ribak2017gapless,klanjvsek2017high,murayama2020effect}\\

\textbf{YbMaGaO$_4$} & $<$40\,mK&$C\propto T^{0.74}$ & Finite $\kappa/T|_{T\rightarrow 0}$ & Semi-Pauli & $\sim$1.5 &  Gapless QSL,  Magnetic continuum by INS, Disorder due to  Mg$^{2+}$/Ca$^{2+}$ site mixing.\cite{shen2016evidence,rao2021survival,li2015gapless,li2016muon,li2017crystalline,zhu2017disorder,parker2018finite}
\\

\textbf{NaYbS$_2$} & $<$50\,mK&Finite $\gamma$ & Zero $\kappa/T|_{T\rightarrow 0}$   & Pauli-like & $\sim$2-4  & Gapless QSL. Magnetic continuum by INS. No structural disorder. \cite{dai2021spinon,li2024thermodynamics}\\

\textbf{NaYbSe$_2$} & $<$50\,mK&Finite $\gamma$ & Zero $\kappa/T|_{T\rightarrow 0}$   & Contro. & $\sim$6.5 &  Gapped or gapless is controversial. Magnetic continuum by INS. No structural disorder. \cite{li2024thermodynamics} \\

\textbf{KYbSe$_{2}$} & 290\,mK & Finite $\gamma$ & Unknown  & Curie-like & $\sim$1-3  & Proximity Gapped $Z_2$ QSL with $J_2/J_1$=0.047. Magnetic continuum by INS, 1/3 magnetization plateau.\cite{scheie2024proximate},\\

\textbf{Na$_2$BaCo(PO$_4$)$_2$} & 148\,mK & \begin{tabular}{c}Finite~$\gamma$ \\ ($>T_N$)\end{tabular} & \begin{tabular}{c}Finite $\kappa/T|_{T\rightarrow 0}$ \\ ($>T_N$)\end{tabular}  & Semi-Pauli & $\sim$1.7  & Gapless spin liquid above $T_N$. Spin supersolid state. Giant magnetocaloric effect. \cite{zhong2019strong,li2020possible,gao2022spin,xiang2024giant,sheng2022two,lee2021temporal}\\
\hline\hline

\end{tabular}
\label{tab:triangle}
	\end{table*}
}


{INS measurements in some of these materials reveal broad continuum excitations, indicating spin fractionalization characteristic of QSL states. Many of these materials exhibit finite spin susceptibility $\chi_s$, which remains nearly temperature-independent (i.e., Pauli-like) at very low temperatures. Additionally, a linear term $\gamma T$ emerges in the low-temperature specific heat. Furthermore, a finite residual linear term in thermal conductivity, denoted as $\kappa_0/T \equiv \kappa/T(T\rightarrow 0)$, has been observed. These findings indicate the emergence of gapless spin excitations reminiscent of those in paramagnetic metals with a Fermi surface, despite these materials being insulators. The mobile and gapless excitations have been interpreted in terms of fermionic spinons forming a spinon Fermi surface \cite{lee2005u,motrunich2005variational,motrunich2006orbital,lee2007amperean,potter2012quantum}}.

Despite these extensive studies, definitive proof of a QSL state in the triangular lattice remains elusive, and the nature of the ground state in many of these materials continues to be a subject of active research and debate.

 \subsection{Kagome lattice}

\begin{table*}
\footnotesize
 \caption{\red{Spin-1/2 quantum spin liquid candidates on frustrated lattices with Heisenberg interactions: 2D Kagome, 2D honeycomb and 3D pyrochlore antiferromagnets.}}
\begin{tabular}{p{2.6cm}p{1.6cm}p{1.6cm}p{2.5cm}p{2.2cm}p{1.5cm}p{5.0cm}}
\hline\hline
	\textbf{Material} &\textbf{Magnetic} &\textbf{Specific } & \textbf{~~Thermal } & \textbf{Susceptibility} & \textbf{Exchange} &  \textbf{~~~~~~~~Notes} \\
	&\textbf{~ordering} &\textbf{~~heat} & \textbf{conductivity} &  & ~~~~(K) &   \\
	\hline 

\multicolumn{7}{l}{\hspace{6cm}\textbf{2D Kagome antiferromagnets}} \\
\textbf{Volborthite
} & $\approx$1\,K &  \begin{tabular}{c}Finite $\gamma$ \\ ($>T_N$)\end{tabular}   & Zero $\kappa/T|_{T\rightarrow 0}$   &\begin{tabular}{c}Pauli-like\\ ($>T_N$)\end{tabular} & $\sim90$ & Cu$_3$V$_2$O$_7$(OH)$_2\cdot$2H$_2$O. Distorted kagome lattice (isosceles triangles). 1/3 magnetization plateau.\cite{hiroi2001spin}\\	

\textbf{V\'esigni\'eite
} & $\approx$9\,K & Unknown  & Unknown  & Complicated & $J_3\sim$9 & BaCu$_3$V$_2$O$_8$(OH)$_2$. Almost perfect kagome lattice ($<$1\% distortion). Orbital frustration. Dominant AFM third-neighbor exchange interaction $J_3$.\cite{okamoto2009vesignieite}\\	

\textbf{Y-Kapellasite
} & 2.2\,K &  \begin{tabular}{c}Finite $\gamma$ \\ ($>T_N$)\end{tabular}  & Unknown  & Unknown & $\sim$65& Y$_3$Cu$_9$(OH)$_{19}$Cl$_8$. Slightly distorted kagome lattice.  Pressure may realize a fully gapped QSL state. \cite{hering2022phase,chatterjee2023spin,chatterjee2025spin}\\ 

\textbf{Herbertsmithite} & ~$<20$\,mK & \begin{tabular}{c}$H/T-$\\ scaling\end{tabular} & Zero $\kappa/T|_{T\rightarrow 0}$ & \begin{tabular}{c}Excitation\\ gap debate\end{tabular} & $\sim$200& ZnCu$_3$(OH)$_6$Cl$_2$. Perfect kagome lattice. Cu/Zn antisite mixing.  $C$ shows $T/H$ scaling collapse.  Magnetic continuum by INS. Gapped vs gapless ground state debated.\cite{mendels2010quantum,norman2016colloquium,mendels2007quantum,helton2007spin,han2012fractionalized,shores2005structurally,freedman2010site,kimchi2018scaling,murayama2022universal,PhysRevB.76.132411,PhysRevLett.104.147201,fu2015evidence,khuntia2020gapless,wang2021emergence}\\	

\textbf{YCOB} & $<50$\,mK & $C\propto T^2$ & Zero $\kappa/T|_{T\rightarrow 0}$   & Pauli-like & $\sim$80& 
YCu$_3$(OH)$_{6}$Br$_2$[Br$_x$(OH)$_{1-x}$]($x\approx$0.5). Perfect kagome lattice. Absence of intersite mixing between  Cu$^{2+}$ and  Y$^{3+}$.  Magnetic continuum by INS.  Gap structure (gapped/gapless/Dirac cone) is controversial. 1/9 and 1/3 magnetization plateaus.\cite{sun2016perfect,chen2020quantum,PhysRevB.105.024418,PhysRevB.105.L121109,PhysRevB.106.L220406,lu2022observation,zeng2024spectral,suetsugu2024emergent,jeon2024,schulenburg2002macroscopic,nishimoto2013controlling,okuma2019series,zheng2023unconventional,suetsugu2024gapless}\\	

	\hline 

\multicolumn{7}{l}{\hspace{6cm}\textbf{2D homeycomb antiferromagnets}} \\

\textbf{YbBr$_3$} & $<100$\,mK & Unknown & Unknown  & Broad peak& $\sim$8& Undistorted honeycomb lattice with competing $J_1$ and $_2$ interactions. Broad continuum by INS exhibits QSL-like features and plaquette fluctuations.
\cite{wessler2020observation} \\

\hline
\multicolumn{7}{l}{\hspace{6cm}\textbf{3D pyrochlore lattice}} \\

\textbf{Pr$_2$Zr$_2$O$_7$} & $<100$\,mK & Complicated &  \begin{tabular}{c}Enhancement \\ of $\kappa/T (T\rightarrow 0)$\end{tabular}  &   \begin{tabular}{c}Nonzero\\   ($T\rightarrow 0$) \end{tabular} & $\approx$0.8&  Pr$^{3+}$ corner-sharing tetrahedra. Quantum spin ice. Strong quantum fluctuations. Exotic excitations from thermal conductivity.\cite{kimura2013quantum,tokiwa2018discovery}\\	

\textbf{Ce$_2$Zr$_2$O$_7$} & $<35$\,mK & Schottky-type anomaly & Unknown  &   \begin{tabular}{c}Weak\,AFM \\correlations \end{tabular}& $\sim$1&  Octupolar $\pi
$-flux quantum spin ice. Magnetic continuum by INS. Spinons: Schottky anomaly in $C$. Visons/photons: INS signatures.\cite{Gao2019NaturePhysics,Sibille2015PRL,Smith2022PRX,Gaudet2019PRL,Gao2024ArXiv,Bhardwaj2022npjQM,smith2023quantum} \\	
		\hline\hline
\end{tabular}
\label{tab:kagome}
	\end{table*}

The AFM spin-1/2 Heisenberg model on a 2D kagome lattice with a corner-sharing triangular structure also represents one of the simplest models, exhibiting strong frustration and quantum fluctuations (Fig.\ref{fig:Frustration}(c)).     In a triangular lattice with AFM interactions, the frustration is partially relieved when two out of the three spins within a triangular unit are aligned antiparallel, while the third spin is neglected.  This configuration in a triangular lattice reduces the frustration compared to a scenario where all three spins are equally frustrated.  On the other hand, in the kagome lattice, all three spins cannot be antiparallel at the same time in units of the smallest triangle.   Even when considered in larger units, frustration always remains. Consequently, due to its geometric features, the kagome lattice is inherently unable to avoid frustration, making it a promising candidate for realizing novel quantum states.  In fact, the synergistic interplay of geometrical frustration, AFM interactions, and the quantum nature of $S = 1/2$ spins is believed to stabilize a highly entangled QSL state.  

Despite longstanding intensive study, elucidating the ground state of the quantum kagome antiferromagnets has been one of the most vexing issues in quantum magnetism \cite{sachdev1992kagome,ran2007projected,yan2011spin,messio2012kagome,jiang2012identifying,depenbrock2012nature,iqbal2013gapless,he2017signatures,lauchli2019s}. The primary reason for the difficulty in identifying the ground state is the existence of a plethora of competing states that are exceedingly close in energy, leading to a wide range of theoretical proposals for the potential ground state. Indeed, the QSL states that may arise are manifold, encompassing the gapped $Z_2$ spin liquid \cite{sachdev1992kagome,yan2011spin,jiang2012identifying,lauchli2019s} and the gapless $U(1)$ Dirac spin liquid \cite{ran2007projected,iqbal2013gapless,he2017signatures}, and the specific QSL state manifested in the kagome AFM continues to elude precise identification.

{The fundamental questions in kagome antiferromagnets concern two key aspects:
\begin{enumerate}
\item Zero-field spin gap:
Whether the spin-1/2 Heisenberg antiferromagnet on the kagome lattice exhibits a gapped or gapless ground state remains controversial. Theoretical proposals include gapped quantum spin liquids and gapless algebraic spin liquids, with numerical studies yielding conflicting results.
\item Fractional magnetization plateaus:
The theoretically predicted plateau at 1/9 saturation magnetization represents a quantum many-body phenomenon where specific spin configurations are stabilized by the interplay of exchange interactions, geometric frustration, and magnetic field.
\end{enumerate}
Experimental verification is challenging due to material imperfections including site dilution, interlayer coupling, and exchange anisotropy that deviate from the ideal kagome Heisenberg model. Some of the most promising candidate materials are catalogued in Table~\ref{tab:kagome}.}



{Herbertsmithite \cite{mendels2007quantum,helton2007spin} and the recently discovered YCOB \cite{chen2020quantum,PhysRevB.105.024418,PhysRevB.105.L121109,PhysRevB.106.L220406,lu2022observation,PhysRevB.105.L121109} represent exceptional platforms for investigating the intrinsic ground state of the kagome QSL. Both materials feature geometrically perfect two-dimensional kagome lattices of Cu$^{2+}$ ions ($S$=1/2) and exhibit no magnetic order down to the lowest accessible temperatures, despite substantial AFM exchange interactions. The broad continuum observed in the magnetic excitation spectra by INS in both systems has been interpreted as a hallmark of fractionalized excitations characteristic of QSL states \cite{han2012fractionalized,zeng2024spectral}.


In contrast, YCOB exhibits negligible intersite mixing between Cu$^{+2}$ and Y$^{+2}$ ions, enabling more direct access to the intrinsic ground state properties and elementary excitations of the kagome QSL.  It is a hot debate for the spin gap excitations, whether it is gapped, gapless \cite{suetsugu2024gapless}, or \red{it has Dirac cones} \cite{zeng2024spectral}. 

 When subjected to strong magnetic fields, YCOB exhibits plateau-like magnetization anomalies at 1/9 and 1/3 of the fully polarized value \cite{suetsugu2024emergent,jeon2024}. While the 1/3 plateau is widely accepted as a genuine plateau representing a quantum phase \cite{schulenburg2002macroscopic}, the origin of the 1/9 anomaly, which is also of purely quantum origin, had been largely unexplored, including debates about its presence or absence \cite{nishimoto2013controlling,okuma2019series}. The emergence of these plateaus, which have not been reported in the kagome material Herbertsmithite \cite{han2014thermodynamic}, demonstrates significantly reduced randomness in YCOB. It should be noted that there is an alternative interpretation for the 1/9 plateau, suggesting it may not be a plateau but rather part of quantum oscillations \cite{zheng2023unconventional}.}

\subsection{Honeycomb lattice}

Focusing on Heisenberg-type interactions, the n.n.\ Heisenberg on the honeycomb lattice (Fig.\,\ref{fig:Frustration}(d)) is not frustrated.  However, quantum fluctuations are inherently stronger in the honeycomb lattice than in the square lattice due to the lower coordination number of each spin.  This geometric feature enhances quantum fluctuations through reduced local magnetic connectivity. Theoretical frameworks predict a quantum phase transition from a N\'{e}el ordered ground state to a quantum entangled state when the ratio of next n.n.\ to n.n.\ exchange interactions surpasses a critical threshold. The precise nature of this quantum entangled ground state remains debated, with theoretical predictions suggesting either a QSL \cite{merino2018role,wang2010schwinger,Ralko20203} or a plaquette valence bond crystal \cite{albuquerque2011phase,ganesh2013deconfined}, each exhibiting distinct magnetic excitation spectra comprising spinons \cite{albuquerque2011phase}, rotons \cite{ferrari2020dynamical}, or plaquette fluctuations \cite{ganesh2013plaquette}. 

{The trihalide compound YbBr$_3$ exemplifies this phenomenon (see Table~\ref{tab:kagome}). Wessler {\it et al.}\ demonstrate that the INS excitation continuum displays features consistent with both QSL behavior and plaquette-type fluctuations, indicating a potential quantum ground state.\cite{wessler2020observation}}


\subsection{3D pyrochlore lattice}
The most well-known example of the 3D geometrical frustrations is the pyrochlore lattice (Fig.\ref{fig:Frustration}(e)).   Rare-earth pyrochlore oxide, where magnetic ions form an array of corner-sharing tetrahedra, is one of the perfect venues to host QSLs.  The magnetic moments residing at the vertices of these tetrahedral units exhibit significant geometrical frustration, producing a range of exotic magnetic excitations.  For classical Ising spins with a strong easy-axis anisotropy, frustration results in macroscopically degenerate spin-ice states \cite{ramirez1999zero}. A spin-flip from the classical spin-ice manifold can be described as a pair creation of magnetic monopoles, which interact with each other through the static Coulomb interaction \cite{castelnovo2008magnetic,jaubert2009signature,morris2009dirac,henley2010coulomb,fennell2009magnetic}.   In pyrochlore quantum magnets, additional spin interactions such as a transverse exchange term endow the spins with quantum fluctuations, which may lead to a QSL state by lifting macroscopic degeneracy { - a state known as quantum spin ice featuring an emergent $U(1)$ gauge field} \cite{hermele2004pyrochlore,ross2011quantum,benton2012seeing,shannon2012quantum,gingras2014quantum,Huang2014QSI}.  

{Among these materials, the Pr$^{3+}$ and Ce$^{3+}$ pyrochlore compounds Pr$_2$Zr$_2$O$_7$ and Ce$_2$Zr$_2$O$_7$, both with effective spin $S=1/2$, have emerged as promising candidates for quantum spin ice (see Table~\ref{tab:kagome}). These are positioned as the first 3D QSL materials with minimal magnetic/non-magnetic ion disorder. Theoretical studies suggest that the ground state of the pyrochlore Ce$_2$Zr$_2$O$_7$ is a rare example of {octupolar} $\pi$-flux quantum spin ice \cite{Gao2024ArXiv,Desrochers2024PRL}. Due to the octupolar nature of the moment, it was suggested that the quantum spin ice is more stable than its counterparts \cite{Bhardwaj2022npjQM}.  


Recently, the key exotic quasiparticles in quantum spin ice states have been reported. The striking enhancement of the thermal conductivity in Pr$_2$Zr$_2$O$_7$ has been attributed to the emergent photon \cite{tokiwa2018discovery}, which is characterized by the linearly dispersing magnetic excitation near zero energy arising from emergent $U(1)$ quantum electrodynamics \cite{gingras2014quantum, hermele2004pyrochlore,benton2012seeing}.  Specific heat and polarized neutron scattering experiments on Ce$_2$Zr$_2$O$_7$ revealed quasielastic magnetic excitations near zero energy at very low temperatures, attributed to possible emergent photons, along with possible signatures of spinons and visons at higher energies.\cite{Gao2024ArXiv, Gao2019NaturePhysics,Gaudet2019PRL,Bhardwaj2022npjQM,smith2023quantum}.}


\section{Kitaev model}
\label{Kitaev}
\begin{figure}[t]
	\includegraphics[clip,width=1\linewidth]{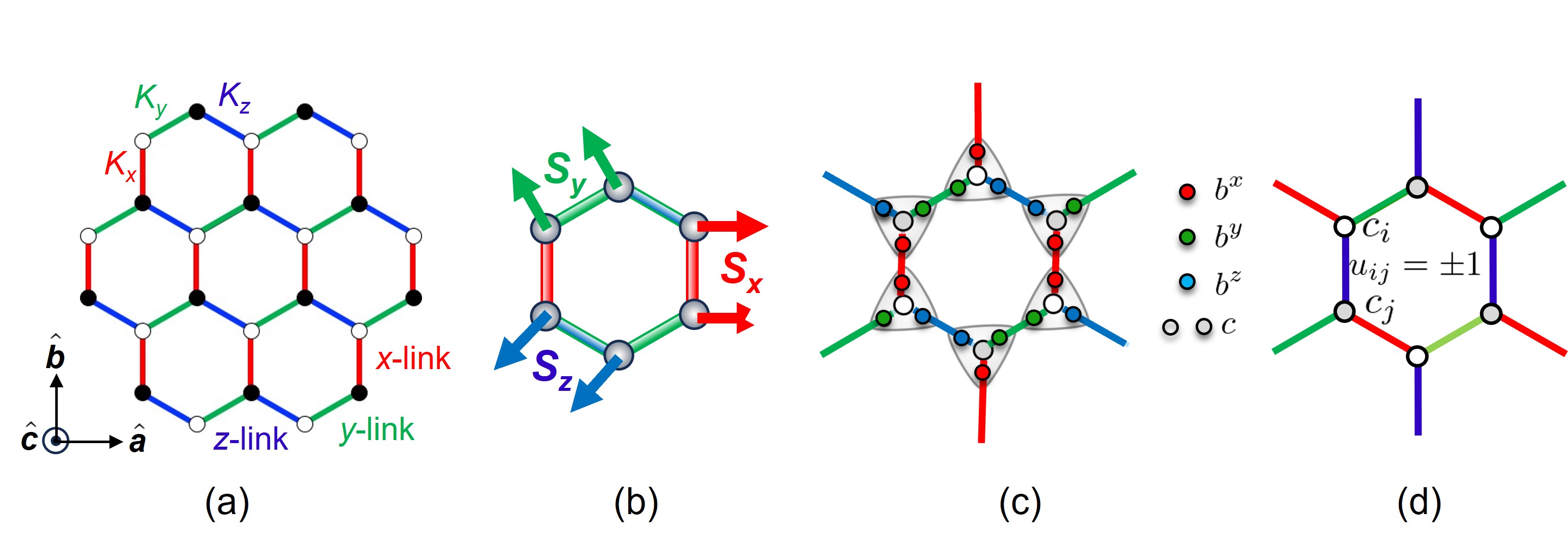}
	\caption{(a) Structural schematic of a honeycomb lattice with crystallographic directions. The lattice comprises two interpenetrating triangular sublattices, where filled and open circles denote Ru atomic sites on distinct sublattices. Bond orientations are designated as $x$-, $y$-, and $z$-links, with corresponding Kitaev exchange interactions $K_x$, $K_y$, and $K_z$ acting along these bonds, respectively. (b) Spin orientations are depicted perpendicular to their respective exchange bonds, illustrating ferromagnetic Kitaev interactions. The system exhibits geometric frustration due to spin-orbit-coupled bond-directional interactions, where Ising-like exchange couplings are constrained by bond orientation. This configuration prevents simultaneous energy minimization across all exchange bonds, a fundamental frustration that persists in both quantum and classical spin systems. (c) Schematic representation of the honeycomb Kitaev model with bond-directional exchange couplings $K_x$, $K_y$, and $K_z$. The model's exact solution is achieved through fractionalization into four Majorana fermions: three localized species (depicted by red, blue, and green circles) and one itinerant species (represented by open and gray circles). (d) The model simplification via gauge field transformation, where itinerant Majorana fermions combine to form a static $Z_2$ gauge field ($u_{ij}$), leaving a system of non-interacting itinerant Majorana fermions (open and gray circles).
 }
 \label{fig:Model}
\end{figure}

Now let us turn to our main topic. We first briefly review the Kitaev model and its ground state. 
The Kitaev model was introduced by A. Kitaev in 2006 \cite{kitaev2006anyons}, which is described by
\begin{align}	\mathcal{H}_K=&-K_x\sum_{<i,j>\in x}S_i^xS_j^x	-K_y\sum_{<i,j>\in y} S_i^yS_j^y \\ \nonumber	&-K_z\sum_{<i,j>\in z} S_i^zS_j^z.
\end{align}
Here,  $x$, $y$, and $z$ denote three different n.n.\ bonds of the honeycomb lattice, as shown in Fig.\,\ref{fig:Model}(a).  $K_x$, $K_y$, and $K_z$ are exchange constant and  $\bm{S}=(S_i^x, S_i^y, S_i^z)$ represents spin-1/2 operator at a site $i$.    Since $S_i^xS_j^x$, $S_i^yS_j^y$, and $S_i^zS_j^z$ do not commute each other,  when the spins are aligned along one direction (say $x$-axis),  the bond energies along the other directions ($y$ and $z$-axes) are not lowered.  Then the spins cannot satisfy three different configurations simultaneously. Thus, the Kitaev interaction gives rise to the exchange frustration (see Fig.\,\ref{fig:Model}(b)).

It was shown that the following plaquette operator $W_p$ defined on a hexagon commutes with the Hamiltonian:
\begin{equation}
    W_p = \sigma_1^x \sigma_2^y \sigma_3^z \sigma_4^x  \sigma_5^y \sigma_6^z,
\end{equation}
where $\sigma_{i}$ are the Pauli matrices on site $i$ ($S_i^\gamma = \frac{1}{2} \sigma_i^\gamma$ where we set $\hbar =1$), $p$ labels the plaqutee, and different $W_p$ operators commute with each other.
This allows to divide the Hilbert space into sectors of $W_p$ eigenspaces, and the problem is greatly simplified.
Replacing the spin operators with four Majorana operators such as 
$\sigma^\gamma_i = i b^\gamma_i c_i$ where $\gamma = x,y,z$,
and Majorana operators $b_i$ and $c_i$ satisfy $c_i^2=1$
and $c_i c_j = - c_j c_i$ if $i \neq j$,
the model is rewritten as:
\begin{equation}
    {\tilde H} = \sum_{\langle ij \rangle \in \gamma} \frac{i}{4} K_\gamma {\hat u}_{ij} c_i c_j,
\end{equation}
where ${\hat u}_{ij} = b^{\gamma}_i b^{\gamma}_j$ for the corresponding n.n.\ $\gamma$-bond (see Figs.\,\ref{fig:Model}(c) and (d)).
Since ${\hat u}_{ij}$ operators also commute with ${\tilde H}$, the Hilbert space splits into eigenspace of ${\hat u}_{ij}$. Once we obtain the eigenspace ${\hat u}_{ij}$, the operator is replaced by a number, and the problem corresponds to free Majorana fermions. However, the Hilbert space of ${\tilde H}$ is enlarged due to four MFs introduced per spin-1/2, and one has to find the physical subspace.  To remove such redundancy and obtain the physical subspace $|\psi_{phy}\rangle$, {one has to project onto the subspace satisfying ${\hat D}_j = b_j^x b_j^y b_j^z c_j = 1$ on each lattice site \cite{kitaev2006anyons}.}

\begin{figure}[t]
	\includegraphics[clip,width=0.8\linewidth]{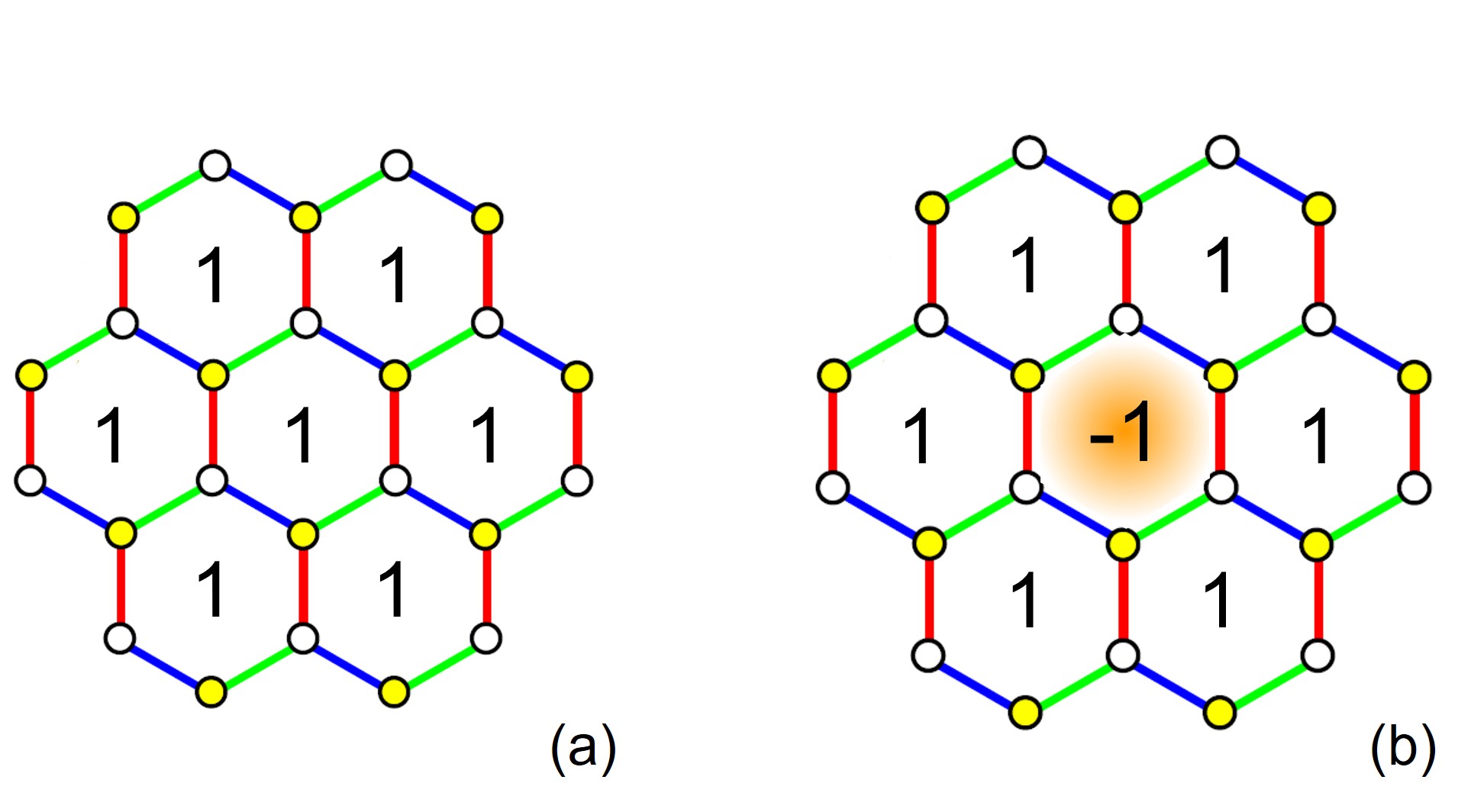}
	\caption{(a) Ground state. $w_p=1$ for all hexagons. (b) A \red{vison} excitation  ($w_p=-1$).  The \red{vison} is a $Z_2$ magnetic flux excitation that occurs when the $Z_2$ gauge field ($u_{ij}$) is flipped from +1 to -1 on a plaquette. It's a gapped, localized excitation that can be understood as a violation of the plaquette operator's ground state condition.
	}
    \label{fig:Vison}
\end{figure}

\begin{figure}[b]
	\includegraphics[clip,width=\linewidth]{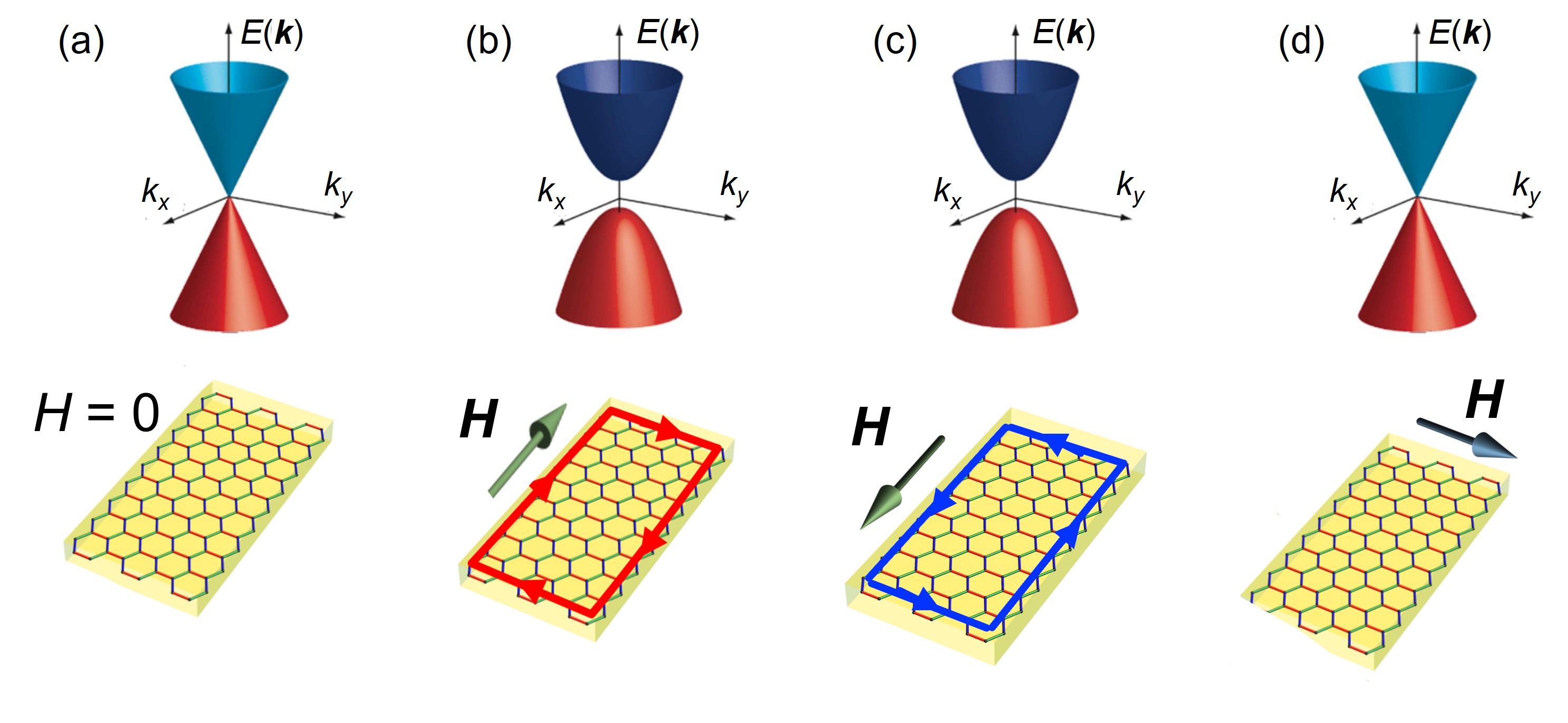}
	\caption{(a) In the absence of an external magnetic field, itinerant Majorana fermions on a honeycomb lattice manifest Dirac cone dispersions at the $K$- and $K'$- points (corners) of the Brillouin zone, preserving particle-hole symmetry. 
    (b)(c) Upon application of a magnetic field, these Dirac points generally develop a topologically non-trivial gap, though this response shows marked directional dependence.  When {\boldmath $H$}  is applied within the 2D plane along the antibonding (zigzag) direction of the honeycomb lattice ($a$-axis in the $C2/m$ notation), the system develops an energy gap.  
    The chirality of the resulting state depends critically on the field orientation: for $\bm{H} || a$,   clockwise chiral thermal edge currents are induced within the honeycomb plane (a), while for $\bm{H} || -a$  counterclockwise circulation is generated (b). 
    (d) The system exhibits markedly different behavior when {\boldmath $H$} is aligned parallel to the bond (armchair) direction ($b$-axis). In this configuration, the Dirac cone structure remains intact, and no chiral edge currents emerge.  These demonstrate the profound influence of field orientation on the system's topological properties.
	}
    \label{fig:Dirac3}
\end{figure}

\begin{figure}[b]
	\includegraphics[clip,width=0.35\linewidth]{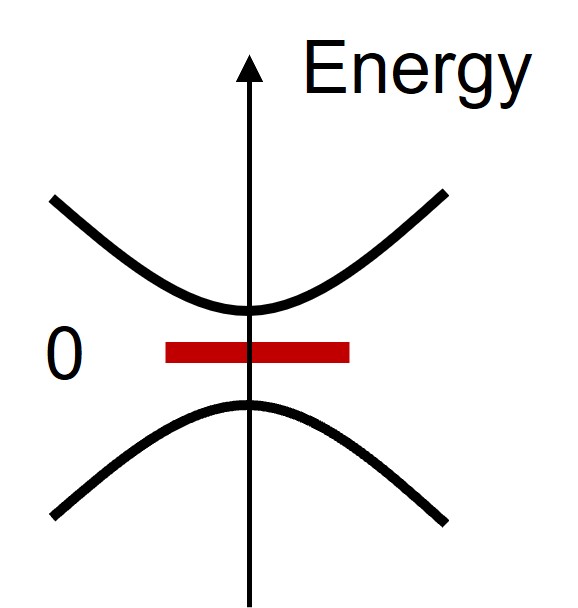}
	\caption{A Majorana zero mode (red line). The non-abelian statistics emerges from the bound states made of vison excitations and Majorana zero modes.
	}
    \label{fig:Majoranazero}
\end{figure}

The ground state corresponds to $w_p=1$ for all hexagons.
This is called the vortex-free field configuration and $w_p =-1$ is regarded as a $Z_2$ vortex, as shown in Figs.\,\ref{fig:Vison}(a) and (b). 
The ground-state phase diagram of Kitaev magnets consists of a gapped $A$-phase and a gapless $B$-phase, depending on the magnetic anisotropy. For isotropic exchange interactions ($K_x \approx K_y \approx K_z$), the gapless $B$-phase is stabilized as the ground state. In this QSL phase, the spin degrees of freedom fractionalize into itinerant Majorana fermions and $Z_2$ vortices. The itinerant Majorana fermions form gapless Dirac cones with linear dispersion at the $K$ points of the honeycomb Brillouin zone. In contrast,  $Z_2$ vortices, termed visons, are gapped.

The application of external magnetic fields induces profound modifications to the excitation spectrum, resulting in the opening of a gap in the gapless Dirac cone dispersion, as shown in Figs.\,\ref{fig:Dirac3}(a)-(d). This time-reversal symmetry-breaking perturbation drives a topological phase transition, giving rise to a chiral spin liquid state characterized by non-zero Chern invariants, termed the Majorana Chern insulator.

This feature can be captured by an Ising topological order, which has two types of fractionalized excitations, non-Abelian anyon $\sigma$ and Majorana fermion \cite{kitaev2006anyons}.
They obey the following fusion rule:
\begin{equation}
    \psi \otimes \psi = {\rm I}, \;\; \sigma \otimes \sigma = {\rm I} \oplus \psi, \;\; \psi \otimes \sigma = \sigma,
\end{equation}
where $\psi$ represents a fermion, I  the vacuum, and Ising anyon $\sigma$ a vortex containing an unpaired zero Majorana mode, as depicted in Fig.\,\ref{fig:Majoranazero}. The $\sigma$ {is a non-Abelian anyon, as the fusion of two $\sigma$'s } produces more than one outcome. The presence of the Majorana zero mode in the vortex gives rise to the emergence of non-Abelian anyonic quasiparticles, which exhibit non-Abelian braiding statistics - a profound manifestation of the intrinsic non-local quantum correlations inherent to the system.

The non-Abelian anyon $\sigma$ can emerge in a variety of systems, including the Kitaev spin model under a small magnetic field in certain directions, $p_x$ + i$p_y$ superconductors, and the Moore-Read Pfaffian state in fractional quantum Hall systems. 
For the $p_x + i p_y$ spin-triplet superconductors, it was shown that a 1/2-quantum vortex carries an unpaired Majorana mode which obeys the non-Abelian statistics \cite{Ivanov2001PRL}. Among the fractional quantum Hall states, Moore-Read Pfaffian state, which can be viewed as the $p_x + i p_y$ paired state of spin-polarized composite fermions \cite{moore1991nonabelions} also exhibits the non-Abelian statistics, as the vortex carries an unpaired Majorana mode. It was shown numerically that the Moore-Read \red{Pfaffian} is a strong candidate for the $\nu=5/2$ quantum Hall state \cite{Morf1998PRL}.

Ising topological order is constrained by the symmetry of the system. This can be understood through the analogy of a spinless $p_x + i p_y$ superconductor after gauging the $Z_2$ fermionic parity of the Bogoliubov quasiparticles. 
 Under the time reversal operator, the pairing function maps to $p_x - i p_y$, and the chirality of the Majorana fermion edge mode is reversed. If the system possesses a two-fold rotational $C_2$ symmetry, say, about 
${\hat y}$-axis, then the pairing function maps to $-p_x + i p_y$ and the chirality of the edge mode is also reversed. Thus when the twofold rotational $C_2$ (or symmetry-equivalent mirror) symmetry is present, even though the time-reversal symmetry is broken, Ising topological order is forbidden unless the $C_2$ symmetry is spontaneously broken. For instance, when the external field is aligned to preserve $C_2$ symmetry, the Ising topological order does not emerge. This suggests that as the magnetic field angle is varied---assuming no spontaneous breaking of $C_2$ symmetry---a phase transition between Ising topological orders with opposite Chern numbers may occur as the field angle crosses the $C_2$ symmetry direction (see Figs.\,\ref{fig:Dirac3}(b)-(d)).

In the chiral spin liquid state in magnetic fields, while the Majorana excitations are fully gapped in the bulk, the 1D Majorana currents that remain gapless emerge at the edge. The chiral Majorana edge current is topologically protected, similar to dissipationless charge currents along the edges of a sample quantum Hall effect in 2D electron gases.  Since the Majorana fermions are charge-neutral, they do not carry charge but can carry heat. Therefore, the Majorana edge currents can be detected by the thermal transport measurements. 
It should be noted that because the degrees of freedom of the Majorana fermion \red{are a half of those of} conventional fermions, the quantized value of the thermal Hall conductance is half of that observed in the integer quantum Hall effect. 
Thus, one can view a chiral edge mode as a consequence of the Ising topological order. The edge mode exhibits a quantized thermal Hall conductance $\kappa_{xy}=1/2$ in units of $(\pi/6) (k_B^2 T/\hbar)$, as shown by arrows in Figs.\,\ref{fig:Dirac3}(b) and (c). This is a smoking gun signature of the Ising topological order. A topological transition is expected to accompany the sign change of $\kappa_{xy}$ as the external field sweeps through the angle aligned with the $C_{2b}$ symmetry (i.e., $b$-axis). We will revisit these symmetry constraints later in our discussion of the thermal Hall conductivity measurements.

\section{Kitaev physics in real materials}
There are review articles covering topics ranging from Kitaev physics to candidate materials \cite{Krempa2014ARCMP,Rau2016ARCMP,winter2016challenges,takagi2019concept,rousochatzakis2024beyond}, and this review article complements these previous works. However, the special focus of this review is on experimental progress on $\alpha$-RuCl$_3$. Before reviewing the experimental developments in recent years, we will first briefly cover the theoretical background to understand how the Kitaev interaction is generated and its key components. Additionally, we will discuss other interactions such as Heisenberg ($J$) and Gamma ($\Gamma$) interactions, which we will refer to as non-Kitaev interactions, and their roles.

\subsection{Jackeli-Khaliullin mechanism and beyond}
The Kitaev QSL is fascinating, but the realization of such a model in solid-state materials has been challenging. In general, the SOC results in entangled spin-orbit wave functions and anisotropic spin interactions. However, in most systems, Jahn-Teller distortions lift the orbital degeneracy, partially inactivating SOC and leading to dominant XY or Ising anisotropic spin models.

Khaliullin noted that anisotropic bond-dependent interactions can appear in systems with partially filled $t_{2g}$ orbitals when spin and orbital degrees of freedom are entangled via SOC, and are separated from higher energy $e_g$ orbitals by octahedral crystal fields \cite{Khal2005Prog}. 
In the presence of SOC, spin and orbital degrees of freedom form total angular momenta (pseudospin $j_{\rm eff}=1/2$), interacting via entangled spin-orbital exchange interactions.
It was also noted that the Heisenberg ($J$) interaction is present together with the Kitaev ($K$) interaction. 
Later, Jackeli and Khaliullin identified the Kitaev interaction in edge-sharing octahedra honeycomb lattices, a key to quantum spin liquids \cite{jackeli2009mott}.
However, due to the presence of the Heisenberg interaction, various magnetic ordering states, depending signs and amplitude of Kitaev and Heisenberg interactions, are found except close to the pure Kitaev limit.

 Later in 2014, Rau, Lee, and Kee uncovered another bond-dependent interaction, known as Gamma ($\Gamma$) interaction \cite{Rau2014PRL}, which was also reported by the quantum chemistry \cite{Katukuri2014NJP} and density functional theory calculations \cite{yamaji2014first}. The $\Gamma$ interaction has the form of $XY$ bond-dependenent interaction leading to further frustration to the Kitaev interaction. This adds significant complexity to understanding Kitaev spin liquid through the interplay of Kitaev and Heisenberg interactions. 
The n.n.\ sites $i$ and $j$ on a bond of type $\gamma$ ($=x,y,z$) in an ideal octahedra environment takes the following form:
\begin{equation}
\renewcommand{\arraystretch}{1.5}
\begin{array}{l}
{\cal{H}}_{\langle ij\rangle,\gamma}\!=\! J \vec{S}_i\!\cdot\!\vec{S}_j\!+\!K_\gamma S^\gamma_i S^\gamma_j
\!+\!\Gamma (S^{\alpha}_iS^{\beta}_j\!+\!S^{\beta}_iS^{\alpha}_j)\\
~~~~~~+\Gamma'\big(S^{\alpha}_i S^\gamma_j + S^\gamma_iS^{\alpha}_j+S^{\beta}_i S^\gamma_j + S^\gamma_iS^{\beta}_j\big)\,,
\end{array}
\label{eq:Hamiltonian1}
\end{equation}
where $(\alpha,\beta)\!=\!(y,z)$, $(z,x)$ and $(x,y)$ for $\gamma\!=\!x$, $y$ (and $y'$), and $z$,  respectively~\cite{Rau2014PRL}, and $\Gamma'$ occurs due to the trigonal distortion \cite{Rau2014-arxiv}.

To understand the link between the spin model and real materials, it is worth 
rewriting Eq.~(\ref{eq:Hamiltonian1}) in the crystallographic $(a,b,c^\ast)$ frame \cite{Chaloupka2015PRB}:
\begin{equation}
\renewcommand{\arraystretch}{1.5}
\begin{array}{l}
{\cal H}_{ij,\gamma}\!=\!J_{ab}(S_{i}^{a}S_{j}^{a}+S_{i}^{b}S_{j}^{b})+J_{c}S_{i}^{c}S_{j}^{c}\\~~
+A\left[c_\gamma(S_{i}^{a}S_{j}^{a}-S_{i}^{b}S_{j}^{b})-s_\gamma(S_{i}^{a}S_{j}^{b}+S_{i}^{b}S_{j}^{a})\right]\\~~
-\sqrt{2}B\left[c_\gamma(S_{i}^{a}S_{j}^{c}+S_{i}^{c}S_{j}^{a})+s_\gamma(S_{i}^{b}S_{j}^{c}+S_{i}^{c}S_{j}^{b})\right],
\end{array}
\label{eq:Hamiltonian2}
\end{equation}
where $c_\gamma\!\equiv\!\cos\phi_\gamma$, $s_\gamma\!\equiv\!\sin\phi_\gamma$, $\phi_{\gamma}\!=\!0$, $\frac{2\pi}{3}$, and $\frac{4\pi}{3}$ for $\gamma\!=\!z$-, $x$-, and $y$-bond, respectively, and 
\begin{equation}
    \label{JabJcABcouplings}
\begin{split}
A=&\frac{1}{3}K+\frac{2}{3}(\Gamma - \Gamma'), \;\; B=\frac{1}{3}K-\frac{1}{3} (\Gamma - \Gamma'),\\
J_{ab}&=J+B - \Gamma',\;\; J_{c}=J+A + 2 \Gamma'\,.
\end{split}
\end{equation}

The Hamiltonian written in the crystallographic $(a,b,c^\ast)$ coordinate offers experimentally relevant insights.
For example, when $K\!=\!0$ and $\Gamma\!=\!\Gamma'$, the Hamiltonian reduces to an anisotropic $XXZ$ model. In this limit, the in-plane vs out-of-plane anisotropy is determined by the combination $J_c\!-\!J_{ab}\!= 3 \Gamma$. When the Kitaev interaction is turned on, the in-plane vs.\ out-of-plane anisotropy is not altered. Furthermore, $\Gamma\! \neq \! \Gamma'$, the anisotropy is determined by $\!\Gamma+2 \Gamma'$.
Thus, a positive $\Gamma\!$ renders the $c^\ast$-axis (i.e, perpendicular to the honeycomb plane) the hard axis \cite{Rau2014PRL}. This was noted in an early study of $\alpha$-RuCl$_3$; the sign of $\Gamma$ interaction was determined positive in the {\it ab-initio} calculation study \cite{kim2015kitaev}, and it was corroborated by the anisotropic magnetic susceptibility data, since the $g$-factor anisotropy alone due to the trigonal distortion cannot account for the large in-plane vs.\ out-of-plane anisotropy observed in $\alpha$-RuCl$_3$ \cite{plumb2014alpha}. 

The symmetry analysis of Hamiltonian Eq.\,(\ref{eq:Hamiltonian2}) is important to understand symmetry-related phenomena.
Without the magnetic field, the Hamiltonian is invariant under the translation operation along the unit vectors, ${\hat a}_1$ and ${\hat a}_2$, and inversion and time-reversal $(\bm{S} \rightarrow -\bm{S})$ operators. In addition, the two-fold rotational symmetry around ${\hat b}$-axis, $C_{2b}$, is preserved \cite{Cen2022CP}.
\begin{equation}
C_{2b}:  \left(S_a, S_b, S_c \right) \rightarrow  \left( -S_a, S_b, -S_c \right), \;\;
\phi_x \leftrightarrow  \phi_y,
\end{equation}
This symmetry is essential for preventing a finite thermal Hall conductivity (including the 1/2-quantized value) when the magnetic field is applied along the ${\hat b}$-direction, as the magnetic field does not break this two-fold rotational symmetry. Figure\,\ref{fig:Dirac3}(d) illustrates the unchanged dispersion when the field is aligned along ${\hat b}$-direction where the Ising topological order is forbidden due to the $C_{2b}$ symmetry inherent in the general ${\cal H}$ (Eq.\,(\ref{eq:Hamiltonian2})), unless the symmetry is spontaneously broken.  

On the other hand, there is no corresponding two-fold rotation symmetry around ${\hat a}$-direction, say $C_{2a}$.
This symmetry is broken due to a finite $B$ term in Hamiltonian, Eq.\,(\ref{eq:Hamiltonian2}) \cite{Cen2022CP}, and thus when the magnetic field is applied along the ${\hat a}$-axis, one expects a finite thermal Hall conductivity in general (see Figs.\,\ref{fig:Dirac3}(b) and (c)). When the $B$ term vanishes, i.e., $K = \Gamma - \Gamma^\prime$, there is no finite thermal Hall conductivity with the field along the ${\hat a}$-axis. Note that a half-quantized thermal Hall conductivity necessitates the Ising topological order, a specific limit of the Hamiltonian, $J_{ab} = J_c = A = B$ in Eq.\,(\ref{eq:Hamiltonian2}), implying that breaking the two-fold rotation symmetry is necessary, but not sufficient for a half-quantized thermal Hall conductivity. 

In the next subsection, we will discuss the effects of the non-Kitaev interactions and present recent studies \red{focusing on the realization of} Kitaev interactions and generalized bond-dependent spin models in a prime candidate, $\alpha$-RuCl$_3$. 

\subsection{Effects of non-Kitaev interactions}
All Kitaev candidates so far display a magnetically ordered phase at low temperatures, suggesting the presence of non-Kitaev interactions. For example, $\alpha$-RuCl$_3$ exhibits a zig-zag ordering pattern below 7-8\,K.
As discussed in the previous section, in an ideal octahedra cage, the symmetry-allowed n.n.\ spin exchange interactions include only Kitaev, Heisenberg, and $\Gamma$ interactions, and thus the non-Kitaev interaction may play an important role in understanding the candidate materials.

Let us first discuss the effects of the Heisenberg interaction. Since the honeycomb lattice is a bipartite lattice, we expect that the Heisenberg interaction results in a magnetic order, and the Kitaev QSL is confined in a narrow window near the pure Kitaev limit. Indeed, it was shown that when antiferromagnetic (AFM) Heisenberg interaction, i.e.\ $J>0$, is larger than $|K|$, AFM order occurs, while $J<0$, ferromagnetic (FM) order sets in. When $J \le K$ with $J, K >0$, the zig-zag order appears, while $|J| \le |K|$ with $J, K <0$, a stripe order is found \cite{chaloupka2013zigzag}.

While the Heisenberg interaction results in various magnetic orders, the $\Gamma$ interaction, the bond-dependent $XY$ interaction, is frustrated. The ground state of the pure Gamma model remains a subject of debate \cite{rousochatzakis2024beyond}. 
It is noteworthy that the $\Gamma$ interaction is highly frustrated, and a QSL has been proposed based on numerical studies \cite{luo2021npjqm}. When the $K$ and $\Gamma$ interactions dominate, a transition between the Kitaev QSL and another QSL phase becomes possible. A two-leg ladder Kitaev-$\Gamma$ model studied using the density matrix renormalization group (DMRG) technique revealed two distinct disordered phases as the ratio of Kitaev and $\Gamma$ interactions was varied \cite{gordon2019theory}.
Interestingly, when $K = \Gamma$, a hidden $SU(2)$ symmetry emerges via a six-site transformation, resulting in a magnetically ordered state \cite{Chaloupka2015PRB}. However, when these interactions have opposite signs, the system becomes highly frustrated. While the quantum model of the Kitaev-$\Gamma$ interaction has been extensively studied, conclusive theoretical results remain elusive except in a few specific limits. Given the complexity of the 2D honeycomb lattice, researchers have turned to 1D chains and ladders for numerical investigations \cite{rousochatzakis2024beyond}.
Furthermore, the classical $\Gamma$ model has been studied, revealing a classical spin liquid phase \cite{Rousochatzakis2017PRL}. Similarly, studies of the classical Kitaev-$\Gamma$ model have identified several large unit cell ground states \cite{Chern2020PRR,Rayyan2021RPB}, adding to our understanding of these complex systems.

Another bond-dependent interaction $\Gamma^\prime$ in Eq.\,(\ref{eq:Hamiltonian1}) is allowed once one of the $C_2$ symmetries is broken either through trigonal distortion or layer stackings. Near the FM Kitaev limit, it was shown that the zig-zag magnetic order is established via the FM $\Gamma^\prime$ \cite{Rau2014-arxiv}.
Additionally, further neighbor interactions are present. In particular, the third n.n.\ Heisenberg interaction $J_3$ is recognized as another interaction, albeit small, that contributes to the zig-zag order \cite{winter2016challenges}.

For a more detailed discussion, we refer to the recent comprehensive review article focusing on theoretical developments beyond the pure Kitaev model \cite{rousochatzakis2024beyond}.

\section{$\alpha$-RuCl$_3$: prime candidate for the Kitaev spin liquid}
 \subsection{$\alpha$-RuCl$_3$}

The first Kitaev QSL candidate suggested were iridium honeycomb oxides $A_2$IrO$_3$ with $A$=Na, Li, where Ir$^{4+}$ leads to $5d^5$ with $j_{\rm eff}=1/2$ moment \cite{jackeli2009mott}. Their Ir$^{4+}$ ions exhibit strong SOC that helps promote the Kitaev interactions.  However, research on the iridates is hampered by difficulties arising from the large lattice distortions away from a pure 2D honeycomb lattice, and significant non-Kitaev interactions. In fact, Na$_2$IrO$_3$ was extensively studied due to the possibility of realizing a Kitaev QSL state \cite{takagi2019concept}. However, due to the trigonal distortion present in Na$_2$IrO$_3$,  the spin-orbit coupled $j_{\text{eff}}$= 1/2 state of the atomic bases is called into questions \cite{mazin20122,foyevtsova2013ab}.  Moreover,  Na atoms may promote non-negligible further neighbor exchange terms additional to the n.n.\ terms \cite{chaloupka2013zigzag,kimchi2011kitaev,plumb2014alpha}.  
In addition, due to the presence of alkali atoms (Na and Li) between IrO$_6$ honeycomb layers, better-cleaved van der Waals materials were considered more suitable candidates. Furthermore, since iridium absorbs neutrons, neutron scattering measurements, one of the most powerful experimental techniques, are less effective. These disadvantages of iridium oxides have motivated both theoretical and experimental communities to search for other candidates. 

In 3D hyper-honeycomb lattice compounds $\beta$- and $\gamma$-Li$_2$IrO$_3$, a high field is necessary to suppress the AFM ordering. However, the spin-polarized state appears immediately after the suppression, and there is no evidence of spin-disordered states.     On the theoretical front, there is still considerable discussion about the appropriate microscopic Hamiltonian describing the magnetism of these Kitaev candidate materials \cite{chaloupka2013zigzag,kimchi2011kitaev,sizyuk2014importance,rousochatzakis2015phase,winter2016challenges,janssen2017magnetization,kim2015kitaev,suzuki2018effective,maksimov2020rethinking,janssen2020magnon}. 

 \begin{figure}[b]
	\includegraphics[clip,width=1\linewidth]{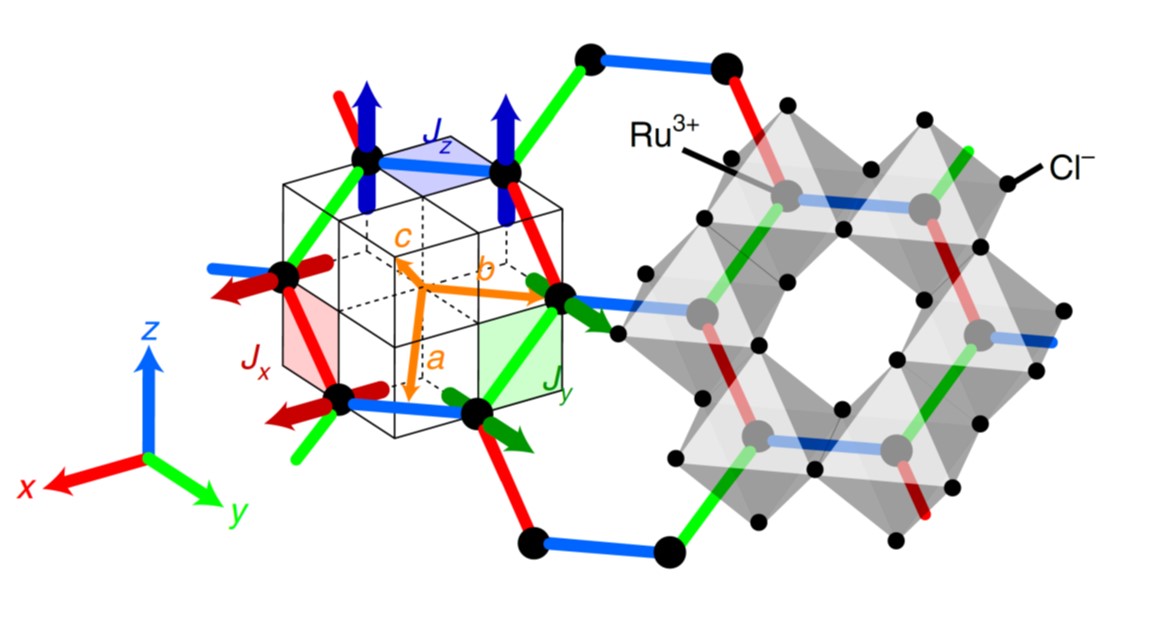}
	\caption{The schematic illustrates the crystal structure of $\alpha$-RuCl$_3$, highlighting both the crystallographic axes ($a$, $b$, $c$) and the Ising spin axes ($x$, $y$, $z$). We define the $a$ axis (along [11$\bar{2}$] in the spin axis) as the zigzag direction and the $b$ axis ([$\bar{1}$10])  as the bond direction, following the room temperature $C2/m$ notation.  For simplicity, the $c$ axis ([111]) is defined perpendicular to the $ab$ plane.
    The bond-dependent Ising interactions ($K_x$, $K_y$, $K_z$) govern the system's magnetic behavior, where each spin axis maintains perpendicularity to a plane containing both a Ru-Ru bond and a shared edge of Cl octahedra (depicted as colored squares) \cite{tanaka2022thermodynamic}.}
	
    \label{fig:RuClcrystal}
\end{figure}

Materials containing 4$d$ electrons had received less attention due to their comparatively weaker SOC relative to 5$d$ systems. However, recent investigations have demonstrated that despite the reduced SOC in 4$d$ systems, the $j_{\text{eff}}$ state can still be realized \cite{plumb2014alpha,kim2015kitaev}. This finding has prompted renewed interest in 4$d$ compounds for exploring novel quantum phenomena.  Of particular note is the black-colored ruthenium trichloride $\alpha$-RuCl$_3$, which has a layered honeycomb structure composed of Ru$^{3+}$. The valence of $d^5$ in the $t_{2g}$ orbitals could, in principle, form a $j_{\rm eff}=1/2$ moment if SOC is significant \cite{plumb2014alpha}. \red{Although the identification of this material as in the low-spin state was made in the 1960s \cite{Fletcher1963,Figgis1966,Fletcher1967},} $\alpha$-RuCl$_3$ gained popularity starting in 2014, following suggestions that it exhibits an SOC-driven Mott insulator where the Kitaev interaction is significant \cite{plumb2014alpha}. 

The low-energy electronic configuration of $\alpha$-RuCl$_3$ is determined by the interplay of four fundamental parameters: the on-site Coulomb repulsion (denoted by $U$), Hund's coupling ($J_H$), SOC ($\lambda$), and crystal-field splitting ($\Delta$).  Initially, it was unclear whether the SOC strength in Ru, which is lighter than Ir, could generate the spin-orbit entangled wavefunction.  Notably,  subsequent studies revealed that the narrow $d$-orbitals band effectively boosts the SOC strength, which then generates the entangled $j_{\rm eff}=1/2$ wavefunction, a critical prerequisite for the Kitaev interaction \cite{kim2015kitaev}. The specific combination of electronic interactions has led to the identification of $\alpha$-RuCl$_3$ as a prominent candidate for realizing Kitaev physics.

\subsection{Crystal structure}

$\alpha$-RuCl$_3$  exhibits a layered structure consisting of RuCl$_6$ octahedra (see Fig.\,\ref{fig:RuClcrystal}).  These planar layers are stacked perpendicular to the honeycomb plane, held together by weak van der Waals interactions. 
The layer comprises 2D networks of 4d$^5$ Ru$^{3+}$ ions coordinated within edge-sharing RuCl$_6$ octahedra.  The Ru$^{3+}$ ions occupy the vertices of the honeycomb lattice. The compound's structural integrity is notable for its minimal lattice distortion.  Crystallographic studies reveal that the Cl-Ru-Cl bond angles deviate by less than  1$^{\circ}$ from the ideal 90$^{\circ}$  configuration, while the Ru-Cl bond lengths exhibit a variance of less than 0.3\%. These results indicate an exceptionally high degree of octahedral symmetry in the  RuCl$_6$ units, approaching the theoretical ideal for such structures.  

A nearly ideal 2D honeycomb network of Ru$^{3+}$ ion makes it a prime candidate for investigating fundamental quantum phenomena in 2D systems, particularly in the context of Kitaev spin liquids and related exotic magnetic states.  Due to the weak van der Waals binding, the crystals are prone to stacking faults, strongly influencing structural and magnetic properties. Therefore, sample preparation and characterization remain important issues.  Its polymorph $\beta$-RuCl$_3$ adopts a different structure, comprising face-sharing RuCl$_6$ octahedra arranged in chains, which shows no magnetic ordering down to the lowest temperatures \cite{kobayashi1992moessbauer}.  The physical properties of $\beta$-RuCl$_3$ are essentially different from those in $\alpha$-RuCl$_3$ \cite{asaba2023growth}.

There have been controversies about the space group of the crystal structure of $\alpha$-RuCl$_3$. At room temperature, the monoclinic $C2/m$ and trigonal $P3_112$ structures have been assigned from X-ray diffraction measurements for single crystals grown by several techniques \cite{johnson2015monoclinic,cao2016low,Park2024,bruin2022origin}. Moreover, $\alpha$-RuCl$_3$ is known to have a first-order structural phase transition at $\sim150$\,K \cite{kubota2015successive,widmann2019thermodynamic,namba2024two}, and three different low-temperature structures have been proposed, $C2/m$, $P3_112$, and rhombohedral $R\bar{3}$ \cite{bruin2022origin}. This confusion comes partly from the van der Waals stacking of the honeycomb layers, which are susceptible to forming multi-domain structures \cite{kurumaji2023symmetry} and stacking faults \cite{cao2016low,banerjee2016proximate,namba2024two}. From the diffraction measurements, it is difficult to distinguish the $P3_112$ structure from $C2/m$ with multiple domains \cite{johnson2015monoclinic}, and the room-temperature structure may be maintained when crystal defects hinder the structural phase transition. One of the most reliable methods to distinguish these three space groups is the nuclear quadrupole resonance (NQR) of $^{35}$Cl site, whose spectra change sensitively with the number of independent sites of Cl, reflecting the crystal structure \cite{morosin1964x,nagai2020two,namba2024two}. In the NQR spectra for high-quality single crystals of $\alpha$-RuCl$_3$, the two-peak structures indicative of the $C2/m$ space group having two Cl sites are observed at room temperature. At low temperatures below the structural transition, the spectra change to a single peak, which identifies that the low-temperature crystal structure is $R\bar{3}$ \cite{nagai2020two}. This transition from $C2/m$ to $R\bar{3}$ is similar to the case of CrCl$_3$ \cite{morosin1964x}.

The low-temperature $R\bar{3}$ structure of clean crystals of $\alpha$-RuCl$_3$ immediately indicates that the crystal has the three-fold rotational symmetry just above the AFM transition. This rhombohedral $R\bar{3}$ structure may have two stacking orders (ABC or CBA) and two lattice settings (obverse and reverse), which are the sources of domain formation that can lead to partial cancellation of the thermal Hall conductivity \cite{kurumaji2023symmetry}. However, upon cooling \red{from} the high-temperature $C2/m$ structure to the low-temperature $R\bar{3}$ structure, there is little change in the RuCl$_3$ geometry and the halogen network cannot rotate \cite{mcguire2015coupling}, and thus single-domain crystals at room temperature are unlikely to generate the low-temperature domain structures relevant for the thermal Hall signal cancellation \cite{kurumaji2023symmetry}. Thus, it is important to use high-quality single-domain crystals to discuss the thermal Hall effect quantitatively.

Although the different crystal structure symmetries have different definitions of crystallographic axes, hereafter we define the $a$, $b$, and $c$ axes as shown in Fig.\,\ref{fig:RuClcrystal}. \red{Note that the $a$ ($b$) axis is perpendicular (parallel) to the Ru-Ru bond direction. The honeycomb plane corresponds to the plane (111) in the spin coordinates.}

\subsection{Electronic structure}

Initial characterization of $\alpha$-RuCl$_3$  based on transport measurements led to its classification as a magnetic semiconductor \cite{binotto1971optical}. However, subsequent spectroscopic studies have revealed a more complex electronic structure. X-ray angle-integrated and ultraviolet angle-resolved photoemission spectra measurements have provided compelling evidence for its classification as a Mott-Hubbard insulator \cite{pollini1996electronic}. Spectroscopic measurements have provided direct experimental evidence confirming the Mott insulating state in  $\alpha$-RuCl$_3$. Optical spectroscopy measurements reveal an optical gap of approximately 200\,meV, corresponding to excitations across the Mott-Hubbard gap from the lower to the upper Hubbard band \cite{sandilands2016spin}.   The atomically resolved real-space study by STM on cleaved $\alpha$-RuCl$_3$ revealed the electronic energy gap of 250\,meV \cite{ziatdinov2016atomic}.   Angle-resolved photoemission spectroscopy studies conducted at 200\,K observe a charge gap of 1.2\,eV. This larger energy scale represents the full Mott-Hubbard gap, which includes both the on-site Coulomb repulsion $U$ and the charge-transfer energy. 

The Mott insulating behavior of $\alpha$-RuCl$_3$ arises from the combined effect of SOC and moderate size electron correlation; Ab initio calculations have elucidated the electronic structure of $\alpha$-RuCl$_3$, revealing it to be a spin-orbit-assisted $j_{\rm eff}$=1/2 Mott insulator \cite{plumb2014alpha}. This finding represents the first computational evidence for the interplay between SOC and electron correlations in determining the insulating nature of this material. Density functional theory (DFT) calculations, incorporating appropriate corrections for strong correlations, demonstrate that the inclusion of SOC is crucial for accurately describing the electronic and magnetic properties of $\alpha$-RuCl$_3$ \cite{kim2015kitaev}.
 A strong cubic crystal field combined with strong SOC gives rise to nearly perfect $j_{\text{eff}}$=1/2 ground state that satisfies the conditions necessary for producing the Kitaev interactions. 

The presence of strong SOC, which is required to generate the Kitaev interaction,  has been confirmed by several measurements.  The optical spectroscopy and Raman scattering report the SOC strength $\lambda\approx 100$\,meV \cite{sandilands2016spin}. The transition between the ground state (low-spin state $j_{\rm eff}=1/2$) and the next excited state ($j_{\rm eff}=3/2$) is separated by $3\lambda/2$ in the octahedral environment.  The INS measurements observed this transition and reported $\lambda\approx 130$\,meV  \cite{banerjee2016proximate}. The resonant inelastic X-ray scattering (RIXS) measurements report   $\lambda\approx 150$\,meV \cite{suzuki2021proximate}, which is very close to the free-ion value of  $\lambda \approx$150\,meV. These experimental findings collectively provide a comprehensive understanding of the electronic structure and SOC in $\alpha$-RuCl$_3$, confirming its classification as a Mott insulator with significant spin-orbit interactions.

\subsection{Magnetic interaction}

The combination of SOC and edge-sharing octahedral structure facilitates bond-dependent Ising interactions with isotropic Kitaev coupling, where $K_x = K_y = K_z = K$ in $R{\bar 3}$ structure via the three-fold rotation symmetry about the ${\hat c}$-axis (out-of-plane), $C_{3c}$ symmetry.
This configuration is conducive to the formation of the gapless Kitaev QSL (B-phase) state that hosts Majorana fermions and $Z_2$ vortex.  However, almost all real Kitaev candidate materials undergo \red{transitions to 3D magnetic orders} at sufficiently low temperatures, primarily due to non-Kitaev n.n.\ interactions which are inevitably present in real materials. These additional interactions include Heisenberg exchange ($J$) that arises mainly from direct electron hopping between adjacent $j_{\rm eff} = 1/2$ magnetic ions \cite{chaloupka2010kitaev} and symmetric off-diagonal couplings (${\it \Gamma}$) that result from the interference of direct and indirect exchange processes
\cite{Rau2014PRL}. As discussed in Sec.\,IV.B, The presence of these non-Kitaev interactions significantly modifies the magnetic behavior of the material, causing deviations from the pure Kitaev model. In the systems with $K$, $J$, and ${\it \Gamma}$ terms, various magnetic orders can be stabilized, depending on the sign and magnitude of these interactions. These orders include, but not limited to, AFM, FM, stripe, zigzag, spiral, or 120$^{\circ}$ structure. The phase diagram of the $JK{\it \Gamma}$ model from the early exact diagonalization study of a 24-site honeycomb cluster has been revealed in \cite{Rau2014PRL}.  In addition to the aforementioned terms, the symmetry-allowed components of the n.n.\ and further n.n.\ exchange matrix emerge. These include a third n.n.\ Heisenberg coupling, denoted as $J_3$, and a fourth symmetry-allowed interaction term, $\Gamma'$. The latter arises in non-ideal octahedral configurations, particularly in materials exhibiting trigonal distortion.

In investigations of the magnetic properties of $\alpha$-RuCl$_3$, anisotropic magnetic susceptibility measurements on single crystals indicate significantly different effective coupling constants for orthogonal field orientations. The magnetic susceptibility $\chi(T)$ can be described by the Curie-Weiss law, $\chi(T)=C/(T-{\it \Theta}_{\rm CW})$, where $C$ is the Curie constant and ${\it \Theta}_{\rm CW}$ is the Curie-Weiss temperature. For the in-plane direction ({\boldmath $H$}$\perp c$), Curie-Weiss analysis yields a positive ${\it \Theta}_{\rm CW}=+37$\,K, indicating an effective FM exchange interaction with $J\sim 37$\,K. Conversely, for the out-of-plane direction {\boldmath $H$}$\parallel c$, a negative Curie-Weiss temperature of ${\it \Theta}_{\rm CW}$ of -150\,K is observed, suggesting an AFM effective exchange. This anisotropy can be attributed to the layered structure of $\alpha$-RuCl$_3$ and the presence of $\Gamma$ and $\Gamma^\prime$ interactions as noted by $J_{c} - J_{ab} = \Gamma + 2 \Gamma^\prime$ from Eq.\,(\ref{eq:Hamiltonian2}).

Initial investigations into the magnetic properties of $\alpha$-RuCl$_3$ utilizing magnetic susceptibility, specific heat \cite{majumder2015anisotropic}, and INS techniques \cite{banerjee2016proximate} reported two distinct magnetic phase transitions at around 7\,K and 14\,K. Subsequent studies, however, have elucidated that the 14\,K transition is not intrinsic to the ideal crystal structure but is instead associated with stacking faults in the layered material. Indeed, high-resolution X-ray diffraction and INS experiments on high-quality single crystals have demonstrated the absence of the 14\,K anomaly. The disappearance of this feature in pristine samples indicates that the true intrinsic magnetic ordering in $\alpha$-RuCl$_3$ occurs only at $\approx 7$\,K.  The sensitivity of crystal structure to mechanical deformation has been identified as a critical factor in the emergence of the 14\,K anomaly. Even minimal bending or strain applied to the crystal can induce stacking faults, which in turn lead to the appearance of the higher-temperature magnetic transition. This structural modification can be described by a change in the stacking sequence of the RuCl$_3$ layers from the ideal ABC order to regions with AB stacking. Given the material's high susceptibility to structural defects and their significant impact on magnetic behavior, extreme caution must be exercised in the preparation, handling, and mounting of $\alpha$-RuCl$_3$ single crystals for experimental studies. 

Below $T_N\approx 7$\,K, $\alpha$-RuCl$_3$ exhibits zigzag AFM ordering typically associated with frustrated honeycomb lattice magnets, as confirmed by multiple experimental techniques. Neutron diffraction studies have revealed a magnetic structure characterized by a doubling of the unit cell along the hexagonal (100) direction \cite{sears2015magnetic} and magnetic propagation vector {\boldmath $q$} =(0.5, 0, 0) in the hexagonal notation, corresponding to the $M$-point of the Brillouin zone \cite{ritter2016zigzag}, consistent with the zigzag AFM ground state. 
This zigzag ordering has been corroborated by X-ray diffraction \cite{johnson2015monoclinic,cao2016low}, INS on powder \cite{banerjee2016proximate} and single crystal \cite{banerjee2017neutron}, and $\mu$SR measurements \cite{lang2016unconventional}.

    
  \begin{table*}
  		\caption{Estimated interaction parameters of $\alpha$-RuCl$_3$ derived from diverse experimental and theoretical approaches \red{based on 2D models}: Kitaev ($K$), Heisenberg ($J$), off-diagonal (${\it \Gamma, \Gamma'}$), and third-neighbor Heisenberg ($J_3$) exchanges.  DFT and ED denote density functional theory and exact diagonalization, respectively. $P3$ and $C2$ denote crystal structures. All parameters are in units of meV. \red{For 3D models, see discussions in \cite{janssen2020magnon,balz2021field,Cen2025}.}}
 
  		\begin{tabular}{cccccc}
        \hline \hline
  			 & $K$ & $J$ & $\Gamma$ & $\Gamma'$ & $J_3$  
  			\\
  			\hline
  			DFT+U, SOC, $U_{\rm eff}$, zigzag \cite{kim2016crystal} & $-9.34$ & $-0.74$ & 3.71 & $-1.04$ & 
  			\\
  			 DFT+U, SOC, $U_{\rm 
     eff}$, zigzag \cite{kim2016crystal} & $-7.64$ & $-1.09$ & 4.38 & $-0.87$ & 
  			\\
Quantum chemistry computations, ED \cite{yadav2016kitaev} & $-5.6$ & 1.2 & $\Gamma_{xy}=-1.2, \Gamma_{zx}=-0.7$ &  & 
  			\\
INS, effective-spin model, \cite{ran2017spin} & $-6.8$ & & 9.5 &  & 
  			\\
DFT, $J_1$-$K$-${\it\Gamma}$-$J_3$-$K_3$ model, $U=3.5$\,eV \cite{hou2017unveiling} & $-10.8 ~(K_3=0.6)$ & $-2.0$ & 3.8 &  & 1.3
  			\\
Five-orbital Hubbard,$P$3 \cite{wang2017theoretical} & $-10.9$ & $-0.3$ & 6.1 &  & 0.03
  			\\
Five-orbital Hubbard,$C$2 \cite{wang2017theoretical} & $-5.5$ & 0.1 & 7.6 &  & 0.1
  			\\
DFT, ED, $C$2 crystal structure \cite{winter2016challenges} & $-7.5$ & $-1.4$ & 5.9 & $-0.8$ & 
  			\\
{\it Ab initio}, INS \cite{winter2017breakdown} & $-5.0$ & $-0.5$ & 2.5 &  & 0.5
  			\\
Spin wave, thermal Hall effect \cite{cookmeyer2018spin} & $-5.0$ & $-0.5$ & 2.5 &  & 0.11
  			\\
Raman, ED \cite{sahasrabudhe2020high} & $-10.0$ & $-0.75$ & 3.75 &  & 0.75
  			\\
Magnetization ({\boldmath $H$}$\parallel b$) \cite{sears2020ferromagnetic} & $-10$ & $-2.7$ & 10.6 & $-0.9$ & 
  			\\
Magnetization ({\boldmath $H$}$\parallel c^*$) \cite{sears2020ferromagnetic} & $-10$ & $-1.5$ & 8.8 &  & 0.4
  			\\
\cite{maksimov2020rethinking} & $-4.8$ & $-2.56$ & 4.08 & 2.5 & 2.42
  			\\
\cite{maksimov2020rethinking} & $-10.8$ & $-4.0$ & 5.2 & 2.9 & 3.26
  			\\
RIXS, ED    \cite{suzuki2021proximate} & $-5$ & $-3$ & 2.5 & 0.1 & 0.75
  			\\
        \hline \hline
  		\end{tabular}
  \end{table*}
  

  It is noteworthy that this zigzag state is theoretically predicted to exist within a broad parameter space adjacent to the pure Kitaev points in the $KJ{\it \Gamma}$ phase diagram \cite{chaloupka2013zigzag,Rau2014PRL,winter2017models}. 
  The zigzag state reported by experiments has offered some insight to the interaction parameters and their signs. 
  When the Kitaev interaction is AFM, i.e, $K > 0$ in Eq.\,(\ref{eq:Hamiltonian1}), the stability of the zigzag order is particularly enhanced by the presence of FM Heisenberg interactions. This can be understood in terms of the frustration relief provided by the $J$ term, which favors collinear spin arrangements. Consequently, increasing $|J|$ tends to drive the system away from the Kitaev QSL phase \cite{kimchi2011kitaev,choi2012spin,rousochatzakis2015phase,rusnavcko2019kitaev}.  It has also been shown that the zigzag phase found in the n.n.\ Kitaev-Heisenberg model has some stability against the ${\it \Gamma}$ term \cite{sizyuk2016selection}. The competition between these different exchange interactions plays a crucial role in determining the ground state of the material. 

  However, when the Kitaev interaction is FM $K< 0$, the zigzag state is not found in the classical and quantum model of the $JK\Gamma$ model \cite{Rau2014PRL} on finite-size clusters near the FM Kitaev region. When a small FM $\Gamma^\prime$ \cite{Rau2014-arxiv} or AFM $J_3$ \cite{winter2016challenges} is introduced, the zigzag state is stabilized. The sign of the Kitaev interaction was debated earlier, and it was noted that it depends on the layer stacking \cite{kim2016crystal}.  
  When the structure is $P{3_12}$, the Kitaev interaction is AFM type, while for the other structures, it is FM. The strength of the Kitaev interaction is highly sensitive to the positions of Cl ion, which is expected, as the Kitaev interaction is dominated by the indirect exchange path via $p$-orbitals \cite{jackeli2009mott}. {\it Ab initio} calculations on other structures corroborate the FM Kitaev and the emergence of zigzag order in $\alpha$-RuCl$_3$ \cite{johnson2015monoclinic,kim2015kitaev,kim2016crystal,yadav2016kitaev} due to a small but finite $\Gamma^\prime$ or $J_3$ interactions, typically employing DFT with various corrections for strong correlations and SOC, and exact diagonalization and density-matrix renormalization group techniques for extended Kitaev-Heisenberg Hamiltonian.  The FM Kitaev interaction was further supported by the magnetic moment direction making the angle of $\sim32^\circ$ within the $ac$-plane \cite{sears2020ferromagnetic} (Fig.\,\ref{fig:Magneticphase}(b)), consistent with the dominant FM Kitaev interaction together with the AFM $\Gamma$ and FM $\Gamma^\prime$ interactions \cite{chaloupka2016magnetic}. 
  
  Theoretical calculations provide estimates of the exchange parameters:$K\approx -5$ to $-15$\,meV, $J\approx -0.5$  to $-2$\,meV,  and ${\it \Gamma}\approx 2$ to 7\,meV.  Microscopic calculations based on the $JK{\it \Gamma}$ model and its variations, such as including longer-range interactions or additional symmetry-allowed terms, consistently predict the zigzag ground state for parameter ranges relevant to $\alpha$-RuCl$_3$ \cite{sizyuk2016selection,winter2016challenges}.  It has also been shown that the momentum direction in the zigzag AFM state is sensitive to the model parameters in real materials \cite{chaloupka2016magnetic}.  

  In addition, it has been suggested that magnetoelastic coupling also plays an important role in the stability of magnetic order in $\alpha$-RuCl$_3$. In frustrated spin systems, the interplay between spin, lattice, and charge degrees of freedom often leads to the stabilization of magnetically ordered states through entropy release. This coupling between spin and lattice is expected to have a particularly pronounced effect on the ground state in systems exhibiting strong SOC. Evidence supporting this hypothesis in $\alpha$-RuCl$_3$ comes from the observation of a notable magneto-dielectric effect within its magnetically ordered state \cite{aoyama2017anisotropic,reschke2017electronic}. 
  
\begin{figure}[b]
	\includegraphics[clip,width=0.5\linewidth]{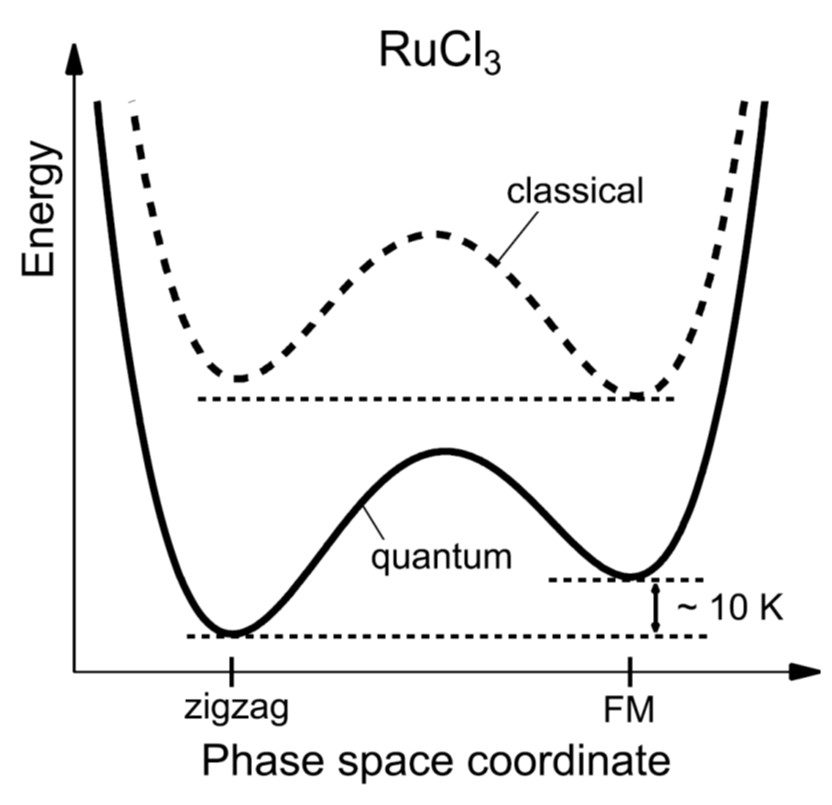}
	\caption{Schematic representation of classical and quantum energy landscapes surrounding the zigzag ground state configuration. Quantum effects serve to stabilize the zigzag-ordered phase, establishing it as the ground state with energy marginally lower than that of the competing metastable FM configuration \cite{suzuki2021proximate}.
	}
    \label{fig:Zigzag}
\end{figure}

In contrast, it has been pointed out that quantum fluctuations stabilize the zigzag state.  To examine the above theoretical models, the spin Hamiltonian of $\alpha$-RuCl$_3$ has been investigated by using RIXS at the Ru $L_3$ absorption edge \cite{suzuki2021proximate}.  In the paramagnetic state, the quasi-elastic intensity of magnetic excitations exhibits a broad maximum around the Brillouin zone center, with no discernible local maxima at the zigzag magnetic Bragg wavevectors. In contrast to the models of Refs.\ \cite{winter2017breakdown} and \cite{sears2020ferromagnetic}, the ferromagnetic state in the RIXS-derived model is lower in energy than the zigzag state on the classical level. This suggests that the zigzag order is inherently fragile and can be readily destabilized by competing ferromagnetic correlations (Fig.\,\ref{fig:Magneticphase}(b)). Therefore, the stabilization of the zigzag order is ascribed to strong quantum fluctuations, which are intrinsic to highly frustrated Kitaev interactions, positioning both ferromagnetism and the Kitaev QSL as energetically proximate metastable states \cite{suzuki2021proximate,kaib2022electronic}, as illustrated in Fig.\,\ref{fig:Zigzag}.  However, this cannot account for the magnetic moment making an angle of 35$^\circ$ in the $ac$-plane \cite{sears2020ferromagnetic}, and the broad maximum is a feature of the dominant Kitaev interaction \cite{gordon2019theory}. 

The ordered moments $\langle\mu\rangle\approx 0.45\mu_B/$Ru$^{3+}$ ($\mu_B$ is a unit Bohr magneton) have been reported to be exceptionally low, which is at most only one-third of the full moment determined from bulk measurements \cite{sears2015magnetic,ritter2016zigzag}, suggesting strong spin fluctuations.  Thus the zigzag magnetic order, highly reduced ordered moments, and strong SOC suggest strongly frustrated systems.   The frustration manifests through $K$ and $\Gamma$ exchange interactions, which collectively contribute to the substantial suppression of the ordered moment \cite{Chern2020PRR,Rayyan2021RPB}.

The construction of a minimal microscopic model for $\alpha$-RuCl$_3$ necessitates the inclusion of several key exchange interactions:  Kitaev ($K$), Heisenberg ($J$), off-diagonal (${\it \Gamma,\Gamma'}$), and third-neighbor Heisenberg ($J_3$) exchanges. Despite its minimalist approach, this model encompasses a five-dimensional parameter space. In  $\alpha$-RuCl$_3$, while consensus indicates an FM Kitaev interaction ($K<0$) as the dominant contribution, estimates of interaction parameters exhibit considerable variation, precluding comprehensive agreement on quantitative values. Numerous \{$K$,$J$,${\it \Gamma}$,${\it \Gamma'}$,$J_3$\} parameter sets have been proposed to describe $\alpha$-RuCl$_3$ using first-principle methods \cite{kim2016crystal,yadav2016kitaev,hou2017unveiling,winter2016challenges,wang2017theoretical} and phenomenological analyses \cite{ran2017spin,winter2017breakdown,cookmeyer2018spin,sahasrabudhe2020high,sears2020ferromagnetic,suzuki2021proximate}. Although estimates of the Kitaev coupling magnitude vary, most values fall within $-4$ to $-11$\,meV. Theoretical considerations suggest a realistic range of $-3.8$ to $-11$\,meV for $K$, with certain observables strongly constraining the minimal model parameters \cite{maksimov2020rethinking}. The ${\it \Gamma}$ term is positive (${\it \Gamma}>0$) and potentially comparable to $|K|$. Most estimations indicate an FM Heisenberg interaction ($J<0$), believed to be subleading. The ${\it \Gamma'}$ term can be negative or positive, with ${\it |\Gamma'|}<{\it |\Gamma|}$. The third-neighbor Heisenberg coupling is antiferromagnetic ($J_3>0$) with $|J_3|<|J|$. Table I presents representative sets of interaction parameters for $\alpha$-RuCl$_3$, constrained to $K$ values within $-4$ to $-11$\,meV.

\subsection{Magnetic field phase diagrams}

 \begin{figure}[b]
	\includegraphics[clip,width=1.0\linewidth]{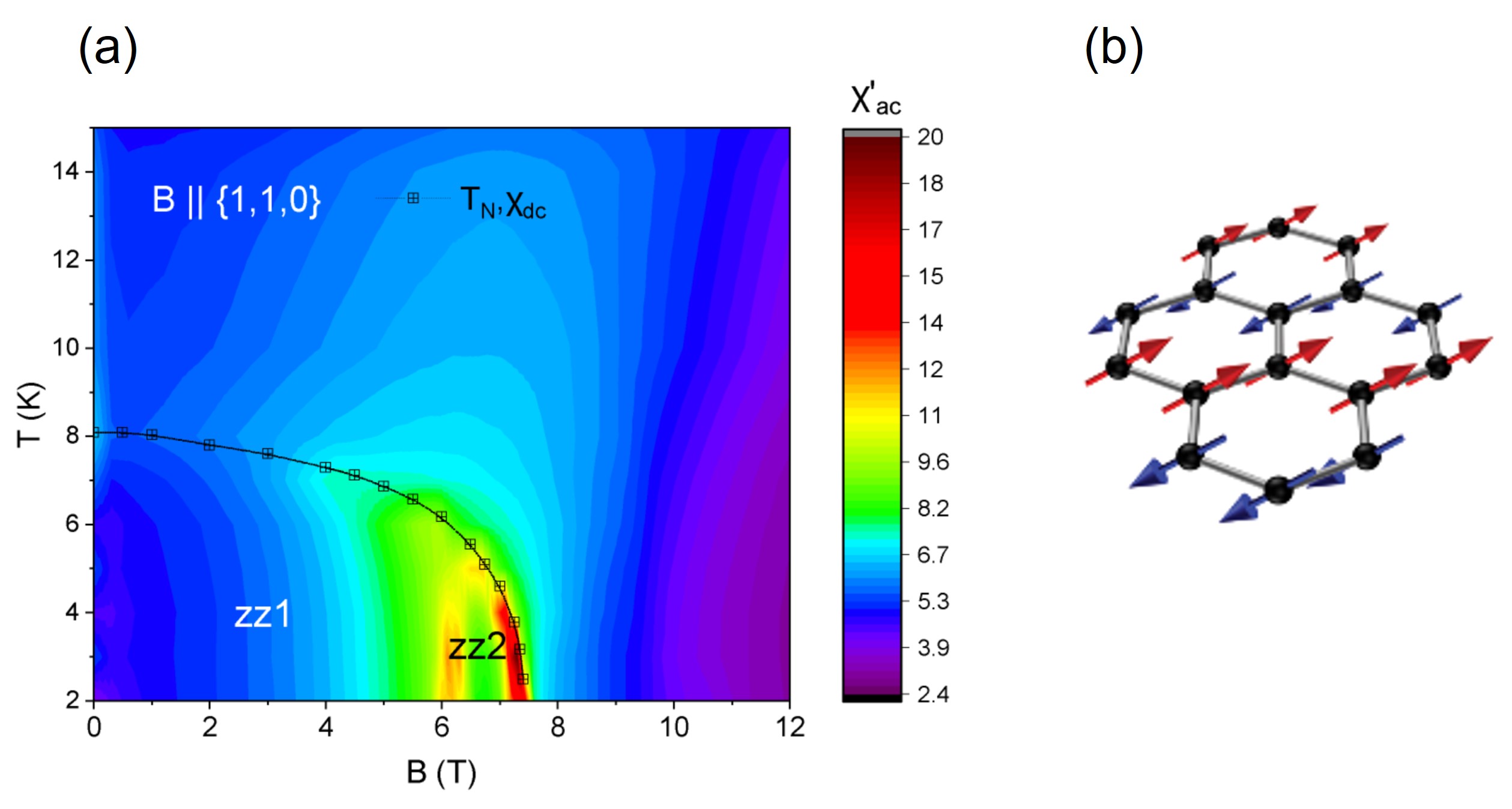}
	\caption{(a) $B$-$T$ phase diagram of $\alpha$-RuCl$_3$ in in-plane magnetic field ($B \parallel a$ axis) constructed from ac susceptibility \cite{balz2021field}.  There are three distinct regions: a low-field ordered phase (zz1), an intermediate ordered phase (zz2), and a high-field FIQD phase.   (b) The zigzag AFM structure in the zz1 phase. The ordered moment size at 4.2 K is determined to be 0.45(5) $\mu_B$/Ru with canting angle 35$^{\circ}$ away from the honeycomb plane.\cite{cao2016low,morgan2024structure}. }
    \label{fig:Magneticphase}
\end{figure}

The application of a sufficiently large in-plane magnetic field suppresses the zigzag AFM order in $\alpha$-RuCl$_3$ \cite{leahy2017anomalous,sears2017phase,wolter2017field,baek2017evidence,banerjee2018excitations,hentrich2018unusual,janvsa2018observation}, inducing a FIQD state. This magnetic order suppression has been confirmed through multiple experimental techniques, including NMR \cite{baek2017evidence,janvsa2018observation}, neutron diffraction \cite{banerjee2018excitations}, and magnetic susceptibility measurements (Fig.\,\ref{fig:Magneticphase}(a)). The critical field for suppression exhibits slight directional dependence.  For both {\boldmath $H$}$\parallel a$ (zigzag direction) and {\boldmath $H$}$\parallel b$ (armchair direction), an intermediate ordered state emerges before the complete suppression of the zigzag-ordered phase. The transition sequence from zigzag AFM (zz1) to the so-called zz2 AFM phases, which are differentiated by their interlayer magnetic ordering patterns (Fig.\,\ref{fig:Magneticphase}(a)), and subsequently to the FIQD  state, has been corroborated by various experimental methods, including neutron diffraction \cite{balz2021field}, $ac$-susceptibility \cite{banerjee2018excitations}, thermal conductivity \cite{suetsugu2022evidence}, heat capacity \cite{tanaka2022thermodynamic}, the magnetocaloric effect \cite{balz2019finite}, and magnetic Gr\"{u}neisen parameter \cite{bachus2020thermodynamic}. \red{The transition between zz1 and zz2 is of first order, which leads to a coexistence of the two phases above 6\,T. The magnetic structure of the zz2 phase} is characterized by a six-layer periodicity perpendicular to the honeycomb planes. Theoretical proposals suggest that the spin structure in the zz2 phase can be accounted for by incorporating spin-anisotropic interlayer couplings in a 3D spin model \red{\cite{Cen2025}}, highlighting the significance of interlayer exchange interactions $\alpha$-RuCl$_3$.

\section{Physical properties of $\alpha$-RuCl$_3$ in zero field}

\subsection{Fractionalization of spin degrees of freedom}

Although $\alpha$-RuCl$_3$ is currently regarded as the most promising candidate for realizing Kitaev physics, the presence of non-Kitaev interactions induces long-range AFM zigzag order at $T_N\approx 7$\,K in zero field.  This indicates that $\alpha$-RuCl$_3$ is only proximate to the Kitaev QSL state. Notably, the Kitaev interaction strength $|K|\approx 5$-11\,meV substantially exceeds the thermal energy scale at the ordering temperature ($k_BT_N\approx 0.6$\,meV). The magnetic properties reveal a characteristic temperature evolution: while the magnetic susceptibility $\chi(T)$ exhibits conventional paramagnetic behavior at high temperatures, it deviates from the Curie-Weiss law below approximately 100-140\,K (Fig.\,\ref{fig:SpecificHeat}(a)). This deviation marks the onset of short-range spin correlations \cite{yoshitake2016fractional}. The temperature scale of this crossover aligns with the estimated Kitaev exchange interaction strength in $\alpha$-RuCl$_3$, suggesting that signatures of the Kitaev QSL may manifest above $T_N$ even without an applied field \cite{rousochatzakis2019quantum}.

A hallmark feature of the Kitaev QSL is the fractionalization of quantum spins into Majorana fermions. Experimental evidence suggestive of fractionalized spin excitations above $T_N$ and below $\sim$100\,K in zero fields has been reported using various techniques, including specific heat \cite{do2017majorana,widmann2019thermodynamic}, Raman spectroscopy \cite{sandilands2015scattering,glamazda2017relation,wang2020range,wulferding2020magnon}, INS \cite{banerjee2016proximate,do2017majorana,banerjee2017neutron}, IXS \cite{metavitsiadis2020phonon,ye2020phonon,li2021giant}, THz spectroscopy \cite{little2017antiferromagnetic,reschke2018sub,wang2017magnetic} and thermal transport \cite{leahy2017anomalous,hirobe2017magnetic,kasahara2018unusual,hentrich2018unusual} studies. 

However, it is important to note that alternative interpretations exist for some of these observations. Some researchers argue that certain experimental results can be accounted for by more conventional phenomena, such as magnon damping \cite{winter2017breakdown}, without invoking spin fractionalization. Furthermore, the spin-phonon interactions may play a more significant role at high temperatures and finite energy excitations, introducing an additional source of decay that warrants further investigation.  These diverse experimental probes offer insights into the nature of the magnetic excitations and their temperature-dependent evolution in $\alpha$-RuCl$_3$. However, the interpretation of these results remains a subject of ongoing debate in the scientific community. This chapter focuses on examining the evidence for and against fractionalized excitations in zero field, a key characteristic proposed for the Kitaev QSL state in $\alpha$-RuCl$_3$.

\subsection{Specific heat}
\begin{figure}[t]
	\includegraphics[clip,width=0.65\linewidth]{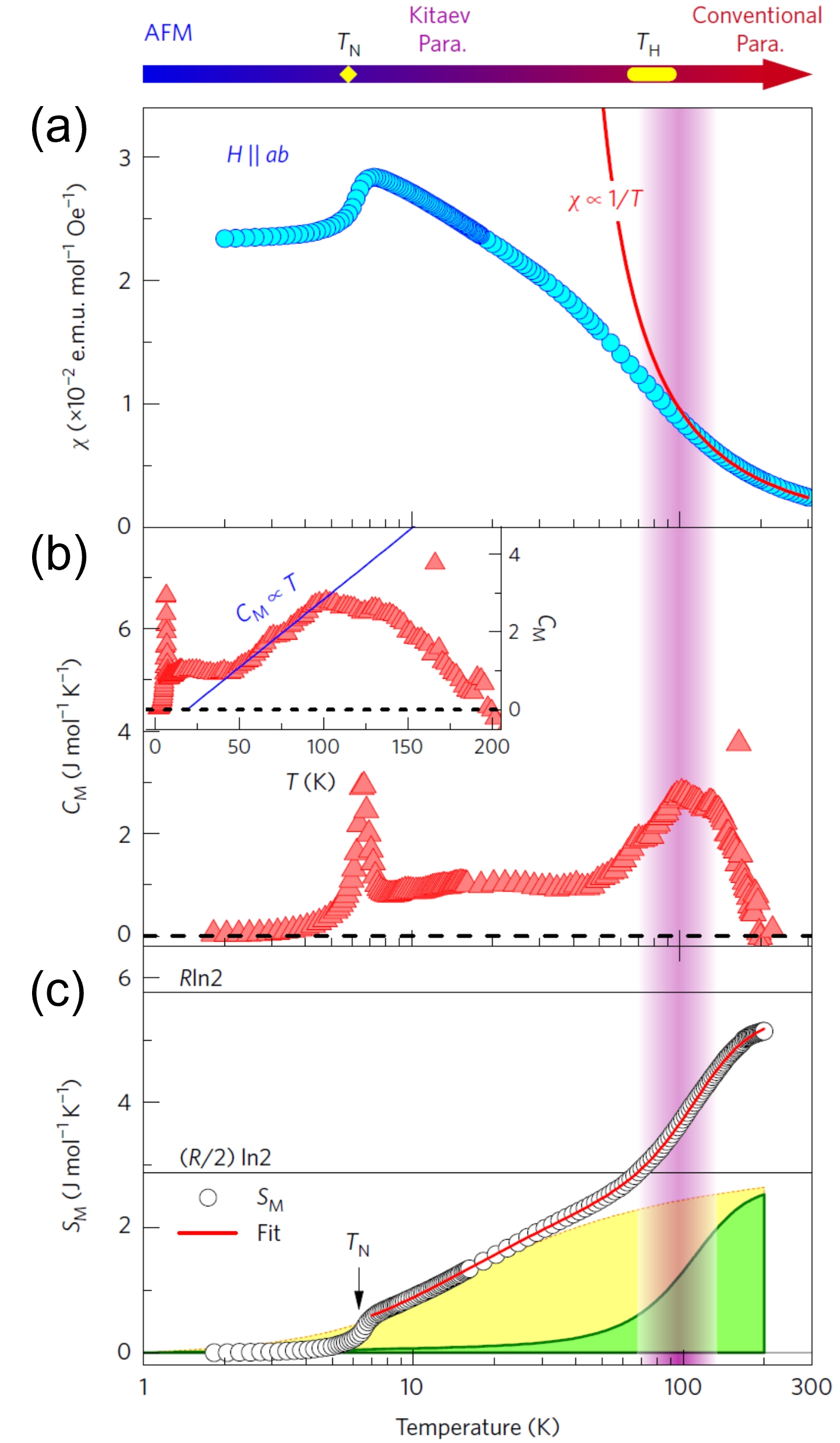}
\caption{(a) Temperature-dependent magnetic susceptibility of $\alpha$-RuCl$_3$ for $\bm{H}\parallel ab$. The susceptibility deviates from the Curie-Weiss behavior (solid red line) below $T\sim 100$\,K.  (b) The magnetic contribution to specific heat ($C_M$), isolated through subtraction of the lattice component. A pronounced AFM peak occurs at $T_N$, followed by two distinct anomalies: broad maxima in the $T$-range $T_N\lesssim T  \lesssim 50$\,K and at $T=T_H\approx 100$\,K (vertical bar), attributed to excitations of localized and itinerant Majorana fermions, respectively. At 50\,K $\lesssim T \lesssim T_H$, $C_M$ exhibits a linear-$T$ dependence, indicative of a metal-like DOS characteristic of itinerant Majorana fermions (inset). A sharp discontinuity at 165\,K corresponds to a structural phase transition.  (c) Temperature dependence of magnetic entropy change, calculated by integrating $C_M$ over the range 2\,K $<T<200$\,K. The horizontal lines denote theoretical benchmarks: the full magnetic entropy $R\ln 2$ and its half-value $(R/2)\ln 2$. A phenomenological fit (solid red line), derived from theoretical simulations, demonstrates that the total entropy release can be decomposed into two distinct fermionic contributions (indicated by yellow and green shaded regions). These components reflect the separation of magnetic degrees of freedom into distinct fermionic excitations \cite{do2017majorana}.
}
\label{fig:SpecificHeat}
\end{figure}

Quantum Monte Carlo simulations \cite{nasu2015thermal,do2017majorana}, exact numerical diagonalization method \cite{suzuki2018effective}, tensor-network approaches \cite{li2020universal}, and thermal pure quantum state method \cite{laurell2020dynamical} have elucidated the finite-temperature thermodynamic properties of the  Kitaev model. These investigations reveal that the fractionalization of quantum spin-1/2 into itinerant and localized Majorana fermions manifests experimentally in the specific heat. This phenomenon is characterized by a bimodal entropy release, resulting in two distinct specific heat peaks at $T_H\approx 0.4|K|$ and $T_L\approx 0.013|K|$. The high-temperature peak corresponds to the formation of Fermi degeneracy of Majorana fermions, associated with the development of n.n.\ spin correlations \cite{motome2020hunting}. Conversely, the low-temperature peak originates from $Z_2$ fluxes composed of localized Majorana fermions, which vanish at low temperatures due to an energy gap.  The entropy release pattern provides further insight into these anomalies. A prominent plateau or shoulder at $(1/2) R\ln 2$ ($R$ is the gas constant) is observed in the intermediate temperature range between the two peaks, representing half the entropy of a spin-1/2 system. 

\begin{figure}[t]
	\includegraphics[clip,width=\linewidth]{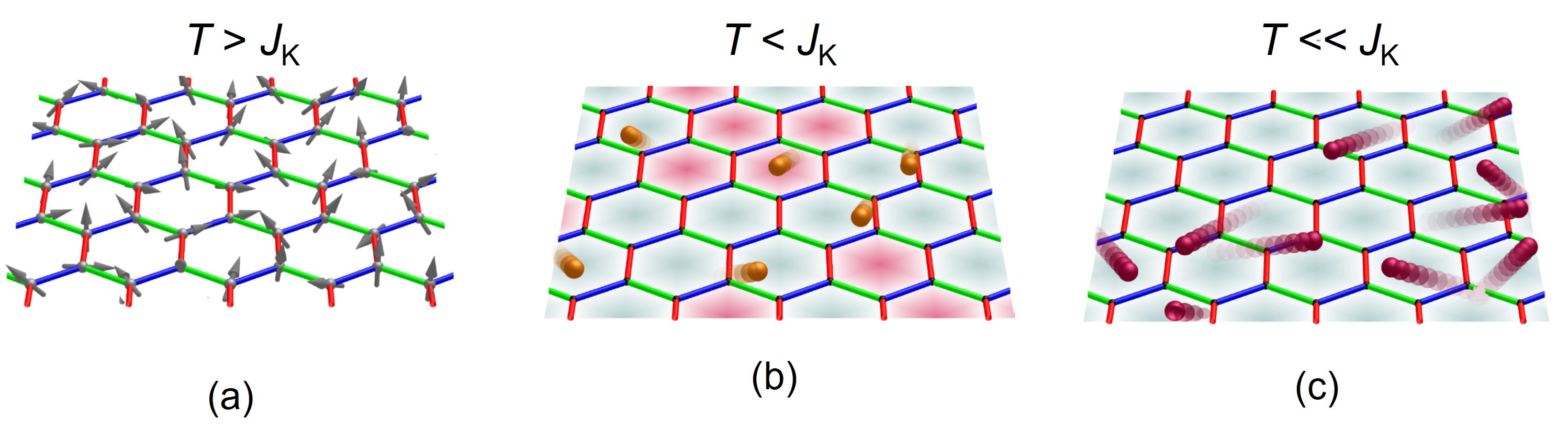}
	\caption{Schematic figures of temperature evolution of Kitaev quantum magnets. (a) High-temperature regime ($T\gg |K|$): conventional paramagnetic phase with thermally randomized spins. (b) Intermediate-temperature regime ($T\lesssim |K|$): fractionalization of spins into Majorana fermions, with itinerant Majorana fermions experiencing scattering from thermally activated visons (gap $\Delta_K$). (c) Low-temperature regime ($T \ll |K|$): the emergence of QSL phase characterized by coherent propagation of itinerant Majorana fermions throughout the crystal lattice, with suppressed vison excitations \cite{motome2020hunting}.
	}
    \label{fig:Excitation}
\end{figure}

In the temperature regime $T_L < T < T_H$, the itinerant Majorana fermion spectrum undergoes significant perturbation due to thermal fluctuations of $Z_2$ gauge fluxes, in addition to the previously discussed bimodal specific-heat structure. This perturbation induces a finite density of states, arising from the scattering of itinerant Majorana fermions with thermally excited $Z_2$ fluxes \cite{nasu2015thermal}. Consequently, a metal-like density of states emerges, manifesting as a temperature-linear specific heat---an anomalous behavior for a system exhibiting a Dirac semimetallic spectrum in the low-temperature limit. The presence of this linear term serves as a potential signature of the fermionic nature of the excitations. This phenomenon elucidates the intricate interplay between distinct quasiparticle excitations within the Kitaev model, emphasizing the unique thermodynamic properties that emerge from the fractionalization of spin-1/2 degrees of freedom. The schematic behaviors at different temperature regimes are shown in Figs.\,\ref{fig:Excitation}(a)-(c).

Thus the fractionalization of spin into two distinct types of Majorana fermions is experimentally manifested in the zero-field specific heat, characterized by the following key features: (1) Bimodal temperature dependence with well-separated peaks at $T_L$ and $T_H$,
(2) An intermediate entropy release of half of $R\ln 2$ between the peaks,
and (3) A linear temperature dependence of the specific heat in the range $T_L<T<T_H$.

In the context of the pure Kitaev model, given the estimated range of $|K|$ values (60-120\,K), theoretical predictions place $T_H$ between 25-50\,K and $T_L$ between 1-2\,K. This distinctive double-peak structure is corroborated by exact diagonalization calculations performed on finite-size clusters using exchange parameters derived from ab initio calculations (refer to Table I in \cite{laurell2020dynamical}).  The observation of these characteristics in experimental systems would provide compelling evidence for the existence of Majorana fermions and substantiate the applicability of the Kitaev model to real materials. Moreover, this specific heat behavior serves as a crucial test for theoretical models and may guide the search for quantum spin liquid states in candidate materials.

In systems where the ground state consists of a Kramers doublet with $j_{\rm eff} = 1/2$, the electronic structure possesses an inherent symmetry that fundamentally constrains its interactions with the crystal lattice. These electronic states interact with phonons through quantum mechanical coupling mechanisms. This coupling becomes important in materials with strong SOC, such as $\alpha$-RuCl$_3$. For such materials, the determination of magnetic specific heat requires careful consideration to properly account for these quantum mechanical interactions. Two groups reported temperature-dependent specific heat measurements on $\alpha$-RuCl$_3$ up to high temperatures \cite{do2017majorana,widmann2019thermodynamic}. Both simply assumed that the total specific heat comprises lattice ($C_{\rm ph}$) and magnetic ($C_{\rm mag}$) contributions, $C=C_{\rm ph}+C_{\rm mag}$. $C_{\rm ph}$ was estimated using the specific heat of an isostructural non-magnetic compound. 

Figures.\,\ref{fig:SpecificHeat}(b) and (c) depict the magnetic specific heat $C_{\rm mag}$ and magnetic entropy $S_M$, respectively, in zero fields on $\alpha$-RuCl$_3$ across a wide temperature range from 2 to 200\,K, reported by Do {\it et al.}\ \cite{do2017majorana}.  To isolate $C_{\rm mag}$, $C_{\rm ph}$ was estimated using the specific heat of ScCl$_3$, scaled by the molecular mass ratio $W_{\rm RuCl_3}/W_{\rm ScCl_3}\approx$1.21. The resulting $C_{\rm mag}$ exhibits a bimodal structure with broad peaks at  $\sim 10$\,K and  $\sim 100$\,K. The higher peak temperature aligns with theoretical predictions for $T_H$ within an order of magnitude.  The magnetic entropy, $S_M=\int (C_{\rm mag}/T) dT$, approaches $R\ln 2$ at $T=200$\,K,  consistent with a spin-1/2 system. Notably, $S_{\rm mag}$ displays a broad kink anomaly at $\sim 50$\,K, indicating a stepwise release of nearly half the entropy. This behavior suggests the decomposition of entropy release into two fermionic components, supporting the fractionalization hypothesis. 

In the intermediate temperature range, 50\,K $\alt T \alt T_H$, $C_{\rm mag}$ exhibits a temperature-linear dependence. This observation aligns with theoretical predictions of a metal-like density of states for itinerant Majorana fermions subjected to significant scattering by $Z_2$ fluxes \cite{nasu2015thermal}. The specific heat profile further reveals a sharp peak at  $T_N\approx 6.5$\,K, preceded by a broad peak at $\sim 10$\,K. The nature of this broad peak remains ambiguous, as $C_M$ in the temperature range just above $T_N$ is influenced by short-range AFM fluctuations, potentially obscuring any fractionalization-related effects.

Widermann {\it et al.}\ report specific heat measurements on  $\alpha$-RuCl$_3$ from 2 to 250\,K in zero field and in-plane fields of 5.5 and 9\,T \cite{widmann2019thermodynamic}. To isolate the magnetic contribution, they employ single-crystalline RhCl$_3$ as a non-magnetic reference and perform {\it ab-initio} calculations of phonon density of states for both RhCl$_3$ and $\alpha$-RuCl$_3$. These calculations reveal that a simple mass-scaling approach is insufficient to determine the intrinsic phonon contribution; substantial renormalization is necessary even for the structurally similar non-magnetic RhCl$_3$. The specific heat exhibits field-independence above 30\,K, and $C_M$ displays a broad peak around 70\,K, lower than the previously reported value \cite{do2017majorana}. 

 In a 9-T in-plane field, sufficient to fully suppress AFM order and induce a quantum disordered state, $C_M$ exhibited a broad maximum at $\approx 10$\,K.  Notably, while the low-temperature anomaly in $C_M$ carried minimal weight, the high-temperature anomaly at $\sim 70$\,K approaches $(R/2)\ln 2$, consistent with the expected entropy release due to itinerant Majorana fermions. However, the low-temperature anomaly's scale exceeded theoretical predictions by an order of magnitude. This discrepancy may be attributed to the significant alteration of Majorana fermion excitation spectra by the applied magnetic field, particularly for itinerant fermions, complicating direct comparison with theoretical models. In contrast to the results by Do {\it et al.}, the study by Widermann {\it et al.}\ does not explicitly report on the presence or absence of temperature-linear specific heat in the $T_L < T < T_H$ range predicted by Nasu {\it et al.}, leaving this aspect of potential Majorana fermion behavior unaddressed.

The analysis of the magnetic contribution of the specific heat anomalies appears to be influenced by the estimation of the phonon background.  Despite slightly different phonon properties, however, the overall features of the anomaly of the specific heat are in good agreement between the two groups.

\subsection{Raman spectroscopy}
 \begin{center}
 \begin{figure*}[t]
	\includegraphics[clip,width=0.8\linewidth]{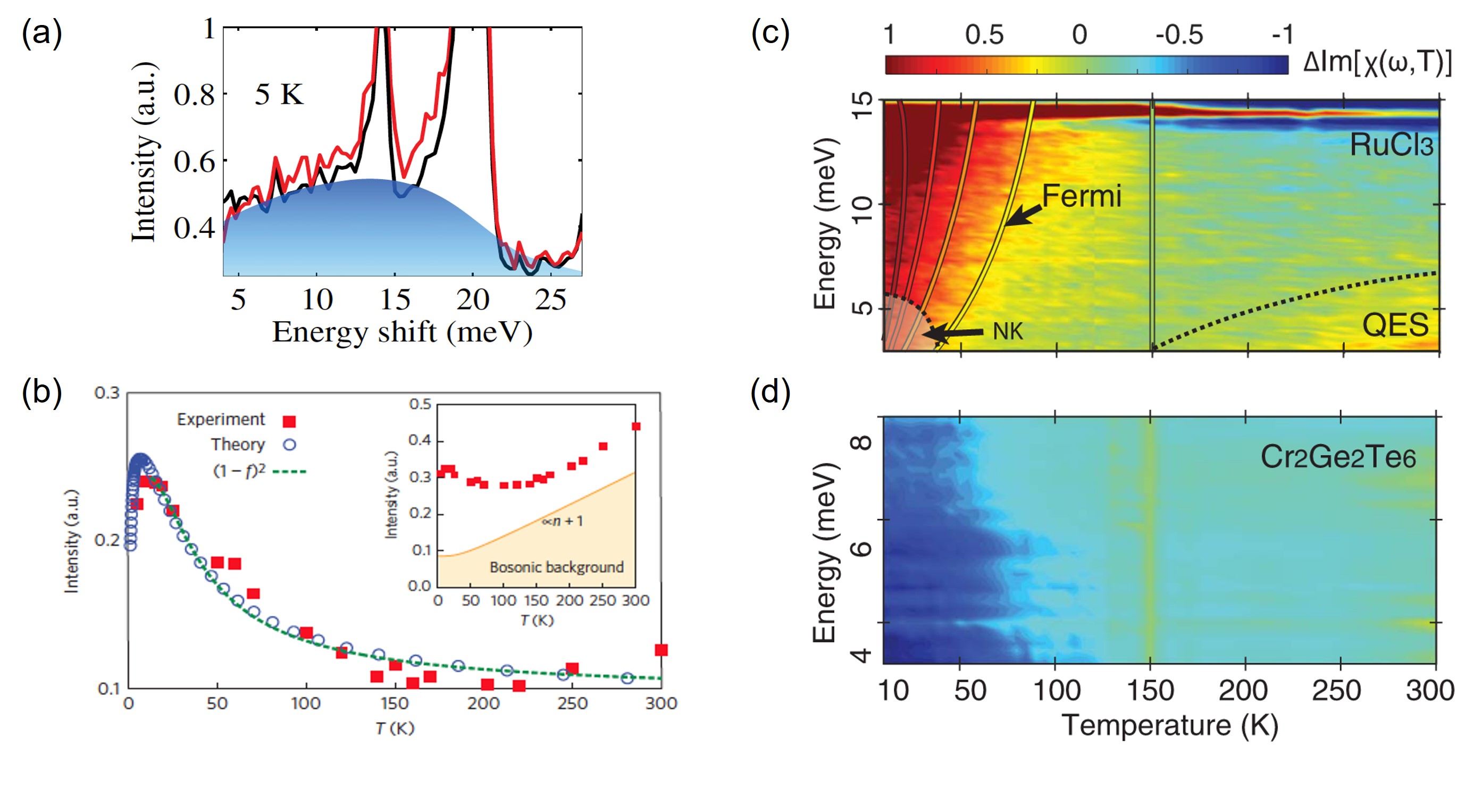}
	\caption{(a) Polarized Raman spectroscopy measurements of $\alpha$-RuCl$_3$ at 5\,K.  The spectrum consists of distinct spectral features: well-defined phonon modes at 14 and 20\,meV, superimposed on a broad, low-intensity magnetic continuum extending to approximately 20\,meV (blue shading) \cite{sandilands2015scattering}. (b) A quantitative analysis of the $T$-dependence is compared with theoretical studies based on fermionic excitations (indicated by blue circles) in Sandilands {\it et al}. The inset depicts the temperature evolution of the integrated Raman spectral intensity. The $T$-dependence of the spectral weight exhibits significant deviation from the conventional bosonic behavior characteristic of traditional insulating magnets, where both magnons and phonons conform to Bose-Einstein statistics. The theoretical model demonstrates exceptional concordance with experimental data across an extensive temperature range, from above $T_N$ to $\sim$150\,K \cite{nasu2016fermionic}.
    (c) The Raman susceptibility analysis of $\alpha$-RuCl$_3$. To examine the fermionic contribution, differential susceptibility, $\Delta {\rm Im}[\chi(\omega,T)]={\rm Im}[\chi(\omega,T)]-{\rm Im}[\chi(\omega,150\,{\rm K})]$ is analyzed. The superimposed contour plots (denoted by black outlines) represent the temperature-dependent Fermi function difference $\Delta n_F(\omega/2,T)=n_F(\omega/2,150\,{\rm K})$-$n_F(\omega/2,T)$. The observed concordance between experimental data and the theoretical contours was interpreted as fermion-pairs excitations. Notably, this fermionic signature persists even in the non-Kitaev (NK) regime. At elevated temperatures and low energies, the Raman intensity exhibits a characteristic upturn, attributable to thermal magnetic fluctuations manifesting as quasi-elastic scattering (QES). \cite{wang2020range} (d) In contrast, another honeycomb lattice system, Cr$_2$Ge$_2$Te$_6$, which exhibits ferromagnetic ordering below 60\,K and possesses a Curie-Weiss temperature comparable to $\alpha$-RuCl$_3$, displays inverse characteristics relative to $\alpha$-RuCl$_3$, with consistently negative susceptibility $\Delta {\rm Im}[\chi(\omega,T)]$ across the entire measured range, which further diminishes with decreasing temperature \cite{wang2020range}.}
\label{fig:Raman}
\end{figure*}
    \end{center}
    
 Raman spectroscopy can serve as a powerful probe for investigating quantum spin systems, revealing the presence of long-range order,  symmetry, and statistics of the quasiparticles.     Although the spins are fractionalized, the spin operator excites itinerant Majorana fermion and $Z_2$ flux simultaneously.  It is usually difficult to observe either one of them separately. 
 Knolle {\it et al.}\ has theoretically proposed that the Raman response of the gapless QSL of the Kitaev-Heisenberg model exhibits characteristic signatures of both excitations \cite{knolle2014raman}.   While the itinerant Majorana fermions give rise to a broad continuum, the $Z_2$ fluxes contribute a sharp feature due to the Heisenberg interaction.   For the direct comparison with experiments, Nasu {\it et al.}\ made the finite temperatures calculations \cite{nasu2016fermionic}.  

Several groups have conducted the spectroscopic studies of spin excitations in $\alpha$-RuCl$_3$ using Raman scattering techniques \cite{sandilands2015scattering,glamazda2017relation,wang2020range,wulferding2020magnon}. Zero-field measurements revealed an unusual broad continuum in the magnetic scattering spectrum. Figure\,\ref{fig:Raman}(a) depicts the magnetic continuum at 5\,K reported by Sandilands {\it et al.}\ \cite{sandilands2015scattering}.  This continuum persists over a wide temperature range above $T_N$, suggesting the presence of frustrated magnetic interactions. Analysis of phonon linewidths, which can exhibit similar anomalies due to spin-phonon coupling, indicated that such coupling does not appreciably damp the magnetic excitations in $\alpha$-RuCl$_3$. Notably, the integrated Raman intensity of the broad magnetic continuum below approximately 100\,K increases as the temperature decreases towards $T_N$, as shown in Fig.\,\ref{fig:Raman}(b). These findings provide valuable insights into the magnetic behavior of $\alpha$-RuCl$_3$ and its potential as a candidate for the Kitaev QSL state.

Nasu {\it et al.}\ \cite{nasu2016fermionic}  provided a theoretical framework to explain both the broad magnetic continuum observed in the Raman spectrum and the unusual temperature dependence of the Raman scattering intensity that were experimentally reported by Sandilands {\it et al}.\ \cite{sandilands2015scattering}, as shown in Fig.\,\ref{fig:Raman}(b) and its inset. For bosonic excitations ($bb^{\dagger}= b^{\dagger}b+1$), the scattering amplitude $\langle bb^{\dagger}\rangle$ is obtained by a simple one-particle photon scattering process,  which is proportional to $n+1$ with $n=1/(e^{\hbar\omega_b/k_BT}-1)$ being the Bose distribution function  ($\hbar\omega_b$ is the energy of bosons).  Therefore, the enhancement of the Raman intensity with decreasing temperature below $\sim$100\,K in $\alpha$-RuCl$_3$ cannot be explained by bosonic excitations, such as phonons and magnons, since the number of bosons decreases monotonically with decreasing temperature.   For fermions,  there are two different Raman scattering processes that correspond to the creation or annihilation of a pair of fermions (process (A)) and a combination of creation and annihilation of fermions (process (B)).   Process (A) is proportional to $[1-f(\omega_1)][1-f(\omega_2)]\delta(\omega-\omega_1-\omega_2)$  with $f(\omega_{1,2})=1/(1+e^{\hbar\omega_{1,2}/k_BT})$, where $\hbar\omega_1$ and  $\hbar\omega_2$  are the energies of fermions and  $\omega~(=\omega_1+\omega_2$) is the Raman shift.  Process (B) is proportional to  $f(\omega_1)[1-f(\omega_2)]\delta(\omega+\omega_1-\omega_2)$.  Process (B) vanishes at $T=0$ due to the absence of fermions in the ground state.   In Process (A), as temperature decreases, the population of high-energy fermions diminishes, facilitating easier excitation of fermions. This leads to an enhancement of Raman intensity. The observed temperature dependence of Raman intensity thus reveals the fermionic nature of the magnetic excitations. However, in real compounds, bosonic scattering must also be considered due to residual interactions that generate incoherent magnetic excitations, including correlated magnons. A study demonstrated that calculations based on Majorana fermions with zero chemical potential and positive frequencies ($\omega_1>0,~\omega_2>0$) well reproduce experimental results when accounting for a bosonic background \cite{nasu2016fermionic}, as shown in the inset of Fig.\,\ref{fig:Raman}(b). 

Glamazda {\it et al.}\ have reported a similar enhancement of integrated Raman intensity with decreasing temperature below $\sim$80\,K \cite{glamazda2017relation}, slightly lower than the temperatures reported by Sandiland {\it et al.}\  \cite{sandilands2015scattering}. The Raman intensity was well reproduced by combining a two-fermion creation or annihilation process with a bosonic background, consistent with Nasu {\it et al.}\ \cite{nasu2016fermionic}. Additionally, Glamazda {\it et al.}\ have reported hysteretic behavior of magnetic excitations over a wide temperature range below the first-order structural transition from monoclinic to rhombohedral at 165\,K \cite{glamazda2017relation}. By comparing Raman results with those of CrCl$_3$, a Heisenberg counterpart that undergoes a similar structural transition, they highlighted that small variations in bonding geometry significantly influence Kitaev magnetism and its associated fractionalized excitations.

Early Raman spectroscopy studies were constrained by weak scattering signals, requiring both energy averaging and integration across wide spectral ranges. Moreover, challenges in accurately determining and removing the bosonic background hinder meaningful comparisons with theoretical predictions derived from the exact Kitaev model. Wang {\it et al.}\ overcame these limitations through innovative high-resolution Raman measurements, achieving substantially enhanced signal intensity, broader temperature range, and superior energy resolution \cite{wang2020range}. Their advanced methodology enables direct extraction of the Raman susceptibility through Stokes and anti-Stokes intensity comparison, eliminating the need for bosonic background corrections.  Through comprehensive analysis of the energy and temperature evolution of the Raman susceptibility in $\alpha$-RuCl$_3$, Wang {\it et al.}\ proposed that Raman scattering generates magnetic excitations comprising fermion pairs \cite{wang2020range}. Their findings reveal remarkable agreement between the measured Raman susceptibility and quantum Monte Carlo calculations for the pure Kitaev model at elevated temperatures and energies (Fig.\,\ref{fig:Raman}(c)). A crucial comparative analysis between the Raman susceptibility of $\alpha$-RuCl$_3$ and the ferromagnetic honeycomb system Cr$_2$Ge$_2$Te$_6$ (which shares a similar Curie-Weiss temperature) underscores the distinctive features of $\alpha$-RuCl$_3$ and strengthens its connection to Kitaev physics (Fig.\,\ref{fig:Raman}(d)). They further observed significant deviations between the measured Raman susceptibility and quantum Monte Carlo calculations at low temperatures (below 40\,K) and energies (below 6\,meV). 

They attributed these discrepancies to non-Kitaev interactions, utilizing the characteristic temperature and energy scales to determine the ratio between Kitaev ($K$) and off-diagonal ($\Gamma$) interactions. The authors reported that these fractional excitations follow Fermi statistics, even in regions where non-Kitaev terms become dominant. However, this conclusion was drawn by comparing with quantum Monte Carlo calculations on the pure Kitaev model. It is unclear, however, how a simple reduction of the measured quantities to the pure Kitaev theory can be applied in general, especially when these interactions are interplaying, and the system deviates from the specific limit where such a comparison is valid.  

Multiple research groups have independently reported quasielastic Raman scattering in $\alpha$-RuCl$_3$, which intensifies at reduced energies and elevated temperatures \cite{sandilands2015scattering,wulferding2020magnon,wang2020range}. This behavior bears a striking resemblance to observations in the quantum spin liquid candidate Herbertsmithite ZnCu$_3$(OH)$_6$Cl$_2$, which realizes a perfect kagome lattice---though disorder effects hamper detailed analysis of Herbertsmithite. These parallel observations provide crucial insights into both the interplay between Kitaev and non-Kitaev interactions in $\alpha$-RuCl$_3$ and the broader characteristics shared among QSL candidates. The distinctive temperature evolution of the quasielastic scattering presents additional evidence for unconventional magnetism in these materials.

Raman spectroscopy studies have consistently demonstrated that the temperature evolution of the spectral weight in $\alpha$-RuCl$_3$ follows a distinctly fermionic pattern, persisting well above $T_N$, rather than exhibiting conventional bosonic behavior \cite{sandilands2015scattering,glamazda2017relation,wang2020range,wulferding2020magnon}. This remarkable observation indicates that the magnetic excitations in $\alpha$-RuCl$_3$ fundamentally differ from traditional bosonic magnons, suggesting an unconventional magnetic ground state. The experimental findings align with theoretical studies \cite{knolle2014raman,nasu2016fermionic,yoshitake2016fractional,yoshitake2017majorana} and suggest that $\alpha$-RuCl$_3$ is in proximity to a QSL.

\subsection{Neutron scattering}

 \begin{figure*}[t]
	\includegraphics[clip,width=0.8\linewidth]{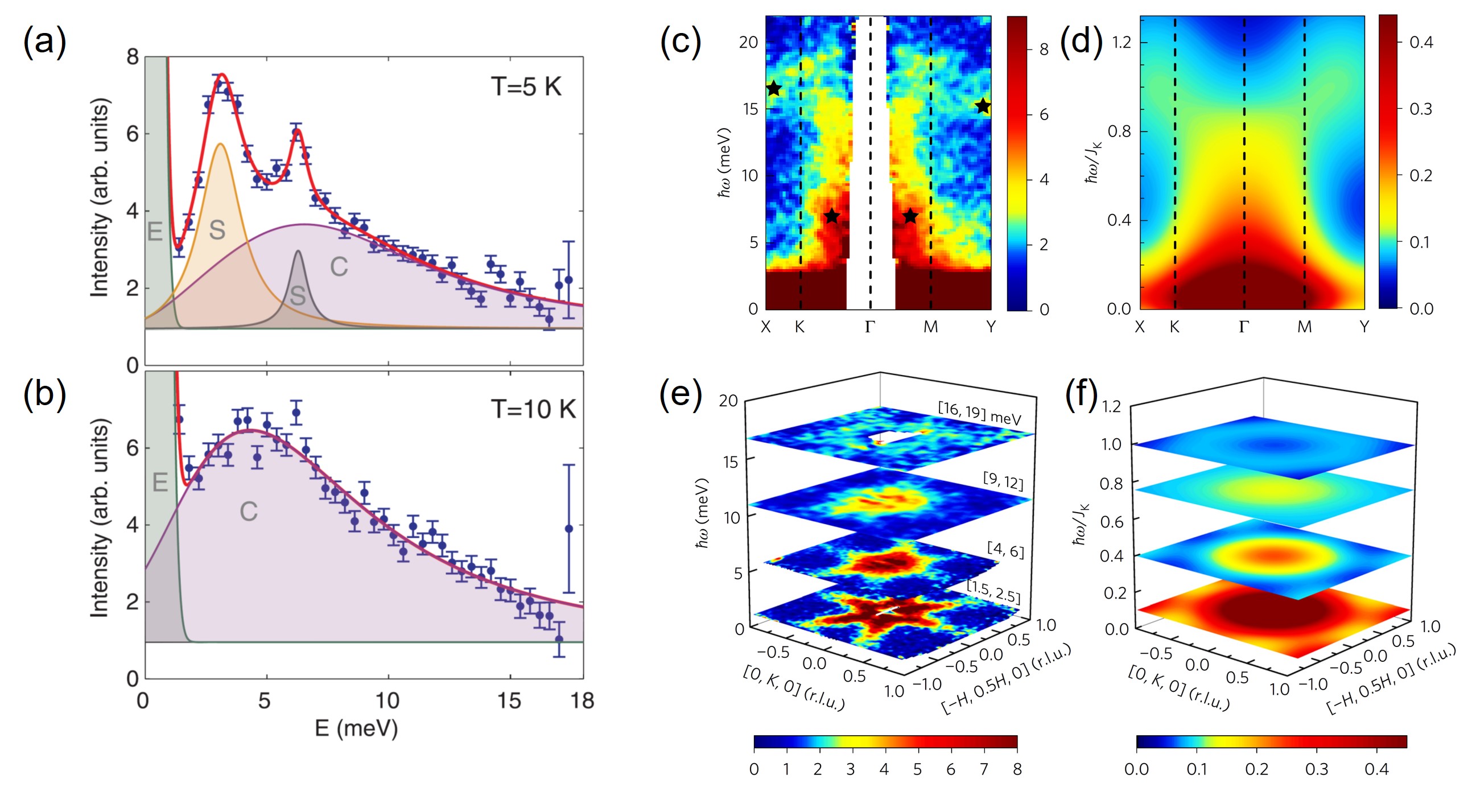}
	\caption{
Inelastic neutron scattering spectra of $\alpha$-RuCl$_3$ (a) in the zigzag AFM state ($T<T_N$) and (b) in the paramagnetic state  ($T>T_N$).   In the paramagnetic phase, a broad continuum of magnetic excitations (designated ``C'') extends to approximately 15\,meV. This spectral signature exhibits remarkable consistency with theoretical predictions derived from the Kitaev model, where the Kitaev interaction is estimated to fall within the range of 5.5 to 8\,meV.  In the zigzag AFM phase, the system manifests a partial redistribution of spectral weight, characterized by the emergence of well-defined spin-wave excitations (denoted ``S''). These sharp features coexist with the remnant continuum, indicating the persistence of quantum fluctuations even in the AFM-ordered state \cite{banerjee2017neutron}. (c)  Total neutron scattering function $S_{\rm tot}({\bm Q},\omega)$, predominantly reflecting magnetic scattering $S_{\rm mag}({\bm Q},\omega)$, measured at $T=10$\,K (above $T_N$) along the $X$-$K$-$\Gamma$-$M$-$Y$ direction, where ${\bm Q}$ and $\omega$ represent momentum and energy transfer, respectively. The spectrum exhibits a distinctive hourglass shape centered at the $\Gamma$-point, extending up to approximately 20\,meV. This pattern is characterized by intense low-energy excitations near the $\Gamma$-point and high-energy excitations forming a Y-shaped pattern. (d) The experimental observations of (c) are well reproduced by theoretical simulations of the isotropic Kitaev model with ferromagnetic Kitaev interactions \cite{yoshitake2016fractional}. The low-energy features correspond to quasielastic responses associated with $Z_2$-flux excitations, while the Y-shaped dispersion in the high-energy regime reflects the itinerant Majorana fermions with energies extending up to $\omega\sim |K|$.
(e) Constant-energy cuts $S_{\rm tot}(Q)$ in the $(hk)$-plane obtained through energy-integrated measurements. (f) Theoretical calculations of $S_{\rm mag}({\bm Q})$ derived from the Kitaev model to compare with (e). The simulations successfully reproduce the primary experimental features, suggesting a robust manifestation of Majorana fermion characteristics. However, the hexagram-shaped ${\bm Q}$-dependence observed at low energies ($\omega \lesssim 6$\,meV) deviates from simulation predictions, likely due to non-Kitaev interactions \cite{do2017majorana}.	
}
\label{fig:Neutron1}
\end{figure*}

Neutron diffraction confirms a zigzag AFM ground state in the honeycomb lattice, characterized by unusually small ordered moments, suggesting strong quantum fluctuations. INS reveals anisotropic magnetic excitations, indicating significant SOC. These findings, interpreted within the $j_{\rm eff}$ = 1/2 framework, align with theoretical predictions for systems near the Kitaev QSL state.

INS measurements provide direct access to magnetic excitations through the neutron scattering function $S({\bm Q},\omega)$, which relates to the imaginary part of the dynamical susceptibility $\chi''({\bm Q},\omega)$ via:
\begin{equation}
S({\bm Q},\omega) \propto [1 - \exp(-\hbar \omega/k_BT)]^{-1} \chi''({\bm Q},\omega).
\end{equation}
This relationship enables the investigation of spin liquid behavior, particularly thermally fractionalized spin excitations, which multiple research groups have explored through comprehensive neutron scattering studies of candidate Kitaev materials. 

Initial INS experiments on powder $\alpha$-RuCl$_3$ samples conducted by Banerjee {\it et al.}\  have revealed an anomalous magnetic excitation spectrum in the AFM zigzag ordered state \cite{banerjee2016proximate}. The INS data elucidate two distinct collective magnetic modes with disparate energy gaps, denoted as $M_1$ and $M_2$, manifesting near the $M$ point in reciprocal space.
The $M_1$-mode, characterized by a low energy gap of approximately 2\,meV, defies reproduction by the conventional linear spin wave theory. Upon heating above $T_N$, the $M_1$-mode exhibits marked softening, with spectral weight redistribution towards $Q=0$.
Conversely, the well-defined $M_2$-mode, featuring a substantial gap of approximately 6\,meV, persists with analogous $Q$-dependence up to at least 70\,K ($\sim 10T_N$).  This temperature and energy dependence has been posited as incompatible with standard spin wave theory predictions.  In fact, spin waves in conventional magnets typically display a pronounced intensity attenuation above $T_N$.
Furthermore, in the magnetically ordered state, the energy width of the $M_2$-mode is significantly broader than linear spin wave theory predictions. This broadening suggests the presence of additional scattering mechanisms or intrinsic lifetime effects not captured by simple spin wave models.  Comparative analysis of these unconventional excitation spectra with theoretical calculations incorporating Majorana fermions and $Z_2$-flux excitations derived from the pure Kitaev model has led to the proposition that $\alpha$-RuCl$_3$ is a prime candidate for the realization of fractionalized Kitaev physics.

In single-crystal INS on $\alpha$-RuCl$_3$, Banerjee {\it et al.}\ and Do {\it et al.}\ presented detailed magnetic excitation spectra under zero-field conditions, as shown in Figs.\,\ref{fig:Neutron1}(a)-(f) \cite{do2017majorana,banerjee2017neutron}. These high-resolution measurements revealed directional anisotropy in momentum space, providing crucial insights into the material's magnetic behavior. These measurements provide a comprehensive determination of the dynamical magnetic susceptibility $\chi({\bm q},\omega)$, both in the zigzag AFM state and well above $T_N$, offering insights previously unattainable through INS on polycrystalline samples or single-crystal Raman scattering studies.
In the zigzag AFM state, low-energy spin wave excitations with well-defined dispersion are observed, exhibiting a minimum at the $M$-point of the Brillouin zone (Fig.\,\ref{fig:Neutron1}(a)). The most striking feature in this regime is the coexistence of magnons with an intense excitation continuum centered at the 2D $\Gamma$-point. This continuum spans a broad energy range of 2-15\,meV, a characteristic incompatible with conventional spin wave responses \cite{do2017majorana,banerjee2017neutron}.

Above the Néel temperature, as shown in Figs.\,\ref{fig:Neutron1}(b) and (c), the spin wave excitations vanish, giving way to a broad scattering continuum in the dynamical structure factor $S({\bm Q},\omega)$ centered at the $\Gamma$-point. This continuum manifests as an hour-glass-shaped spectrum extending to approximately 20\,meV, featuring pronounced low-energy excitations near the $\Gamma$-point. Remarkably, the continuum persists across an extensive temperature range up to 120\,K ($\gg T_N$), corresponding to the characteristic energy scale of the Kitaev interaction. The scattering intensity, distributed across a substantial portion of the Brillouin zone, indicates short-range correlations typical of a spin liquid state. The rod-like scattering profile along the out-of-plane momentum direction further emphasizes the two-dimensional character of the magnetic response. At higher energies (above 15\,meV), $S({\bm Q},\omega)$ exhibits distinctive Y-shaped, horn-like excitations \cite{do2017majorana}, as shown in Figs.\,\ref{fig:Neutron1}(c) and (d). As the system transitions into the paramagnetic phase with increasing temperature, the low-energy spectral weight diminishes significantly while the high-energy features largely persist, albeit with enhanced broadening. This characteristic temperature evolution provides vital insights into the nature of magnetic correlations across different phases in $\alpha$-RuCl$_3$.

The magnetic excitation spectrum of $\alpha$-RuCl$_3$ shows key features predicted for a ferromagnetic Kitaev model, as initially reported by multiple groups \cite{yoshitake2017temperature,yoshitake2017majorana,knolle2015dynamics}. Laurell and Okamoto later showed that including non-Kitaev terms improves agreement with experimental data \cite{laurell2020dynamical}. The high-energy Y-shaped $({\bm Q}, \omega)$ dependence, observed by Do {\it et al.}, reflects itinerant Majorana fermion dispersion following $\omega$ $\sim$ $|{\bf k}|$. Meanwhile, Yoshitake {\it et al.}\ pointed out that the low-energy features represent quasi-elastic responses from $Z_2$-flux excitations \cite{yoshitake2017temperature,yoshitake2017majorana}. The $\Gamma$-point energy continuum, persistent both above and below $T_N$, matches theoretical predictions for a Kitaev QSL at zero temperature, suggesting the presence of the dominant Kitaev interaction, a hallmark of Kitaev physics.

\red{The experimental data also reveal features that cannot be fully accounted for by the pure Kitaev model. Notably, as shown in Fig.\,\ref{fig:Neutron1}(e), a six-pointed star shape in the ${\bf Q}$-dependence of $S({\bf Q},\omega)$ is observed at low energies ($\omega \lesssim 6$ meV) around the $\Gamma$-point \cite{do2017majorana,banerjee2017neutron}. Below $T_N$ in zero field, this star-shaped pattern originates from low-energy spin waves at the $M$ points, which are visible at all six $M$ points due to the presence of different magnetic domains in the zigzag ordered phase. While the general star shape is a consequence of the zigzag order and is not specific to any particular term in the Hamiltonian, the detailed energy scale and intensity distribution of this feature depend on the microscopic interactions.}

\red{It should be noted that the interpretation of color representations in time-of-flight INS measurements requires careful consideration of the integration limits chosen in $Q$-$E$ space, as broad integration ranges can potentially conflate phonon and magnon scattering. Furthermore, the INS measurements by Balz {\it et al.} \cite{balz2019finite} revealed that spin-wave excitations in the ordered phase exhibit a dependence on the out-of-plane momentum, indicating that the underlying magnetic interactions possess some 3D character.}

\red{This star-shaped pattern is not observed in theoretical calculations based on the pure Kitaev model, as shown in Fig.\,\ref{fig:Neutron1}(f). Gohlke {\it et al.}\ showed that when the AFM $\Gamma$ interaction is introduced to a FM-like Kitaev system, the dynamical structure factor exhibits enhanced star-shaped scattering intensity at low energy scales \cite{gohlke2018quantum}, which is consistent with the experimental observations. These results indicate that the interplay between Kitaev and non-Kitaev interactions, together with the three-dimensional nature of the magnetic couplings, shapes the low-energy physics and magnetic behavior of $\alpha$-RuCl$_3$.}

These observations collectively provide strong evidence for the proximity of $\alpha$-RuCl$_3$ to a Kitaev QSL state, while also emphasizing the importance of considering additional interactions for a complete description of its low-energy physics. The coexistence of Kitaev-like features with deviations from the pure model underscores the richness of the physics in this material and its potential as a platform for studying QSL physics in real systems.

The INS spectra reveal distinctive features: a broad, continuous excitation spectrum spanning $\sim 1$-15\,meV, and temperature-dependent spectral weight redistribution that persists well above the magnetic ordering temperature. These characteristics align with theoretical predictions for the Kitaev QSL state and find support in complementary Raman spectroscopy measurements, which also show a broad magnetic excitation continuum. While this congruence between INS, Raman data, and theoretical models suggests proximity to the Kitaev QSL state, the influence of non-Kitaev interactions requires careful consideration.
The interpretation of the broad magnetic continuum remains debated. The Kitaev QSL framework attributes it to multiple excitations of coherent fractional particles. However, Winter {\it et al}.\ proposed an alternative view \cite{winter2017breakdown} that does not require spin fractionalization. In this perspective, the continuum represents incoherent excitations arising from magnetic anharmonicity, driven by frustrated anisotropic interactions. Such interactions, inherent to systems with strong SOC and significant off-diagonal $\Gamma$-terms, cause conventional magnon descriptions to break down even within the magnetically ordered phase.

Applying magnetic fields in the 2D honeycomb plane suppresses the zigzag AFM order,  leading to the FIQD state above $\mu_0H\approx 8$\,T. 
A detailed analysis of the high-field INS results will be presented in a dedicated section.

\subsection{Inelastic X-ray scattering}

 \begin{figure}[b]
	\includegraphics[clip,width=1.0\linewidth]{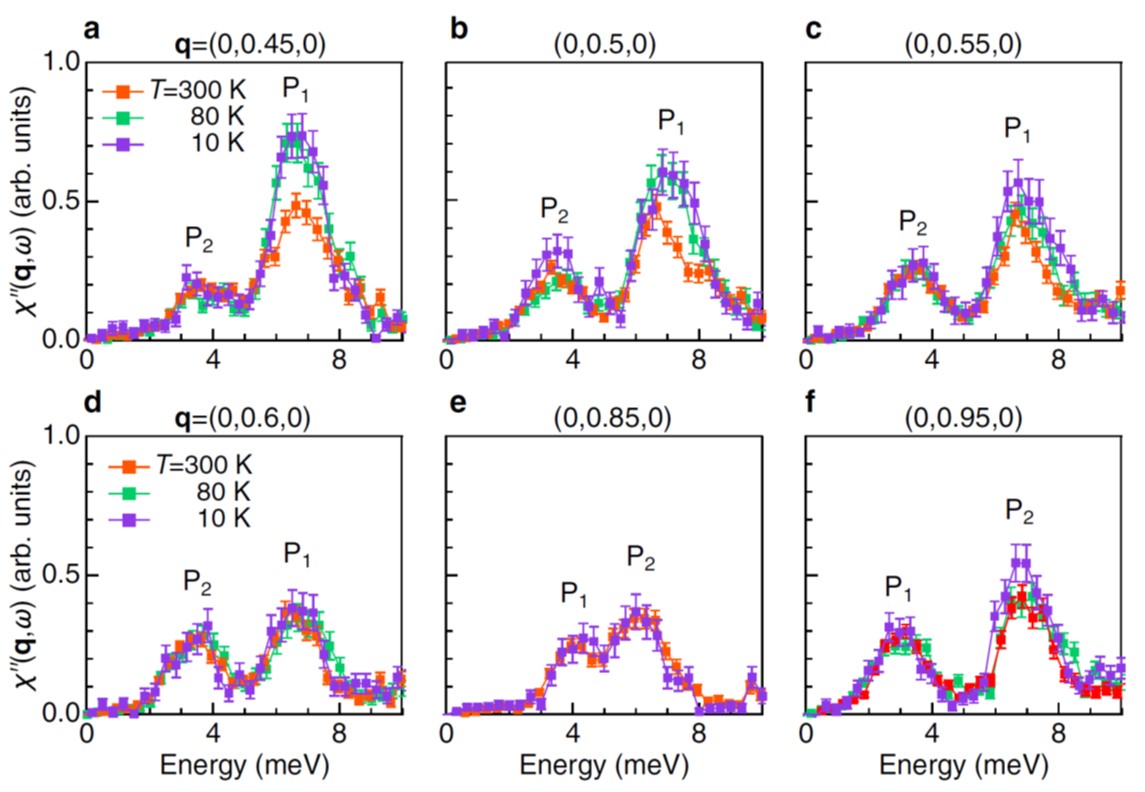}
	\caption{Optical phonon spectra of $\alpha$-RuCl$_3$ as a function of reduced momentum transfer {\boldmath $q$}, with the $M$ point defined at $\bm{q} = (0, 0.5, 0)$. Two optical phonon branches (P$_1$ and P$_2$) are observed. The relative peak positions switch at $\bm{q} = (0, 0.75, 0)$. The $\chi''(${\boldmath $q$}, $\omega$)  exhibit a characteristic frequency of $\omega \approx$ 7 meV, comparable to the estimated Kitaev interaction energy $K$. Upon cooling from 300\,K to 10\,K, the spectra show significant enhancement in spectral weight near the $M$ point, while maintaining constant intensity at other momentum transfers \cite{li2021giant}.
}
\label{fig:InelasticXray}
\end{figure}

The coupling between lattice vibrations and exotic quasiparticles in QSLs can reveal signatures of fractional excitations in phonon spectra. Theoretical predictions by Metavitsiadis {\it et al.}\ and Ye {\it et al.}\ suggest that interactions between phonons and fractionalized excitations (Majorana fermions and $Z_2$-gauge fluxes) in Kitaev QSLs produce distinctive zero-field phonon signatures \cite{metavitsiadis2020phonon,ye2020phonon}. Specifically, the dynamical phonon susceptibility,
$\chi_{\rm ph}''({\bm Q},\omega)\propto S_{\rm ph}({\bm Q}, \omega)\times(1-e^{-\hbar\omega/k_BT})$,
provides crucial insights into the Majorana-phonon coupling through the phonon anomalies., where $\chi_{\rm ph}''({\bm Q},\omega)$ is the imaginary component and $S_{\rm ph}({\bm Q}, \omega)$ is the dynamical phonon structure factor directly measured by IXS. 

In $\alpha$-RuCl$_3$, the Ru-Cl-Ru bond angles of approximately 90$^{\circ}$ combine with moderate SOC to favor Ising-type magnetic interactions oriented perpendicular to the Ru-Cl-Ru plane. Magnetic interactions are perturbatively modified by lattice vibrations, resulting in coupling between phonons and spins of the system. All bond-dependent interactions, including $\Gamma$ and $\Gamma'$, are sensitive to such vibrations.  The Kitaev interaction magnitude $K$ is estimated to range from 5 to 11\,meV, which notably corresponds to the upper limit of certain optical phonon bands, suggesting the possibility of detecting the fractionalization through phonon anomalies. Theoretical work by Metavitsiadis and Brenig demonstrated that the Majorana-phonon coupling in the pure Kitaev model exhibits momentum dependence, reaching its maximum near the $M$ point \cite{metavitsiadis2020phonon}. \red{They suggested that the anomalous dispersion of Majorana fermions (Dirac cone), scattering from thermally excited $Z_2$ gauge fluxes, and enhanced contribution from particle-particle channels lead to completely different phonon broadening behavior compared to conventional metals and semiconductors.} However, to make a direct comparison, studies on the generic model, including $\Gamma$ and $\Gamma'$ bond-dependent interactions, are needed.

Li {\it et al.}\ conducted IXS measurements under zero-field conditions and confirmed the above predictions, revealing anomalous effects in honeycomb-plane optical phonons \cite{li2021giant}.  Figure \ref{fig:InelasticXray} shows the optical phonon spectra measured at different reduced momentum transfers {\boldmath $q$}, where the $M$ point corresponds to $\bm{q} = (0, 0.5, 0)$. The spectra reveal two optical phonon branches, labeled P$_1$ and P$_2$.  The relative peak positions switches at $\bm{q}=(0, 0.75, 0)$. The temperature dependence of $\chi''(${\boldmath $q$}, $\omega$) shows a spectral weight enhancement at a phonon frequency of $\omega \approx 7$\,meV, which is close to the Kitaev interaction energy $K$. As the temperature is lowered from 300\,K to 10\,K, the optical phonons display a marked enhancement in spectral weight near the $M$ point. Notably, this enhancement is localized to the vicinity of the $M$ point, with no corresponding increase observed at momentum transfers far from this point. The authors claim that these observations are consistent with the theoretically predicted coupling between itinerant Majorana fermions and phonons, demonstrating both momentum and energy-dependent behavior.

Li {\it et al.}\ also observed a significant softening of acoustic phonons below 2\,meV that intensifies below $\sim 100$\,K, coinciding with the Kitaev interaction's temperature scale. They detected a phonon peak shift of $\sim 0.3$\,meV below $\sim 100$\,K, comparable to shifts seen in well-established electron-phonon coupled systems and notably occurring near the Kitaev interaction's energy scale. Two mechanisms have been proposed to explain this phonon softening: a) The intersection of dispersionless $Z_2$  flux excitations ($\Delta_K \sim 0.065|K|$) with linearly dispersing acoustic phonons near $\hbar\omega \sim 0.065|K| \sim 0.5$\,meV leads to a phonon anomaly at this energy scale.
b) The acoustic phonons and itinerant Majorana fermions share nearly identical linear dispersions as $\bm{q} \rightarrow 0$, resulting in phonon dispersion renormalization below the Kitaev interaction temperature.

The observation of phonon anomalies at two distinct energy scales---corresponding to the Kitaev interaction and $Z_2$ flux excitation energies---is intriguing.  The authors claim that this is consistent with fractionalization. \red{However, theoretically discussed phonon broadening \cite{metavitsiadis2020phonon} has not been observed in IXS experiments, and the development of higher-resolution IXS experimental techniques may be crucial.}
The experimental findings require further systematic investigation to quantitatively understand the phonon modes' origins and definitively establish their connection to fractional excitations in the Kitaev materials, including $\Gamma$ and $\Gamma'$ interactions. Notably, these phonon anomalies display well-defined energy scales with sharp features, in apparent contrast to the broad magnetic continuum observed in inelastic neutron scattering experiments.

\subsection{Terahertz spectroscopy}

Terahertz spectroscopy has emerged as a powerful experimental probe for investigating the quantum spin dynamics of $\alpha$-RuCl$_3$ in zero magnetic field \cite{little2017antiferromagnetic,reschke2018sub,wang2017magnetic,shi2018field,wu2018field,reschke2019terahertz}. This technique's unique capabilities in accessing the relevant energy scales have been crucial in understanding the material's magnetic excitations, with theoretical frameworks \cite{suzuki2018effective,bolens2018mechanism} providing essential insights into the observed phenomena.

The experimental configuration employs near-vertical incidence of terahertz radiation to optimize the alignment of the probing field with the 2D honeycomb plane, thereby maximizing sensitivity to dynamical responses near the $\Gamma$-point. Multiple studies consistently reveal a broad continuum of magnetic quasiparticle excitations spanning approximately 2-15\,meV at temperatures above $T_N$. This continuum exhibits remarkable thermal evolution, persisting up to $\sim 200$\,K while showing rapid development and a broad maximum as temperature decreases.

 \begin{figure}[b]
	\includegraphics[clip,width=0.85\linewidth]{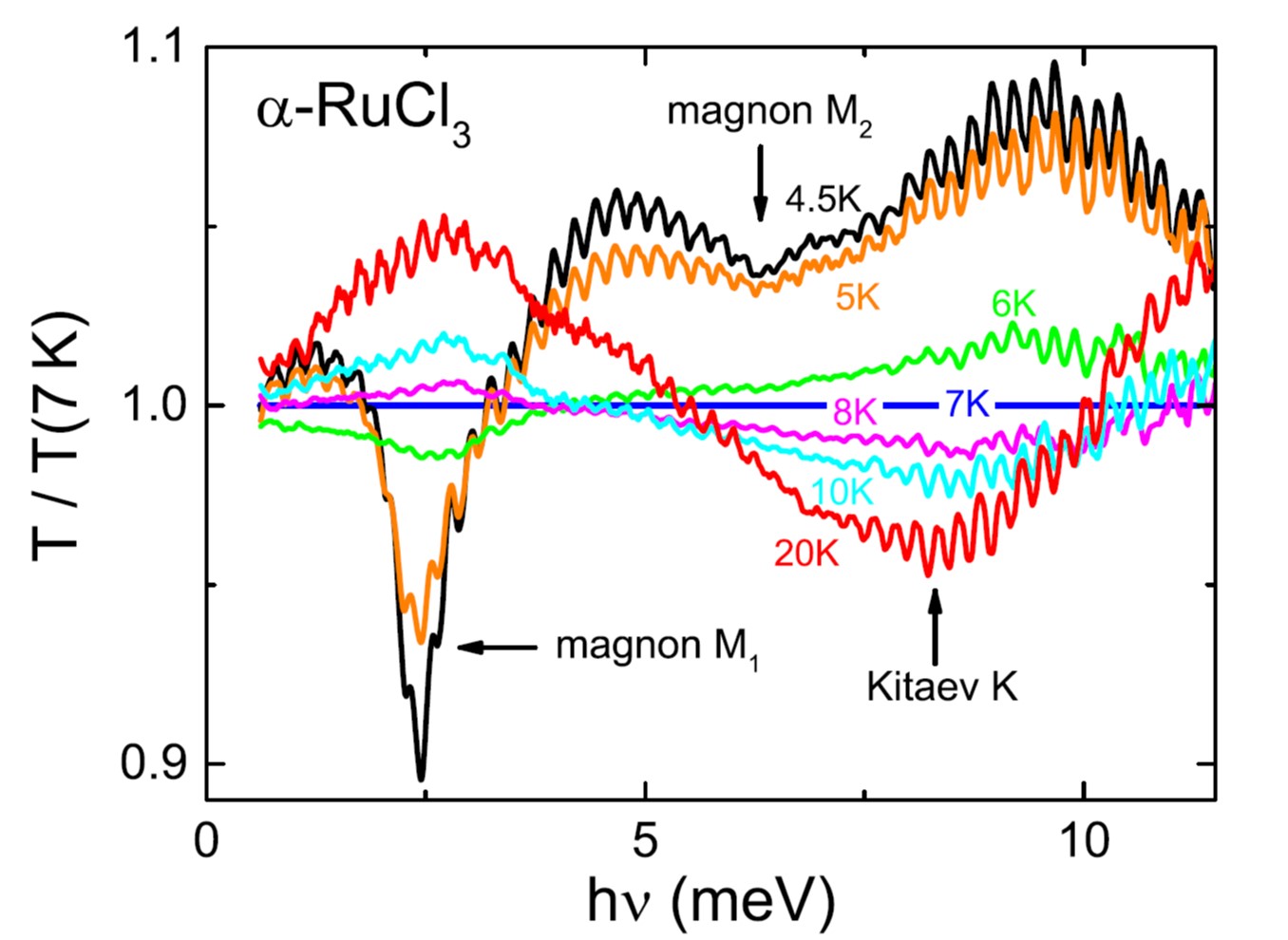}
	\caption{THz transmission spectra of $\alpha$-RuCl$_3$ from 1-12\,meV, normalized to spectra at  7\,K ($\approx T_N$). Below $T_N$, the spectral weight redistributes into two distinct magnetic excitations: a sharp magnon mode (M$_1$) at 2.5\,meV and a broader mode (M$_2$) at 6.5\,meV.  M$_1$ represents an AFM resonance from single-magnon excitations, while M$_2$ arises from two-magnon processes. A broad magnetic continuum persists even in the AFM state, consistent with INS data. Above $T_N$, both magnetic modes vanish as a broad continuum (K) emerges around 8.5\,meV, interpreted as evidence for fractionalized excitations characteristic of Kitaev QSLs \cite{reschke2019terahertz}.
	}
    \label{fig:THz1}
\end{figure}

 Figure\,\ref{fig:THz1} presents the normalized THz transmission spectra of $\alpha$-RuCl$_3$ measured at temperatures above and below $T_N$, as reported by Reschke {\it et al.}\ \cite{reschke2019terahertz}.   Below $T_N$, the spectral weight undergoes a redistribution, manifesting in two distinct magnetic excitations: a sharp mode (magnon M$_1$) at around 2.5\,meV and a broader mode (magnon M$_2$) at around 6.5\,meV. Similar spectra have been reported in previous THz spectroscopy studies \cite{little2017antiferromagnetic,wang2017magnetic,shi2018field,wu2018field}.  These observations corroborate INS measurements, demonstrating spectral weight transfer from a continuum to well-defined magnon excitations below $T_N$\cite{banerjee2016proximate,do2017majorana}. The 2.5\,meV mode has been identified as an AFM resonance corresponding to single-magnon excitations \cite{little2017antiferromagnetic}, while the 6.5\,meV feature is attributed to two-magnon processes \cite{wang2017magnetic,reschke2019terahertz}, consistent with INS data.
 
Above $T_N$, both magnetic modes are completely suppressed, and a broad continuum emerges centered at approximately 8.5\,meV (Kitaev K) in the paramagnetic phase, significantly reducing transmission in this energy range. Reschke {\it et al.}\ present this as the first THz spectroscopic evidence that the high-temperature continuum originates from fractionalized spin excitations characteristic of a Kitaev spin liquid, rather than conventional magnetic excitations \cite{reschke2019terahertz}.  The temperature evolution of the spectrum, particularly at 8, 10, and 20\,K, demonstrates the complete disappearance of magnetic modes just above $T_N$ and the emergence of a continuum at higher energies. The observation of persistent fractionalized quasiparticles above $T_N$ aligns with theoretical predictions for Kitaev spin liquids \cite{rousochatzakis2019quantum}.  Notably, a broad magnetic continuum above $T_N$ persists even in the AFM-ordered state, consistent with INS observations \cite{banerjee2016proximate}.
  
Winter {\it et al.}\ have proposed that this continuum above $T_N$ represents incoherent excitations arising from strong magnetic anharmonicity, primarily through single-magnon decay processes \cite{winter2017breakdown}. However, several inconsistencies challenge this magnon anharmonicity scenario. First, single magnons become unstable due to terms of anharmonic interaction. Furthermore, magnon fluctuations persist at temperatures exceeding 10\,$T_N$, a phenomenon unexpected within conventional magnon theory. Additionally, the energy range of the observed continuum significantly surpasses that of the one-magnon response. Studies beyond single-magnon decay is needed. 

Wang {\it et al.}\ and Reschke {\it et al.}\ have reported that terahertz experiments further reveal a step-like decrease in optical weight upon the first-order transition from the low-temperature rhombohedral phase to the high-temperature monoclinic phase \cite{wang2017magnetic,reschke2018sub,reschke2019terahertz}.  The authors suggest that this indicates a significant enhancement of Kitaev interactions in the rhombohedral phase with its near-ideal honeycomb lattice structure.  A similar large effect on the excitation spectra at the first-order structural transition is observed in Raman experiments \cite{glamazda2017relation}. 

It is worth noting that the terahertz observations show remarkable consistency with complementary experimental techniques, particularly INS and Raman spectroscopy. INS measurements reveal a magnetic continuum centered at the $\Gamma$-point, extending to energies of approximately 15\,meV and persisting up to temperatures of roughly 150\,K \cite{banerjee2016proximate,banerjee2017neutron,do2017majorana}. Raman spectroscopy complements these findings by detecting continuum features at even higher energies (up to approximately 30\,meV) that remain discernible even at room temperature \cite{sandilands2015scattering,glamazda2017relation,wang2020range,wulferding2020magnon}. 

The striking convergence of evidence from the above four distinct spectroscopic methods---Raman, INS, IXS, and terahertz---suggests a common frustrated magnetism origin of the observed continua. This correlation between multiple spectroscopic techniques appears to strengthen the hypothesis that $\alpha$-RuCl$_3$ exhibits properties consistent with proximity to Kitaev QSL behavior, despite the appearance of AFM order at low temperatures.  Although this may be a characteristic magnetic behavior in quantum magnets with dominant Kitaev interactions, the effect of other bond-dependent interactions on these experimental quantities requires further studies.

\subsection{Thermal transport properties}
 
Both localized and itinerant excitations influence the specific heat of a system, whereas the longitudinal thermal conductivity tensor component $\kappa_{xx}$ is exclusively determined by itinerant excitations. Notably, thermal conductivity measurements are immune to the Schottky anomaly, enabling the extension of experimental investigations into ultra-low temperature regimes. Consequently, $\kappa_{xx}$ serves as a sensitive probe for low-energy itinerant excitations within the system. 

Fractionalized excitations may play a crucial role in mediating thermal transport in quantum magnets. Several research groups have conducted comprehensive thermal conductivity measurements on $\alpha$-RuCl$_3$ in zero field, spanning a wide temperature range from below $T_N$ to significantly above it \cite{leahy2017anomalous,hirobe2017magnetic,kasahara2018unusual,hentrich2018unusual,lefranccois2022evidence}. Despite some variations between studies, the overall behavior of $\kappa_{xx}$ exhibits consistent trends.

  \begin{figure}[b]
	\includegraphics[clip,width=0.9\linewidth]
    {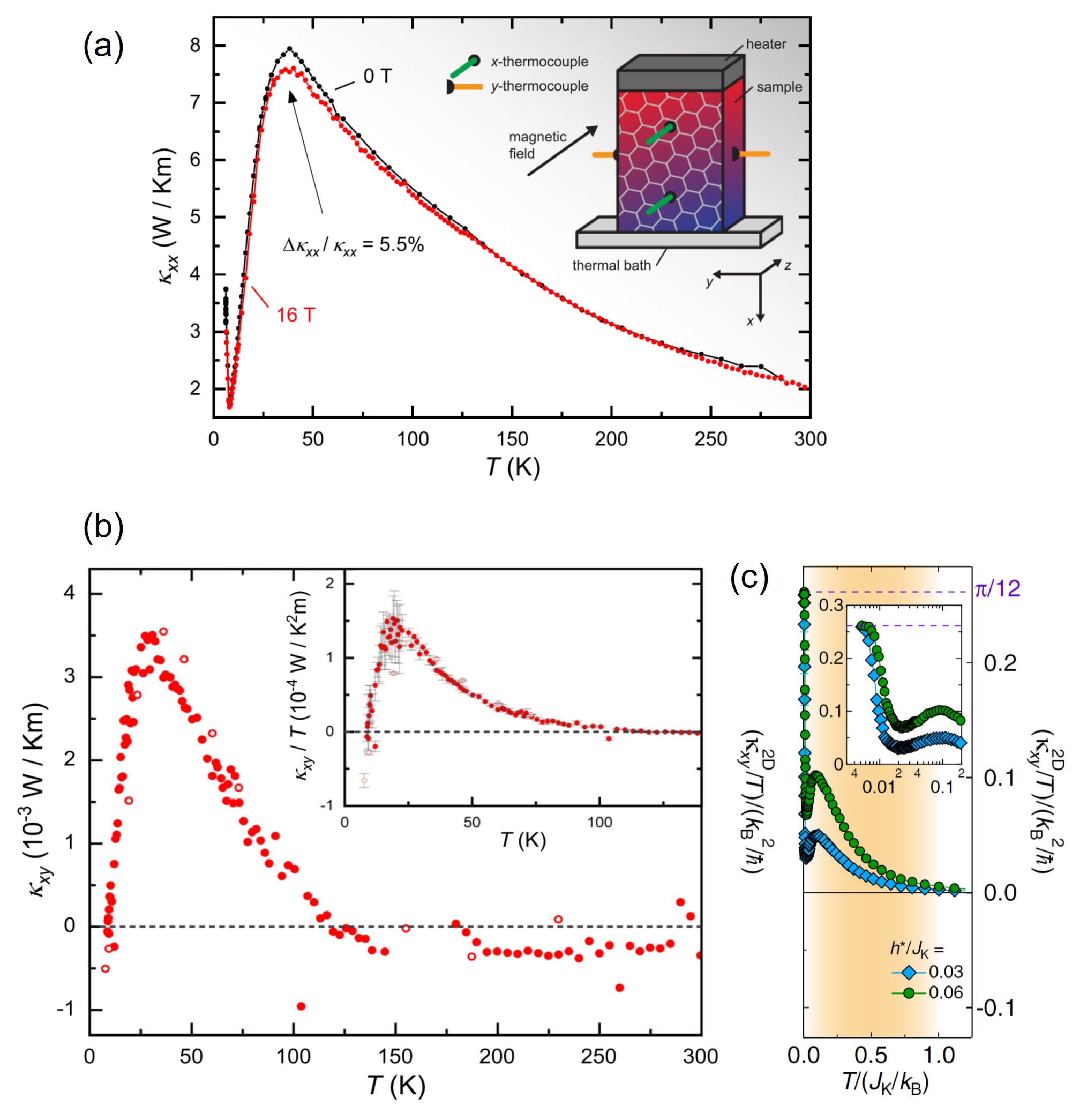}
	\caption{Temperature dependence of (a) thermal conductivity $\kappa_{xx}$ and (b) thermal Hall conductivity $\kappa_{xy}$. Upon application of a magnetic field $H$ perpendicular to the honeycomb lattice planes, a significant positive $\kappa_{xy}$ is observed, exhibiting linear scaling with field strength $H$. The temperature dependence of $\kappa_{xy}/T$ manifests below approximately 100\,K, coinciding with the energy scale of Kitaev interactions (inset). $\kappa_{xy}/T$ reaches a broad maximum at approximately 25\,K, followed by a steep decrease below this temperature. Notably, $\kappa_{xy}$ undergoes a sign reversal at or marginally above the Néel temperature $T_N$ \cite{hentrich2019large}. (c) Temperature dependence of the normalized thermal Hall conductivity $(\kappa_{xy}^{2D}/T)/(k_B^2/\hbar)$ calculated theoretically for the pure Kitaev model under an effective magnetic field $h^* \propto H^3$. Results are plotted versus reduced temperature $T/(J_K/k_B)$, where $J_K$ is the Kitaev interaction strength. Inset: Same data shown on a semilogarithmic temperature scale \cite{nasu2017thermal}.}
    \label{fig:Thermalconductivity}
\end{figure}
 
  As shown in Fig.\,\ref{fig:Thermalconductivity}(a),  with decreasing temperature, $\kappa_{xx}$  increases steadily up to a distinct maximum at around 40\,K, and decreases rapidly at lower temperatures \cite{hentrich2019large}.   At $T_N$, a sharp kink anomaly is observed in $\kappa_{xx}$, followed by a steep increase as the temperature further decreases \cite{kasahara2018unusual,hentrich2018unusual,lefranccois2022evidence}. The broad peak at around 40\,K is caused by the phonon Umklapp process, which is a typical behavior of conventional phononic heat conductors.   The steep enhancement of $\kappa_{xx}$ below $T_N$ is attributed to the suppression of the AFM fluctuations due to the long-range ordering, which results in the rapid increase of phonon mean free path.

In addition to the broad peak at around 40\,K, Hirobe {\it et al.}\ have reported a secondary broad peak in thermal conductivity at around 100\,K \cite{hirobe2017magnetic}.  This temperature scale is comparable to that of the Kitaev couplings in this system.  This additional feature becomes apparent after subtracting the calculated phonon thermal conductivity from the total measured thermal conductivity.  Although the estimation of phonon contribution should be scrutinized, Hirobe {\it et al.}\ \cite{hirobe2017magnetic}  claim that this higher-temperature broad peak originates from the magnetic contribution and shares a common origin with the broad peak observed in specific heat measurements \cite{do2017majorana,widmann2019thermodynamic}.   Hirobe {\it et al.}\ claimed that zero-field thermal conductivity measurements at elevated temperatures align with specific heat data, with both experimental techniques providing evidence supporting the presence of spin fractionalization. 

The anisotropic field dependence of $\kappa_{xx}$ above $T_N$ suggests strong spin-phonon coupling in $\alpha$-RuCl$_3$. For perpendicular magnetic fields $\bm{H} \parallel c$), where $T_N$ remains relatively constant, $\kappa_{xx}$ shows minimal field dependence. In contrast, parallel fields ($\bm{H} \parallel ab$) exhibit a striking enhancement of $\kappa_{xx}$ above the critical field $\mu_0H_c \approx 8$ T, where $T_N$ vanishes. Notably, Hentrich {\it et al.}\ reported a similar increase in the out-of-plane thermal conductivity ($\kappa_{zz}$), observing an enhancement of approximately one order of magnitude above $H_c$ \cite{hentrich2018unusual}. These observations indicate that phonons dominate the heat transport in $\alpha$-RuCl$_3$, with magnetic excitations significantly influencing phononic heat transport along all crystallographic directions through scattering mechanisms. The field dependence of $\kappa_{xx}(H)$ for $\bm{H} \parallel ab$ is particularly interesting above the critical field, which remains one of the most studied phenomena in $\alpha$-RuCl$_3$. This behavior will be discussed in a separate section.

The thermal Hall conductivity ($\kappa_{xy}$) under magnetic fields applied parallel to the $c$ axis ($\bm{H} \parallel c$) has been extensively studied by several research groups \cite{kasahara2018unusual,hentrich2019large,lefranccois2022evidence}. Figure\,\ref{fig:Thermalconductivity}(b) and its inset show the temperature dependence of $\kappa_{xy}$ and $\kappa_{xy}/T$ at $\mu_0H=16$\,T, as reported by Hentrich {\it et al.}\ \cite{hentrich2019large}. The thermal Hall conductivity exhibits distinctive features across different temperature ranges. At high temperatures, $\kappa_{xy}/T$ remains negligible. However, when the temperature drops below approximately 100\,K---corresponding to the energy scale of Kitaev interactions---$\kappa_{xy}/T$ begins to increase steadily.  The positive sign of $\kappa_{xy}$ persists throughout most of the temperature range. As the temperature decreases, $\kappa_{xy}/T$ gradually increases, reaching its maximum near 20\,K, followed by a dramatic decrease. A sign reversal in $\kappa_{xy}/T$ occurs either slightly above \cite{hentrich2019large} or at $T_N$ \cite{kasahara2018unusual}. Kasahara {\it et al.}\ observed a distinct kink anomaly in $\kappa_{xy}/T$ just below $T_N$ \cite{kasahara2018unusual}. While the absolute magnitude of $\kappa_{xy}$ varies among studies---with Kasahara {\it et al.}\ reporting values 1-2 times larger than those of Hentrich {\it et al.}---the temperature dependence remains qualitatively consistent across different samples.

The temperature dependence of thermal Hall conductivity in the pure Kitaev model has been analyzed through quantum Monte Carlo simulations by Nasu {\it et al.}\ \cite{nasu2017thermal}, as shown in Fig.\,\ref{fig:Thermalconductivity}(c). Their numerical results reveal several distinct temperature regimes. Figure\,\ref{fig:Thermalconductivity}(c) shows the calculated thermal Hall conductivity per layer, $\kappa_{xy}^{2D}=\kappa_{xy}/d$ (where $d$ represents the interlayer distance), normalized by $(k_B^2/\hbar)$. In the high-temperature paramagnetic phase ($T > K/k_B$), $\kappa_{xy}^{2D}/T$ remains negligible. As the temperature drops below $K/k_B$ and the system enters the spin-liquid regime, $\kappa_{xy}/T$ becomes finite due to the emergence of itinerant Majorana fermions scattering off thermally excited $Z_2$ gauge fluxes. At intermediate temperatures ($T \approx 0.1K/k_B$), $\kappa_{xy}/T$ exhibits a characteristic broad maximum, arising from the interplay between itinerant Majorana fermions and $Z_2$ fluxes. With further cooling, $\kappa_{xy}^{2D}/T$ shows a rapid enhancement before converging to $(\pi/12)(\kappa_{xy}^{2D}/T)(k_B^2/\pi)$ as $T$ approaches zero, corresponding to half-integer quantum thermal Hall conductance---a phenomenon discussed in detail later.

A critical consideration in these numerical simulations is that they exclude non-Kitaev interactions, which are known to induce magnetic ordering in real materials. This limitation becomes particularly significant when comparing theoretical predictions with experimental measurements. The magnetic transition at $T_N$ fundamentally modifies the low-temperature thermal Hall response, thereby precluding direct observation of the predicted quantum regime.
Kasahara {\it et al.}\ proposed that several key experimental observations support the applicability of the Kitaev model. The peak temperature of $\kappa/T$, observed at approximately 20\,K, demonstrates remarkable agreement with theoretical calculations. Moreover, although non-Kitaev interactions may reduce the observed $\kappa/T$ peak magnitude to approximately 50\% of the pure Kitaev model prediction, the temperature evolution of the thermal conductivity above $T_N$ remains consistent with the behavior predicted for an ideal Kitaev magnet. However, Hentrich {\it et al.}\ identified notable discrepancies between experimental measurements and theoretical predictions, particularly regarding the field dependence of $\kappa_{xy}$.

Despite the constraints imposed by magnetic ordering and non-Kitaev terms, thermal Hall measurements in $\alpha$-RuCl$_3$ offer valuable insights into the nature of unconventional dynamics in these materials.  Recent studies have introduced additional complexity to the interpretation of experimental results. Lefran\c{c}ois {\it et al.}\ proposed that the thermal Hall conductivity of $\alpha$-RuCl$_3$  contains phonon contributions \cite{lefranccois2022evidence}. The phonon contribution to the thermal Hall conductivity merits detailed examination and will be addressed comprehensively in a dedicated section.

\section{Pressure induced state}

\begin{figure}[b]
	\includegraphics[clip,width=0.6\linewidth]
    {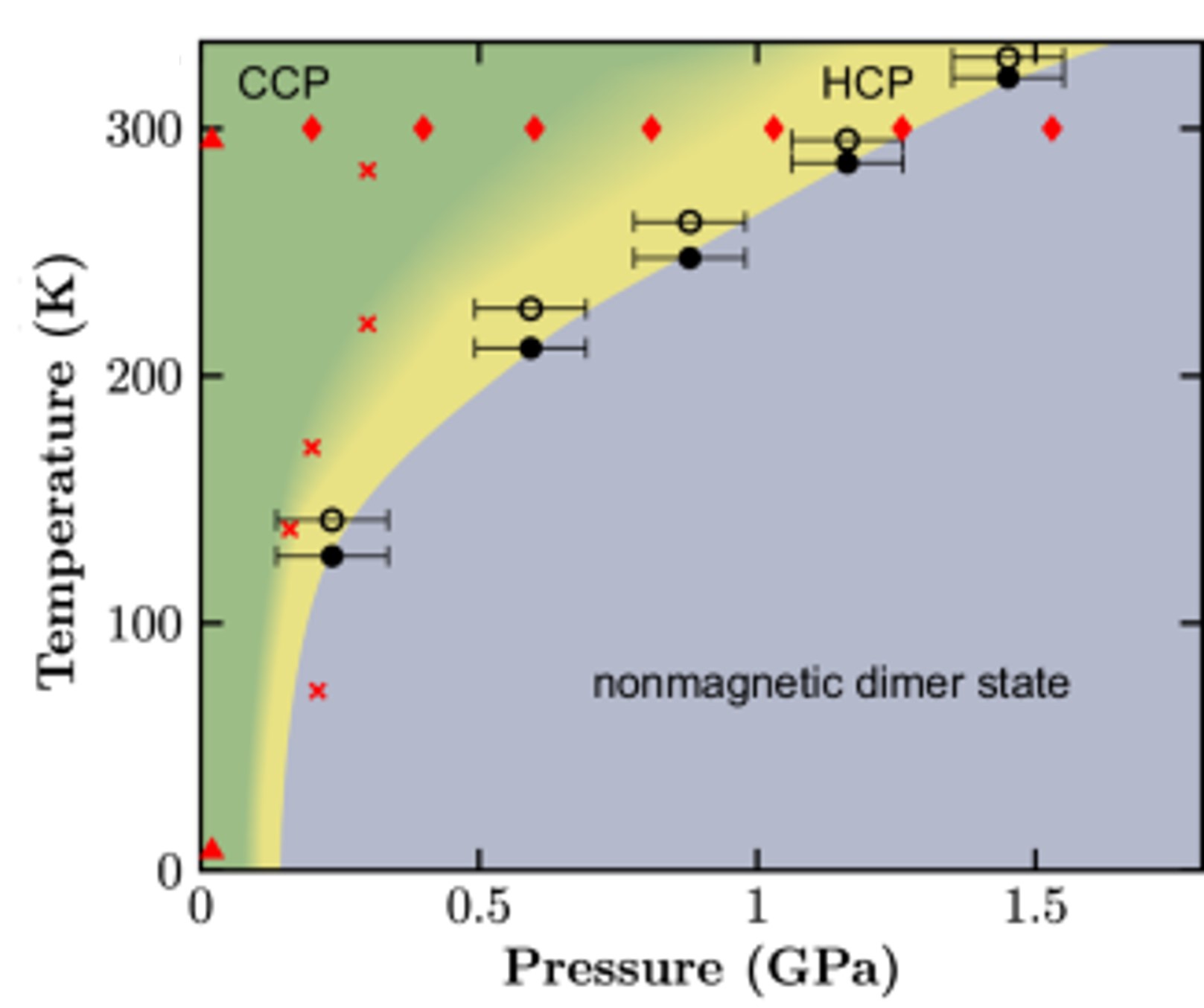}
	\caption{The pressure-temperature ($P$-$T$) phase diagram of $\alpha$-RuCl$_3$. At ambient pressure (depicted in green), the system maintains a cubic close-packed (CCP) chlorine arrangement down to 3\,K. As pressure increases, the material undergoes a transition to a phase characterized by hexagonal close-packed (HCP) chlorine coordination (shown in yellow). Further compression induces a transition to a nonmagnetic dimer state. These structural transitions are verified through single-crystal X-ray diffraction measurements \red{(red symbols).} 
    The magnetic phase boundary, determined via susceptibility measurements during cooling (solid black circles) and warming (open black circles), exhibits thermal hysteresis---a characteristic signature of a first-order phase transition \cite{stahl2024pressure}.}
    \label{fig:Pressure}
\end{figure}

At ambient pressure, deviations from the ideal Kitaev honeycomb geometry disrupt the formation of the QSL ground state, preventing its realization at low temperatures.   The long-range AFM order must be suppressed to realize a QSL state in $\alpha$-RuCl$_3$. Pressure was used to explore such possibilities.  To date, three groups have reported on the pressure effects on $\alpha$-RuCl$_3$ \cite{cui2017high,wang2018pressure,stahl2024pressure}. The pressure-temperature ($P$-$T$) phase diagram of $\alpha$-RuCl$_3$ reveals three distinct structural phases, as illustrated in Fig.\,\ref{fig:Pressure}.

Cui {\it et al.}\ conducted high-pressure magnetization and $^{35}$Cl NMR measurements on $\alpha$-RuCl$_3$ up to pressures of 1.5\,GPa. Their findings reveal that the N\'{e}el temperature remains nearly pressure-independent up to $P \sim 0.5$\,GPa, beyond which the magnetic volume fraction becomes progressively suppressed, exhibiting phase separation-like behavior. The magnetic volume completely vanishes at pressures exceeding 1.05\,GPa \cite{cui2017high}. Their measurements indicate that both the static susceptibility and low-energy spin fluctuations are nearly absent in the high-pressure regime, which they attribute to a magnetically spin-disordered state.

Wang {\it et al}.\ investigated the specific heat of $\alpha$-RuCl$_3$ up to $P \approx 1.6$\,GPa \cite{wang2018pressure}.  The pressure-temperature phase diagrams reported by their groups share similarities but differ in some aspects to Cui {\it et al.}  Wang {\it et al.}'s measurements show that the N\'{e}el temperature increases with pressure up to $\sim 0.7$\,GPa. At 1\,GPa, the specific heat measurements reveal no anomaly. Based on these observations, the authors propose that the AFM order dissolves at a critical pressure slightly below 1\,GPa. Additionally, they demonstrate that the high-pressure phase remains robust up to pressures of $\sim$140\,GPa. 

Recent synchrotron radiation studies by Stahl {\it et al.}, combining Bragg and diffuse scattering techniques, revealed a fundamental pressure-induced reorganization in $\alpha$-RuCl$_3$ \cite{stahl2024pressure}, as shown in Fig.\,\ref{fig:Pressure}. Their main claim is the emergence of a high-symmetry phase with nearly ideal conditions for Kitaev QSL behavior. This structural transformation establishes high trigonal symmetry within individual layers, yielding an unprecedented Kitaev-to-Heisenberg interaction ratio ($K/J=124$) - the highest reported in any material system to date.  Importantly, they pointed out that this high-symmetry phase can be stabilized using modest biaxial pressure, enabling diverse experimental investigations of both the phase itself and its potential QSL state. 

While the precise nature of the nonmagnetic ground state in pressurized  $\alpha$-RuCl$_3$ remains to be determined, Stahl {\it et al.}\ reported that pressure is a powerful tuning parameter for the quantum phase transitions between ordered and disordered states.   These findings open additional avenues for exploring the material's pressure-temperature phase diagram and its underlying physics.

  \begin{figure}[t]
 \includegraphics[clip,width=0.8\linewidth]{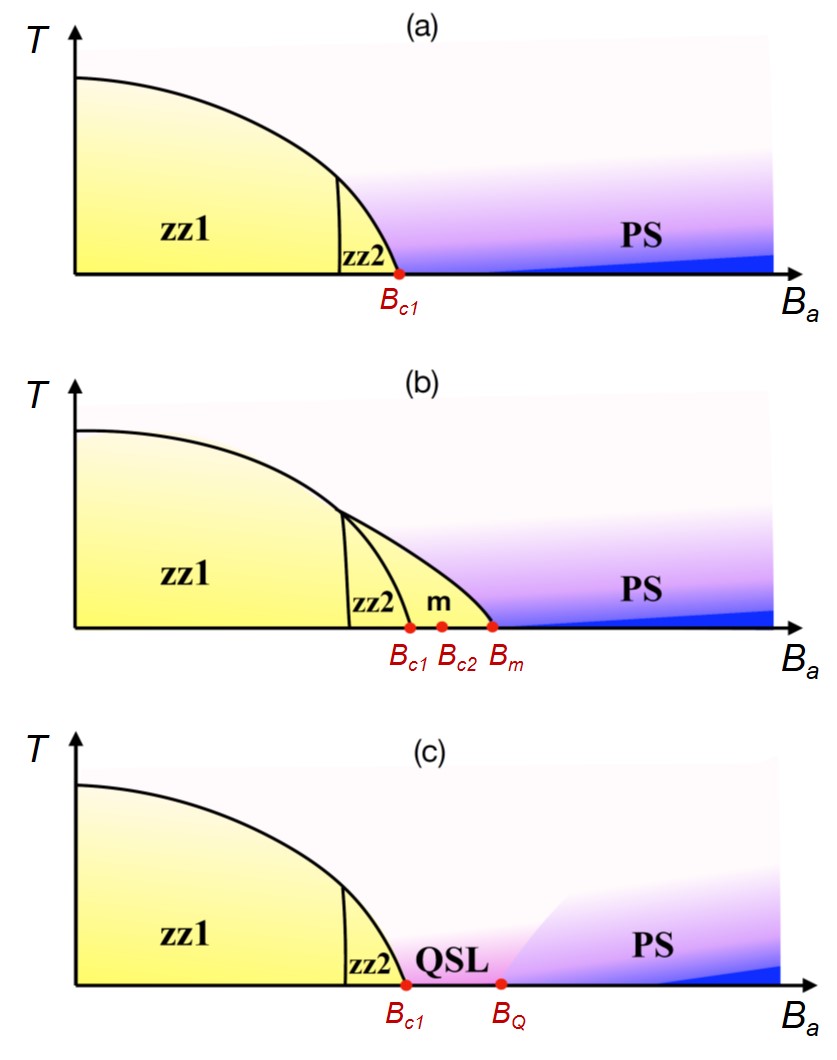}
\caption{The magnetic field-temperature phase diagram of $\alpha$-RuCl$_3$ exhibits three distinct scenarios under applied along the $a$-axis magnetic fields ($B _a$). The system demonstrates the following possible phase transitions:
(a) Scenario A: At the critical field $B_{c1}$ ($=\mu_0H_{ab}^c$), the system undergoes a direct quantum phase transition from zigzag AFM ordered ground states (zz1 and zz2) to a quantum paramagnet characterized by a partially polarized state (PS). The transition is marked by the suppression of long-range magnetic order.
(b) Scenario B: Multiple magnetic phase transitions occur at intermediate fields ($B_{c2}$) within a region denoted as $m$. Above a threshold field $B_m$, the system enters the PS. 
(c) Scenario C: This scenario features an intermediate QSL phase bounded by two distinct quantum critical points: (1) at $B_{c1}$, marking the transition from zigzag order to the QSL phase, and (2) at $B_Q$, corresponding to the transition from QSL to PS. The QSL phase is characterized by the absence of magnetic order while maintaining quantum entanglement. These temperature-field phase diagrams demonstrate fundamental differences in ground-state topology between the three scenarios, while sharing qualitatively similar finite-temperature behavior characterized by continuous thermal crossovers.
}
\label{fig:Scenario3}
\end{figure}

\section{Magnetic field-induced quantum-disordered state}

 \subsection{Magnetic field tuning of zigzag AFM order}

 Upon application of an in-plane magnetic field with a critical magnitude of $\mu_0 H_{ab}^c\approx 7$\,T, the zigzag AFM order is suppressed, giving rise to a FIQD state.   In the high-field limit ($H\gg H_{ab}^c$), the system enters a fully spin-polarized state \cite{kubota2015successive,zhou2023possible}.

A fundamental question in this field centers on whether distinct intermediate phases exist between the AFM ordered state and the spin-polarized state. Three competing theoretical scenarios have been proposed to explain the system's behavior above the critical in-plane magnetic field $H^c_{ab}$ (Fig.\,\ref{fig:Scenario3}), see, for example, \cite{trebst2022kitaev}. For details of the numerical calculations supporting these scenarios, we refer to \cite{li2021identification}.
 
 \begin{description}	
 	\item[Scenario A] Direct transition to a conventional spin-polarized phase  (Fig.\,\ref{fig:Scenario3}(a)). 
    	This scenario posits that a conventional spin-polarized phase emerges immediately above $H_c^{ab}$, featuring multiparticle magnetic excitations with integer quantum numbers.

    	\item [Scenario B]   
 	Transition to a series of magnetic orders before it enters a polarized phase (Fig.\,\ref{fig:Scenario3}(b)).
    One or more magnetic transitions may occur before it polarizes. In this scenario, stacking fault could have a serious impact.
    
     	\item [Scenario C]   
 	Emergence of a QSL including Kitaev QSL  (Fig.\,\ref{fig:Scenario3}(c)).
In scenario C, 
a chiral QSL phase may manifest within an intermediate field range above $H_{ab}^c$. This state is characterized by Majorana fermionic excitations of a non-Abelian nature, exhibiting 1/2-integer quantum numbers.

 \end{description}

The primary challenge in understanding the Kitaev physics in $\alpha$-RuCl$_3$ lies in differentiating between these scenarios and elucidating the nature of the FIQD states. This is crucial for determining whether one can tune materials to a Kitaev QSL phase starting from the AFM-ordered state. The FIQD state in $\alpha$-RuCl$_3$ has been the subject of extensive experimental investigations employing a diverse array of techniques. These studies aim to probe the magnetic, thermodynamic, and thermal transport properties of the material under applied magnetic fields.
 
The Kitaev model subject to an applied magnetic field has also been extensively investigated through theoretical approaches \cite{janssen2019heisenberg}. Multiple research groups have utilized a diverse array of computational and analytical techniques to study this system, including exact diagonalization \cite{hickey2019emergence,kaib2019kitaev,zhu2018robust,jiang2019field,patel2019magnetic} density-matrix renormalization group \cite{zhu2018robust,jiang2019field,patel2019magnetic,gohlke2018dynamical}, tensor-network methods \cite{lee2020magnetic}, quantum Monte Carlo techniques \cite{yoshitake2020majorana}, and slave-particle mean-field theories \cite{berke2020field}. Some of these approaches give the same conclusion that the ferromagnetic Kitaev model has a single transition into a polarized phase when the field is applied along the $a$ axis \cite{zhang2022theory}.
 
The phase diagram under magnetic fields applied perpendicular to the honeycomb plane (${\bm H} \perp ab$-plane) has not been as thoroughly investigated compared to the configuration. This is primarily because the zigzag AFM state shows remarkable robustness against fields applied along the c-axis, requiring fields exceeding 20\,T to suppress the AFM phase. Such anisotropy aligns with anisotropic susceptibility, due to AFM $\Gamma$ interaction, as discussed by \cite{plumb2014alpha}. Recall that $J_c-J_{ab} = \Gamma + 2 \Gamma'$ where the $\Gamma'$ is due to the trigonal distortion and/or the layer stacking and is assumed to be small. Even in an ideal octahedra environment with the isotropic $g$-factor, the magnetic anisotropy between the $ab$-plane and $c$-axis occurs due to the $\Gamma$ interaction \cite{Rau2014PRL}.   Modic {\it et al.}\ performed the measurements of magnetotropic coefficient--the thermodynamic coefficient associated with magnetic anisotropy--up to 35\,T, revealing an anomaly at $\mu_0 H_{c}^{\ell}\approx 30$\,T at 1.3\,K, suggesting a phase transition from the AFM state to a different phase \cite{modic2021scale}. This transition field decreases rapidly as the field direction is tilted away from the $c$-axis. Subsequently, Zhou {\it et al.}\ conducted magnetization measurements up to 102\,T, observing double-peak anomalies in $dM/dH$: one at 35\,T $\approx \mu_0 H_{c}^{\ell}$ (consistent with Modic {\it et al.}) and another at $\mu_0 H_{c}^{h}\approx80$\,T \cite{zhou2023possible}. These observations suggest the emergence of an intermediate phase between the AFM and fully polarized states. Zhou {\it et al.}\ claimed that this intermediate phase is a possible Kitaev QSL state. These experimental findings align with theoretical predictions by Gordon {\it et al.}, who showed that for magnetic fields perpendicular to the honeycomb plane, the Kitaev QSL state should be stabilized over a wide field range above $H_{c}^{\ell}$ when ferromagnetic Kitaev interactions coexist with antiferromagnetic off-diagonal symmetric interactions ($\Gamma$-term) \cite{gordon2019theory}.

\subsection{Inelastic neutron scattering}

  \begin{figure}[b]
	\includegraphics[clip,width=0.8\linewidth]
    {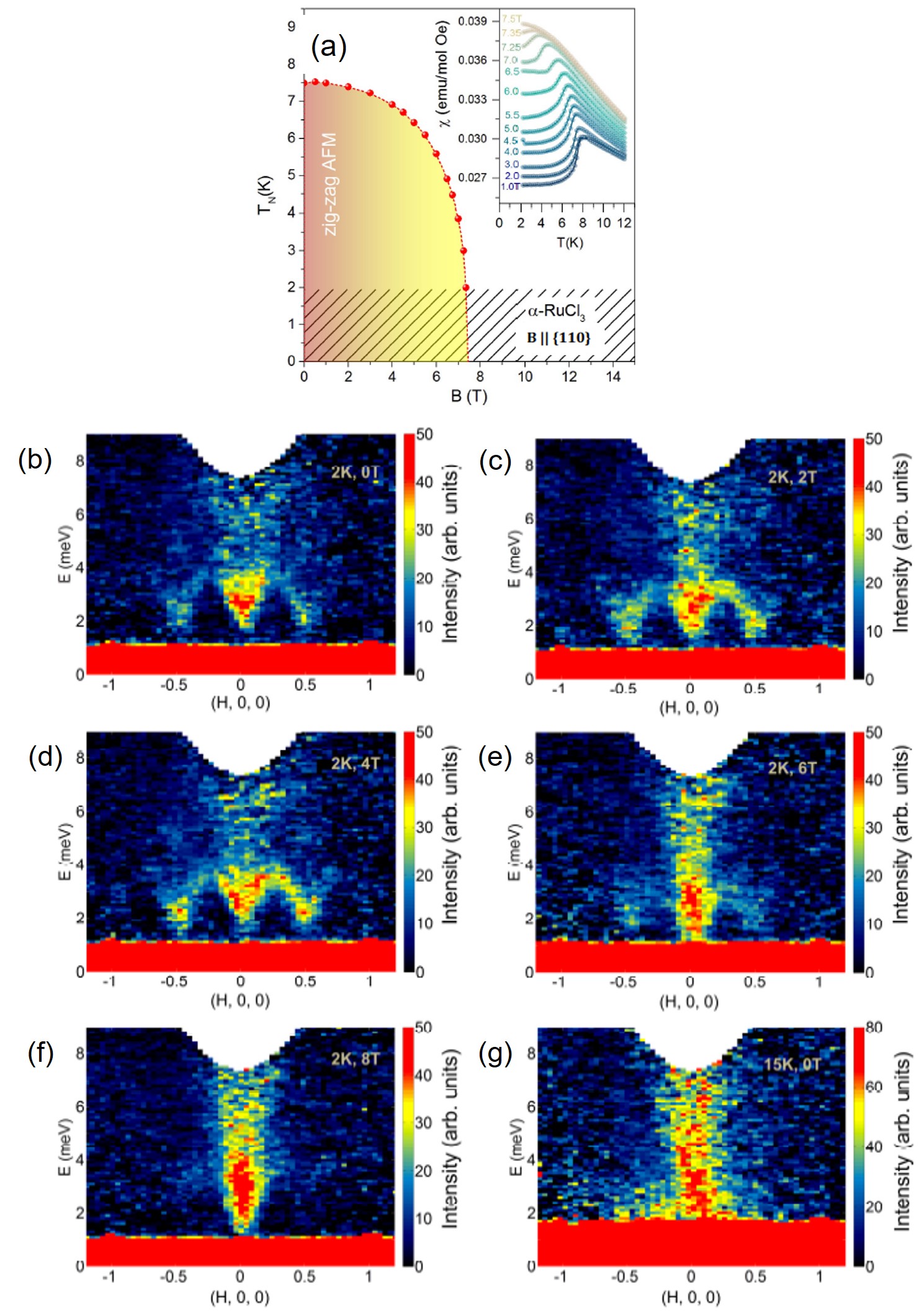}
	\caption{(a) $H$-$T$ phase diagram of $\alpha$-RuCl$_3$ for $\bm{H}\parallel a$ axis (zigzag direction) determined by magnetic susceptibility curves in fixed fields (inset). The zz1-zz2 phase boundary within the magnetically ordered region shown in Fig.\,\ref{fig:Magneticphase}(a) is omitted. Magnetic excitations in magnetic fields ((c)-(f)) and in zero field ((b) and (g)) along the ($H$, 0, 0) direction revealed by INS. The measurements are performed within ((b)-(e)) and outside ((f) and (g)) the AFM ordered phase. At $T=2$\,K with fields applied along $\eta= (-1, 2, 0)$ ($\bm{H}\parallel a$ axis), the measurements span 0-8\,T, revealing complex evolution of the magnetic structure. A persistent energy continuum at the $\Gamma$-point characterizes all field strengths. The spin-wave spectrum exhibits clear dispersion along $Y$-$M$-$\Gamma$-$M$-$Y$ points of the 2D honeycomb Brillouin zone at lower fields (0-6\,T) ((b)-(e)) at $T=2$\,K, with distinct minima at these high-symmetry points. This coherent excitation signature vanishes at 8\,T (f), suggesting a phase transition. Notably, the 8\,T spectrum shown in (f) bears remarkable similarity to the zero-field data collected at $T$ = 15\,K (above $T_N$) shown in (g), indicating the potential emergence of a QSL state in the FIQD regime \cite{banerjee2018excitations}.
	}
    \label{fig:Neutron2}
\end{figure}

Banerjee {\it et al.}\ compared the two spectra outside the AFM ordered phase (Figs.\,\ref{fig:Neutron2}(f) and (g)). Notably, they observed that the 8\,T spectrum at 2\,K (Fig.\,\ref{fig:Neutron2}(f)) shows remarkable similarity to the zero-field spectrum observed at 15\,K above $T_N$ (Fig.\,\ref{fig:Neutron2}(g)). The correspondence between these spectra suggests an intriguing connection between the high-energy excited states in these distinct regions of the phase diagram. This similarity indicates that the magnetic field may drive the system toward a state reminiscent of the paramagnetic phase above $T_N$, potentially approaching a field-induced Kitaev QSL state. The observed broad continuum of excitations and the absence of well-defined magnon modes align with theoretical predictions for fractionalized excitations in a QSL phase.

The application of stronger parallel magnetic fields beyond the zigzag order threshold dramatically modifies the INS spectra \cite{balz2019finite}. This evolution is marked by a transformation from a magnetic continuum to sharp magnon peaks, suggesting an additional phase transition from the putative QSL state at high field strengths, as depicted in Fig.\,\ref{fig:Scenario3}(c). Theoretical support comes from unbiased numerical studies of the pure Kitaev model under an applied magnetic field by Yoshitake {\it et al.}\ \cite{yoshitake2020majorana}. While these calculations do not capture the zigzag ordering, they demonstrate that both flux and Majorana excitations develop dispersive behavior, transitioning from momentum-independent modes in zero field to acquire distinct wave-vector dependence. Furthermore, the calculations predict the emergence of a gap at the $\Gamma$-point, consistent with magnon formation.

However, it is important to note that the field-induced transition between $zz1$ and $zz2$ phases involves different interlayer ordering patterns. This observation suggests that even weak interlayer interactions are sensitive to modest magnetic fields, potentially giving rise to complex 3D ordered states with large unit cells before eventual polarization. This aspect warrants further investigation to fully understand the field-induced phase diagram.

\subsection{Nuclear Magnetic Resonance}

Nuclear magnetic resonance (NMR) spectroscopy is a bulk probe that is sensitive to magnetic ordering and low-energy excitations. The Knight shift, $K$, probes the static spin susceptibility $\chi({\bm q})$ at ${\bm q}$=0.  The spin-lattice (or longitudinal)  relaxation rate $1/T_1$ measures a {\boldmath $q$} average of the imaginary part of the dynamical spin susceptibility, $(T_1T)^{-1}\propto \sum_{\bm q} A_{hf}^2({\bm q}) \chi''({\bm q},\omega)/\omega$ where $A({\bm q})$ is the hyperfine coupling constant and $\omega$ is the resonance frequency.  To date, NMR measurements on the $^{35}$Cl nucleus (nuclear spin $I = 3/2$) in $\alpha$-RuCl$_3$ have been independently performed by four research groups, each exploring the material's response as a function of applied magnetic field and temperature \cite{baek2017evidence,zheng2017gapless,janvsa2018observation,nagai2020two}. 

\begin{figure}[b]
	\includegraphics[clip,width=1.0\linewidth]
    {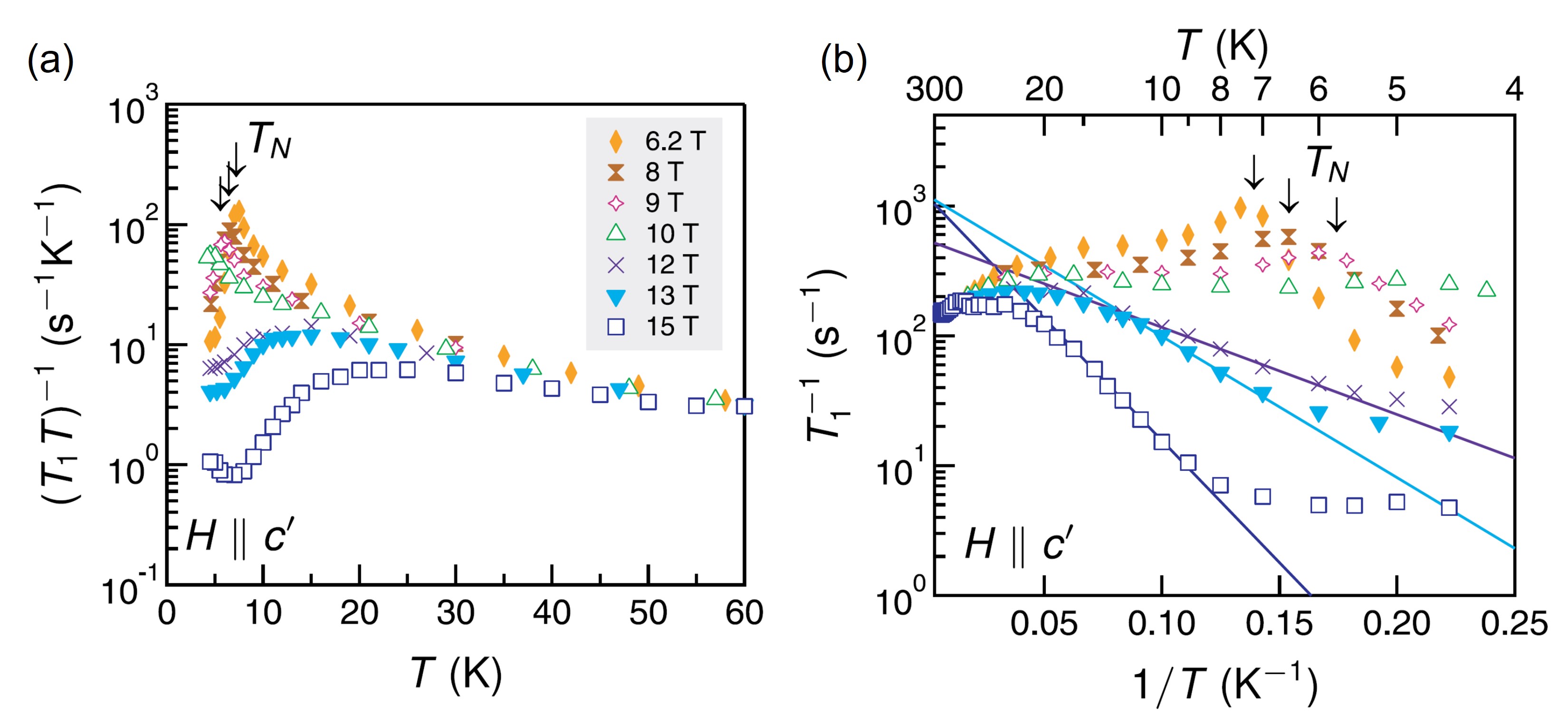}
	\caption{(a) The temperature-normalized NMR spin-lattice relaxation rate $(T_1T)^{-1}$ shows strong field dependence at low temperatures with $H$ applied along $c'$ ($30^\circ$ from crystallographic $c$ axis). Below the critical field $\mu_0H_c = 9$\,T, clear signatures of antiferromagnetic (AFM) ordering emerge.
(b) A semilogarithmic analysis of $T_1^{-1}$ versus $1/T$ reveals spin gap behavior characterized by $T_1^{-1} \propto \exp{(-\Delta/T)}$ \cite{baek2017evidence}. 
	}
    \label{fig:NMR}
\end{figure}

Baek {\it et al.}\ performed NMR measurements \red{in crystals with a single transition at $T_N=6.2$\,K} with the magnetic field oriented at $\theta \sim 30^{\circ}$ from the crystallographic $c$-axis within the $ac$ plane (denoted as $c'$ axis) to achieve optimal $^{35}$Cl spectral resolution \cite{baek2017evidence}. As shown in Fig.\,\ref{fig:NMR}(a), the temperature-normalized spin-lattice relaxation rate $(T_1T)^{-1}$ exhibits pronounced field dependence at low temperatures with $H\parallel c'$. Below $\mu_0H = 9$\,T, sharp peaks in $(T_1T)^{-1}$ clearly indicate the presence of an AFM ordered phase. A semilogarithmic analysis of $T_1^{-1}$ versus $1/T$ (Fig.\,\ref{fig:NMR}(b)) reveals activated behavior below 30\,K characterized by $T_1^{-1} \propto \exp{(-\Delta/k_BT)}$, with $\Delta \approx 50$\,K at $\mu_0H = 15$\,T. The spin gap $\Delta$ exhibits linear field dependence, decreasing systematically with reducing $H\parallel c'$ until its complete closure at 10\,T. Baek {\it et al.}\ claimed that these observations, combined with evidence of local spin fluctuations in the quantum disordered state, support the existence of a field-induced gapped quantum phase distinct from a conventionally spin-polarized state \cite{baek2017evidence}.

Jan\v{s}a {\it et al.}\ \cite{janvsa2018observation} conducted NMR measurements with the magnetic field applied in a plane perpendicular to the $ab$ plane, at a 15$^{\circ}$ angle from the $b$-axis. The sample exhibits two distinct AFM transitions at approximately 14\,K and 7\,K. The spin gap, estimated from the temperature dependence of $T_1^{-1}$ at high temperatures, displays markedly different behavior compared to the results by Baek {\it et al.}\  \cite{baek2017evidence}. The authors posit that the spin gap exists even in zero field and increases rapidly, following a cubic dependence on the applied magnetic field. They attribute the zero-field gap to $Z_2$ gauge flux excitations and the field-dependent gap to Majorana fermion excitations.

Nagai {\it et al.}\  performed NMR measurements in a parallel field ($\bm{H} \parallel b$) down to temperatures nearly three orders of magnitude lower than the Kitaev interaction energy scale \cite{nagai2020two}. They report the evolution of two distinct gapped spin excitations with different field dependencies in the FIQD state \red{in a crystal with a single zero-field transition at 6.5\,K}. The temperature dependence of the gap can be decomposed into a large gap $\Delta_H = \Delta_0 + bH^3$ and a small gap $\Delta_L \propto (H-H_c)^{0.3}$.
They interpret these results as indicative of field-induced hybridization between fractionalized quasiparticles.
However, their findings are inconsistent with specific heat measurements, which indicate the preservation of gapless Dirac cone excitations for $\bm{H} \parallel b$ \cite{tanaka2022thermodynamic,imamura2024majorana}.

In contrast to the findings of the aforementioned three groups, Zheng {\it et al.}\ reported \red{for a high-quality crystal with a single transition at $T_N\sim7.5$\,K} that the spin-lattice relaxation rate ($1/T_1$) exhibits a power-law decay upon entering the FIQD state, persisting up to 16 T. This behavior follows the relation $T_1^{-1} \propto T^{\delta}$, where $\delta$ ranges from 3 to 5 \cite{zheng2017gapless}. Such a power-law dependence suggests the presence of gapless spin excitations over a wide field range in the FIQD phase. However, it is important to note that the exact direction of the applied magnetic field within the $ab$ plane is not clearly specified in their study, which may have significant implications for the interpretation and comparison of results.

Future NMR investigations of $\alpha$-RuCl$_3$ require systematic angular-dependent studies with precise field orientation control both within the $ab$ plane and along the out-of-plane direction to comprehensively map the material's anisotropic magnetic response. Such measurements would help resolve existing experimental discrepancies in the literature regarding field-induced quantum phases. Additionally, theoretical frameworks need to be developed to establish quantitative relationships between NMR relaxation rates and the characteristic energy scales of Kitaev quantum spin liquids, specifically addressing how the Majorana fermion and $Z_2$ flux gaps manifest in the NMR response functions.

\subsection{High frequency spectroscopy}

 \begin{figure}[b]
	\includegraphics[clip,width=0.7\linewidth]
    {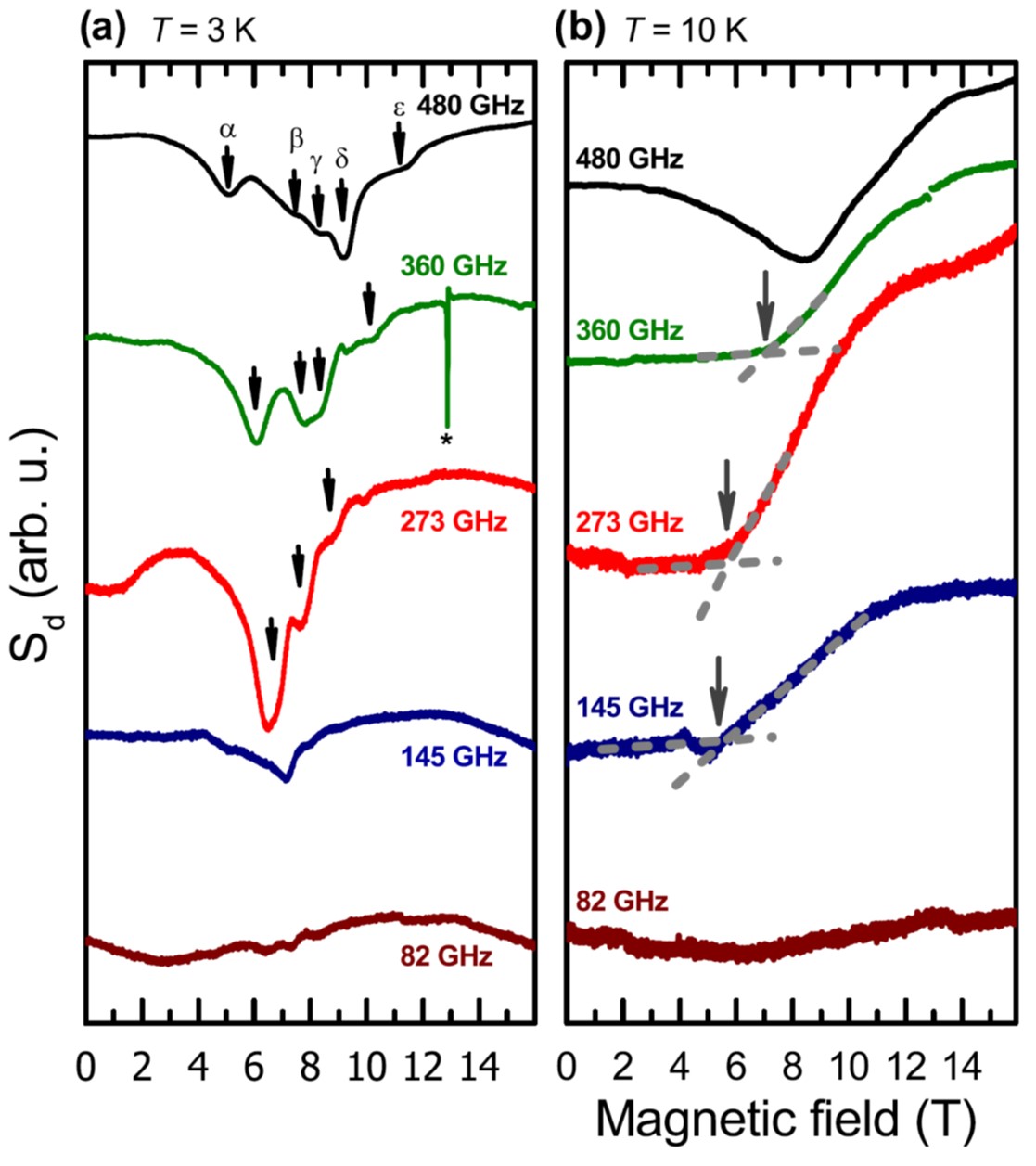}
	\caption{Microwave absorption spectra analysis. 
(a) The detector signal $S_d$ measured at 3\,K across various microwave frequencies. Magnon modes are labeled as $\alpha$, $\beta$, $\gamma$, $\delta$, and $\varepsilon$. (b) At 10\,K, the detector signal $S_d$ exhibits a broad excitation continuum with distinctive features. This continuum appears at energies lower than the lowest sharp magnon mode and persists even above $T_N$. Notably, the continuum develops an energy gap when subjected to magnetic fields exceeding the critical field of $\sim 7$\,T. Wellm {\it et al.}\ interpret these observations as evidence of fractionalized excitations, consistent with theoretical predictions for Kitaev QSL. Furthermore, they propose that the gapped phase observed above 7\,T demonstrates unconventional behavior that deviates from typical AFM systems \cite{wellm2018signatures}.}
\label{fig:Microwave}
\end{figure}

The quantum spin dynamics of $\alpha$-RuCl$_3$ under parallel magnetic fields have been extensively investigated using multiple spectroscopic techniques, including THz spectroscopy \cite{wang2017magnetic,little2017antiferromagnetic,wu2018field,ponomaryov2017unconventional,ponomaryov2020nature} and microwave absorption spectroscopy \cite{wellm2018signatures}. 

THz spectroscopic measurements reveal the evolution of magnetic correlations and excitations across temperature regimes and magnetic phases in $\alpha$-RuCl$_3$. In the zero-field regime above the N\'{e}el temperature, these measurements show a broad continuum extending to approximately 10\,meV. As the temperature decreases, this continuum develops rapidly and reaches a broad maximum. Upon transitioning to the zigzag-ordered AFM state, the continuum intensity diminishes while the spectral weight transfers to well-defined spin-wave excitations.

Wang {\it et al.}\ performed THz spectroscopy measurements on $\alpha$-RuCl$_3$ with magnetic fields applied parallel to the $ab$ plane, investigating regions both below and above the critical field $\mu_0H^c_{ab} \approx 7$\,T. Their findings reveal that at $H^c_{ab}$, the THz spectra are characterized by a broad continuum spanning an extensive spectral range, with no discernible sharp excitations. As the applied field strength is increased beyond $H_{ab}^c$, the broad continuum undergoes a shift towards higher energies, concurrent with the emergence of two distinct excitation peaks at 2.7\,meV and 4.4\,meV slightly above $H_{ab}^c$. These peaks exhibit a near-linear increase in energy with respect to the applied field strength. The authors attribute the lower-energy peak to the opening of a spin gap. Notably, the linear field dependence of this gap opening is consistent with theoretical predictions by Song {\it et al.}\ derived from the Kitaev-Heisenberg model \cite{song2016low}.

Ponomaryov {\it et al.}\ conducted high-field THz ESR experiments on $\alpha$-RuCl$_3$ at 1.5\,K, extending up to 16\,T for magnetic field orientations  $\bm{H} \parallel [100]$ and $[110]$ \cite{ponomaryov2020nature}. Their investigation primarily focused on magnon excitations across different magnetic phases. In the zigzag AFM phase, they observed two conventional AFM resonance modes excited at the $\Gamma$-point, corroborating previous findings by Wu {\it et al.}\ \cite{wu2018field}. As the applied field strength increased, both modes exhibited pronounced softening, accompanied by a significant decrease in intensity as the critical field was approached. Upon transitioning into the FIQD phase, several distinct magnetic resonance modes emerged. These modes demonstrated a positive correlation between their frequencies and the applied magnetic field strength. Notably, some of the observed resonance frequencies showed quantitative agreement with results obtained from INS experiments \cite{balz2019finite}.  The authors propose that the high-field spin dynamics in $\alpha$-RuCl$_3$  can be interpreted in terms of one- and two-particle excitations, which they identify as magnons.

Wellm {\it et al.}\ conducted microwave absorption measurements on $\alpha$-RuCl$_3$ across a wide range of frequencies (70-660 GHz, corresponding to 0.3-2.8\,meV) in parallel magnetic fields and temperatures, covering the vicinity of the critical field \cite{wellm2018signatures}. This frequency range is particularly significant as it probes the low-energy excitations below 1.5\,meV, a regime inaccessible to INS experiments. Below $T_N\approx 8$\,K (Fig.\,\ref{fig:Microwave}(a)), they observed several resonant magnon modes, corroborating previous findings from terahertz spectroscopy \cite{wang2017magnetic,little2017antiferromagnetic,wu2018field} and ESR measurements \cite{ponomaryov2017unconventional}. Above the AFM critical field at low temperatures, the study revealed continuum excitations at the wave vector $q=0$, spanning energies between 0.4 and 1.2\,meV. These excitations persist at temperatures significantly above $T_N$ (Fig.\,\ref{fig:Microwave}(b)) and occur at energies well below the magnon-like modes. The authors also report that the progressive gapping of this continuum as the field strength increases beyond the AFM critical field. The emergence of this field-induced gap aligns with observations from other experimental techniques, including specific heat measurements \cite{wolter2017field}, thermal conductivity studies \cite{hentrich2018unusual}, and NMR investigations \cite{baek2017evidence,janvsa2018observation,nagai2020two}. However, the magnitude of the gap observed in this study is smaller than that reported by other methods.

The magnetic excitation continuum observed above the parallel critical field bears remarkable similarities to the spectrum measured above the N\'{e}el temperature in zero field. The characteristics of this continuum correspond closely with inelastic neutron scattering measurements \cite{banerjee2018excitations,balz2019finite}, suggesting the persistence of spin fractionalization in the QSL state. Meanwhile, the distinct magnon excitations detected above the critical magnetic field provide crucial insights into the magnetic behavior of $\alpha$-RuCl$_3$ within the FIQD phase.

These excitations provide valuable insights into the fundamental excitation spectrum under high magnetic fields. The observed magnon excitation energy of approximately 0.5\,THz persists near the critical magnetic field, corresponding to a temperature scale of $\sim 30$\,K. This high-energy scale stands in marked contrast to the low-temperature regime where QSL behavior is typically expected. A significant conceptual gap exists between these ``high-energy" magnon excitations and the hypothesized ``low-energy" quasiparticles thought to govern the topological properties of a potential Kitaev QSL at low temperatures. This pronounced separation of energy scales demands systematic investigation to establish the relationship between the observed magnon excitations and the proposed QSL state. Future experimental and theoretical studies should address this energy-scale hierarchy and explore how high-energy excitations might influence the quantum magnetic properties of the system at low temperatures.

\subsection{Ultrasound spectroscopy}

It has been theoretically suggested that itinerant Majorana fermions and thermally induced static $Z_2$ gauge fluxes generate a continuous spectrum of scattering processes. These interactions modify phonon dynamics and its energy dissipation \cite{metavitsiadis2020phonon,ye2020phonon,feng2021temperature}. Ultrasonic spectroscopy emerges as a powerful experimental technique for probing magnetoelastic coupling through systematic analysis of phonon propagation and scattering.  

Hauspurg {\it et al.}\ report that peculiar behavior of the sound attenuation in $\alpha$-RuCl$_3$ over a wide range of temperatures can be understood as a fingerprint of the fractionalized excitations of a Kitaev magnet with comparable Fermi and sound velocities. They reveal distinct temperature and in-plane field dependencies of acoustic properties, specifically the sound velocity and attenuation coefficients, across multiple longitudinal and transverse phonon modes propagating along principal crystallographic axes \cite{hauspurg2024fractionalized}.  From the comprehensive data analysis, they show experimental evidence for phonon-Majorana fermion scattering processes, providing the results that phonons are scattered off fractionalized excitations.  Specifically, the scattering cross-section exhibits strong dependence on the ratio between phonon propagation velocities and the characteristic velocity of low-energy fermionic excitations that govern the spin dynamics. Furthermore, they observe a marked reduction in sound attenuation anisotropy, consistent with theoretical predictions for systems hosting thermally excited $Z_2$ \red{visons}. These findings establish ultrasonic measurements as a powerful experimental technique for detecting and characterizing fractionalized excitations in QSLs.

When magnetic fields are applied parallel to the $a$ axis, measurements reveal that the relative sound velocity variation ($\Delta v/v$) of the longitudinal acoustic mode $c_{11}$  exhibits distinctive minima as functions of both $T$ and $H$ \cite{hauspurg2024fractionalized}. This acoustic mode is characterized by the wave vector ({\boldmath $q$}) and polarization vector ({\boldmath $u$}) both aligned along the $a$ axis. These minima are observed in the paramagnetic phase, specifically at magnetic fields exceeding the critical field where the AFM order is suppressed. The field dependence of the minimum position shows a positive correlation with field strength, following a trend similar to the evolution of the spin gap energy scale previously identified through specific heat \cite{sears2017phase,wolter2017field,tanaka2022thermodynamic} and thermal conductivity \cite{hentrich2018unusual} measurements.

\section{Topological signature of Kitaev quantum spin liquid in magnetic fields}

\subsection{Topological properties}
 
In zero field, the Kitaev model exhibits a QSL where the spin excitations are fractionalized into non-interacting Majorana fermions and localized plaquette fluxes which couple to the Majoranas via a static $Z_2$ gauge field. The quasiparticle excitations are characterized by a gapless linear Dirac dispersion, $\varepsilon(\bm{k})\propto |\bm{k}|$, at low energies much smaller than the flux gap $\Delta_K$.   

The system's behavior undergoes significant modifications when applying an external magnetic field. The primary effect is the introduction of a Zeeman term to the Hamiltonian,
\begin{equation}
	\mathcal{H}_Z=-\sum_jh_x\sigma_j^x+h_y\sigma_j^y+h_z\sigma_j^z,
\end{equation} 
which splits the energy levels of the spins. Here, $h_x$, $h_y$, and $h_z$ are the $x$, $y$, and $z$ components of the applied magnetic field with respect to the spin axis, respectively. It should be noted that the spin quantization axes are defined with respect to the octahedral coordinate system (see Eqs.\,(\ref{eq:Hamiltonian1}) and (\ref{eq:Hamiltonian2})). The application of a magnetic field breaks both time-reversal and $C_2$ symmetries in a direction-dependent manner, consequently leading to gap formation at the Dirac points.

According to the perturbative analysis, the first-order contribution of the Zeeman term vanishes due to the absence of a net magnetic moment in the spin liquid state. While the second-order term is non-zero, it preserves time-reversal symmetry. Consequently, it does not induce an energy gap in the excitation spectrum, maintaining the gapless nature of the Majorana fermions.

The third-order contribution of the Zeeman term gives rise to a three-body spin correlation, which has profound implications for the physical properties of the system. This term can be expressed as, 
\begin{align}
	\mathcal{H}_h^{3rd}&\sim-\frac{h_xh_yh_z}{\Delta_K^2}\sum_{j,k,l}\sigma_j^x\sigma_k^y\sigma_l^z \\ \nonumber
	&\sim -\frac{i}{2}\frac{h_xh_yh_z}{\Delta_K^2}\sum_{j,k}c_jc_k,
\end{align}
where $\Delta_K$ is a \red{vison} gap. The second line of this equation reveals that this three-body spin correlation can be rewritten in terms of Majorana fermions. Specifically, it describes a second n.n.\ hopping term for the Majorana fermions  $c_jc_k$. The imaginary nature of this hopping term is crucial, as it breaks time-reversal symmetry. This imaginary second n.n.\ hopping term has a significant effect on the energy spectrum: it opens a gap at the Dirac points of the Majorana fermion dispersion when $h_xh_yh_z\neq0$. This situation bears a striking resemblance to the mechanism proposed by Haldane in his seminal work on the quantum Hall effect without a net magnetic field, often referred to as the Haldane model \cite{haldane1988model}. The Haldane model demonstrates that a quantum Hall state can be realized in a system with complex hopping amplitudes, even in the absence of an external magnetic field. In the present case, the third-order Zeeman term induces a similar effect in the Kitaev model, potentially driving the system into a topologically non-trivial phase analogous to the quantum Hall state.

\begin{figure}[t]
\includegraphics[clip,width=0.7\linewidth]{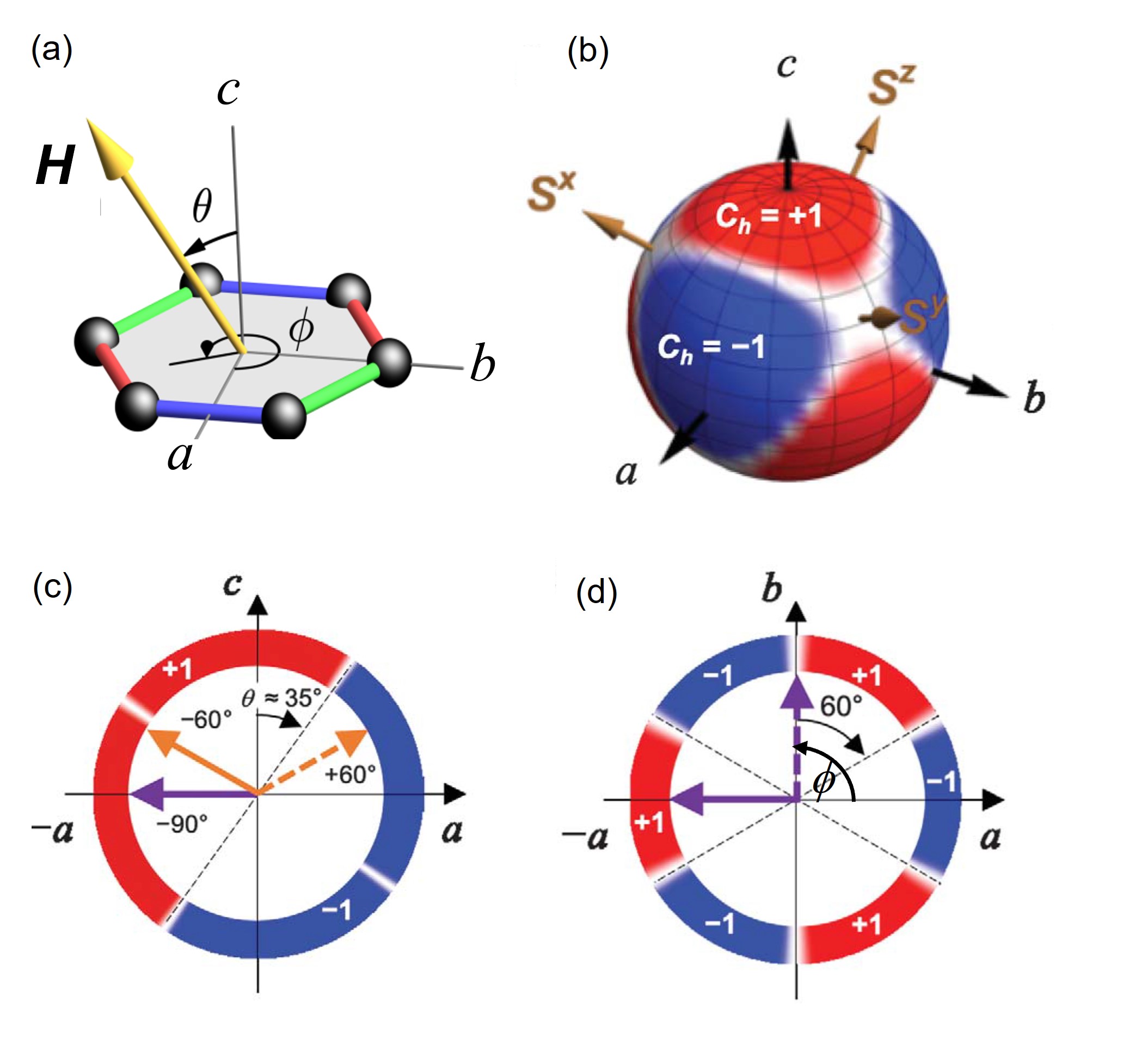}
	\caption{Field-angular dependence of the Chern number in the pure Kitaev model. (a) Schematic representation of the honeycomb lattice structure showing crystallographic directions and applied magnetic field ({\boldmath $H$}). The field orientation is defined by $\theta$, the polar angle between {\boldmath $H$} and the $c$-axis, and $\phi$, the azimuthal angle between the projection of {\boldmath $H$} and the $a$-axis in the $ab$-plane.
(b) Theoretical calculations demonstrate the field-angular dependence of the Chern number ($C_h$), which alternates between $\pm 1$ in the pure Kitaev model with respect to the spin configuration and crystallographic axes.
(c) Evolution of $C_h$ within the $ac$-plane. The solid and dashed orange arrows denote magnetic field orientations at $\theta=-60^{\circ}$ and $+60^{\circ}$ relative to the $c$-axis, respectively. The purple arrow indicates the field direction along the $-a$ axis ($\theta=-90^{\circ}$). The Chern number vanishes ($C_h = 0$) at critical angles $\theta=35^{\circ}, 125^{\circ}, -60^{\circ},$ and $-150^{\circ}$.
(d) Evolution of $C_h$ within the $ab$-plane. The solid and dashed purple arrows represent field directions along the $-a$ axis ($\phi=180^{\circ}$) and $b$ axis ($\phi=90^{\circ}$), respectively, where $\phi$ denotes the azimuthal angle.\cite{yokoi2021half}
	}
\label{fig:Chern}
\end{figure}

 Thus, the application of magnetic field changes the low-energy gapless linear dispersion of Majorana fermions to a gapped one (see Figs.\,\ref{fig:Dirac3}(a)-(d)), leading to the two gap structure with $\Delta_{K}$ and Majorana excitation gap $\Delta_M$. $\Delta_M$ is given as
\begin{equation}
  \Delta_M\propto \frac{|h_xh_yh_z|}{\Delta_{K}^2} .
  \label{eq:Majoranagap}
\end{equation}
  Equation\,(\ref{eq:Majoranagap}) immediately indicates that the Majorana gap is strongly field-angle dependent and has nodes at directions where the product $h_xh_yh_z$ is zero. When the field-induced gap is finite, the chiral QSL has edge states that are topologically protected by a non-zero Chern number (Figs.\,\ref{fig:Chern}(a)-(d)), which is determined by the sign of the product $h_xh_yh_z$, given as 
 \begin{equation}
   C_h={\rm sgn}(h_xh_yh_z).
   \label{eq:Chern}
 \end{equation}

The topological properties of the chiral QSL are manifested through the quantized thermal Hall effect, with the Chern number directly correlated to the sign of this effect. While the above analysis only works when other non-Kitaev interactions are small (in the Kitaev QSL) and the applied field is weak (perturbative approach), the generic model described by Eqs.\,(\ref{eq:Hamiltonian1}) and (\ref{eq:Hamiltonian2}) shows that Ising topological order is possible when the field breaks the $C_2$ symmetry in addition to time-reversal symmetry, which goes beyond \red{perturbation} theory.  Recall that \red{${\bm H}\parallel {\bm b}\parallel [\bar{1}10]$ gives} the perturbation result $h_xh_yh_z=0$, leading to the absence of the Ising topological order.   These topological characteristics, in conjunction with the quasiparticle excitation spectra in the FIQD state, provide rigorous criteria for validating the Kitaev QSL model in $\alpha$-RuCl$_3$, if Kitaev QSL is found by tuning a quantum parameter. The high-energy quasiparticle excitation spectra are obtained through spectroscopic methods, while low-energy excitations are probed via thermodynamic and thermal transport studies.

\subsection{Chiral edge mode}

Thermal Hall conductivity provides crucial insights into quantum magnetic systems due to its sensitivity to the non-trivial Berry phase of quasiparticles. 
Recent investigations have demonstrated that electrically insulating materials can also exhibit a non-zero thermal Hall conductivity, suggesting the presence of topologically protected edge states of spin excitations (for comprehensive reviews, see \cite{mcclarty2022topological} and references therein). The remarkable aspect of this phenomenon is that the thermal Hall conductivity can quantify the non-trivial Berry curvature of charge-neutral quasiparticles, thereby providing experimental access to various topological excitations \cite{katsura2010theory,matsumoto2011theoretical,matsumoto2011rotational,lee2015thermal}.
\begin{equation}
	\kappa_{xy}=\frac{k_BT}{\hbar V}\sum_{\bm k}\sum_n c_2 [g(\epsilon_{n{\bm k}})]\Omega_{n{\bm k}},
 \label{eq:thermalHall}
\end{equation}
where $\epsilon_{n{\bm k}}$ and  $\Omega_{n{\bm k}}$ are energy and Berry curvature of the elementary excitations and  $c_2 [g(\epsilon_{n{\bm k}})]$ is a distribution function. 
Thermal Hall conductivity measurements have emerged as a powerful probe of fractional excitations in QSLs, offering unique insights where conventional electrical transport techniques fail due to the charge-neutral nature of the relevant quasiparticles. 

 Recent experimental advances have successfully resolved finite thermal Hall conductivity in a diverse array of insulating magnets on geometrically frustrated lattices, significantly expanding our understanding of exotic magnetic states. In ferromagnetic insulators, the observation of finite $\kappa_{xy}$ has been interpreted as evidence of the predicted topological thermal Hall effect arising from bosonic magnon excitations, as demonstrated in seminal studies by Onose {\it et al.}\ \cite{onose2010observation} and Ideue {\it et al.}\ \cite{ideue2012effect}. This phenomenon has been further explored in paramagnetic states, with noteworthy results obtained in both kagome lattices \cite{hirschberger2015thermal,watanabe2016emergence,doki2018spin} and spin ice system \cite{hirschberger2015large}.  Across these diverse magnetic structures, the consistent detection of finite $\kappa_{xy}$ has been predominantly interpreted within the theoretical framework of topological thermal Hall effects originating from bosonic spin excitations, particularly magnons. 

Despite significant research progress, several unresolved questions remain regarding topological phases in charge-neutral bosonic systems. A notable example involves the Shastry-Sutherland magnet  SrCu$_2$(BO$_3$)$_2$, which exhibits an exactly solvable ground state. Romh{\'a}nyi {\it et al.}\  proposed that this material should behave as a magnetic analog of a Chern insulator, with bosonic magnetic excitations called triplons \cite{romhanyi2015hall}. Their theoretical analysis predicted a large thermal Hall conductivity. However, subsequent experimental work by Suetsugu {\it et al.}\ failed to detect any measurable thermal Hall effect \cite{suetsugu2022intrinsic}. This discrepancy may be explained by nonperturbative damping of magnon modes, as suggested by Chernyshev and Maksimov \cite{chernyshev2016damped}.

 \begin{figure}[t]
\includegraphics[clip,width=0.8\linewidth]{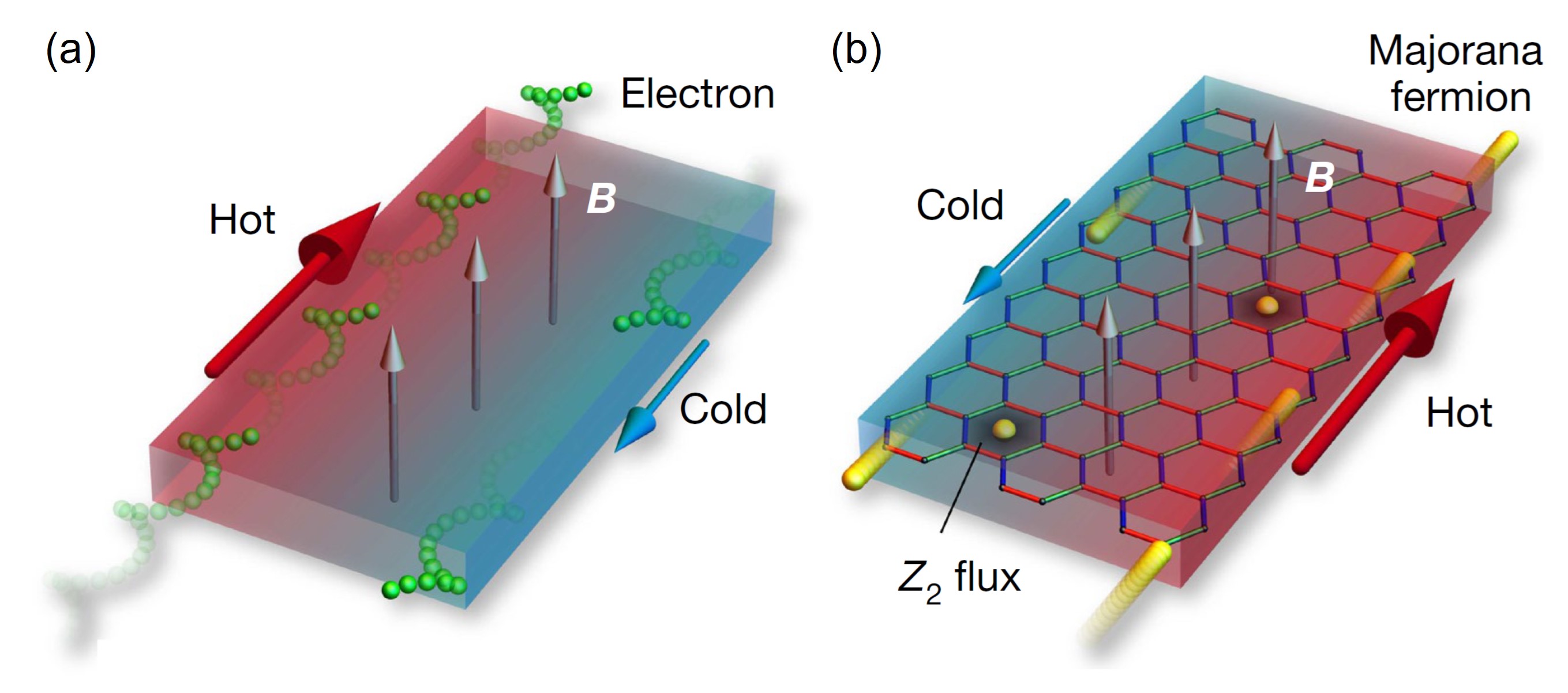}
	\caption{Schematic illustrations of heat conduction in quantum states.
(a) Integer quantum Hall state in a 2D electron gas:
Heat transport occurs when a magnetic field is applied perpendicular to the sample plane (indicated by gray arrows). In regions of temperature gradient, the higher-temperature area is shown in red and the lower-temperature area in blue, with thermal flow represented by red and blue arrows. Heat conduction occurs through skipping orbits of electrons (depicted as green spheres) at the edge, forming one-dimensional edge channels.
(b) Kitaev QSL state:
Under a magnetic field (gray arrows), the system exhibits spin fractionalization into two distinct components: Majorana fermions (shown as yellow spheres) and $Z_2$ fluxes (represented by hexagons). The thermal transport occurs through chiral edge currents composed of charge-neutral Majorana fermions, with heat flowing between high-temperature (red) and low-temperature (blue) regions.
}
\label{fig:Edgecurrent}
\end{figure}

The thermal Hall effect has also been observed in phonon systems \cite{strohm2005phenomenological,sugii2017thermal,hirokane2019phononic,li2020phonon,grissonnanche2020chiral,akazawa2020thermal,li2023phonon} and analyzed through theoretical frameworks \cite{zhang2010topological,qin2012berry,saito2019berry,zhang2019thermal,chen2020enhanced}. However, the theoretical understanding of phonon-mediated thermal Hall mechanisms remains less developed compared to that of topological magnons, and quantitative predictions continue to pose significant challenges. We will examine the phonon thermal conductivity of $\alpha$-RuCl$_3$ in a later section.

\subsubsection{Thermal Hall  effect of Kitaev quantum spin liquid}

In the integer quantum Hall (IQH) state, a 1D dissipationless chiral edge current of electrons manifests, resulting in a quantized electrical Hall conductance of $\sigma_{xy}^{2D}=\nu (e^2/h)$, where $\nu=1,2,3,\ldots$ denotes integer values, and $e^2/h$ is the quantum of conductance (Fig.\,\ref{fig:Edgecurrent}(a)).
Notably, Banerjee {\it et al.}\ demonstrated that the thermal Hall conductance in the IQH state exhibits concurrent quantization according to:
\begin{equation}
\kappa_{xy}^{2D}=\nu \mathcal{K}_0,
\end{equation}
where $\mathcal{K}_0=(\pi^2k_B^2/3h)T=(9.464\times10^{-13}{\rm W/K}^2)T$ is the quantum of thermal conductance at temperature $T$ \cite{banerjee2017observed}. 

In contrast, the Kitaev QSL hosts a chiral edge heat current comprised of itinerant, charge-neutral Majorana fermions  (Fig.\,\ref{fig:Edgecurrent}(b)). The fundamental distinction lies in the Majorana fermions possessing half the degrees of freedom of ordinary fermions like electrons.  
In the Kitaev QSL state,  thermal Hall conductance, thermal Hall conductivity per layer,  $\kappa_{xy}^{2D}=\kappa_{xy}/d$ ($d$ is the interlayer distance), is quantized to half-value of $\mathcal{K}_0$ and its sign is determined by the Chern number $C_h(=\pm 1)$ as,
\begin{equation}
	\kappa_{xy}^{2D}=C_h\frac{\mathcal{K}_0}{2}.
	\label{eq:ThermalHall}
\end{equation}
Here it was assumed that layers are decoupled due to the van der Waals nature of the interlayer coupling. The field-angular variation of $C_h$ with respect to the spin and crystal axis is shown in Fig.\,\ref{fig:Chern}(b).  

According to conformal field theory, the thermal Hall conductance in such systems can be generally expressed as $\kappa_{xy}^{2D}=c\mathcal{K}_0$, where $c$ is called the central charge. Notably, when $c<1$, it signifies the presence of non-Abelian anyons in the system \cite{kitaev2006topological}. Non-Abelian Kitaev QSLs host a gapless chiral Majorana edge mode described by a chiral Ising conformal field theory with central charge $c=1/2$ \cite{liu2022dynamical}. Thus half-integer thermal Hall conductance with $c=1/2$ confirms the existence of non-Abelian anyonic excitations and underscores the topological nature of this exotic quantum state of matter.

The Kitaev QSL states and the fractional quantum Hall (FQH) states exhibit compelling topological and phenomenological similarities, warranting a comparative analysis. In the FQH states,   the collective behavior of electrons gives rise to emergent quasiparticle excitations with fractional electric charge. These fractionally charged quasiparticles are responsible for carrying the edge current. The FQH states can have multiple edge modes, each corresponding to different types of excitations. In the FQH states, the Hall conductance exhibits quantization described by $\sigma_{xy}^{2D}=\nu(e^2/h)$, where $\nu=p/q$ represents a fractional value. Typically, $\nu$ takes the form of $1/2, 2/5, 3/7,$ or similar fractions, with $q$  being an odd integer. Concurrently, the thermal Hall conductance in this state demonstrates quantization according to the equation:
\begin{equation}
	\kappa_{xy}^{2D}=cn\mathcal{K}_0,
\end{equation}
where $c=1$ and  $n=1,2,3,\ldots$ denotes integer values. A distinctive feature of conventional FQH states is the contrasting quantization behavior observed between $\sigma_{xy}^{2D}$ and $\kappa_{xy}^{2D}$. While $\sigma_{xy}^{2D}$ exhibits fractional quantization, $\kappa_{xy}^{2D}$ maintains integer quantization, suggesting that electron-electron interactions, which are crucial for the formation of the FQH states, do not affect the quantum of thermal conductance. This remarkable disparity in quantization patterns provides compelling evidence for the abelian anyonic nature of the fractional quasiparticles in the FQH state.

In the domain of FQH physics, Banerjee {\it et al.}\ reported a significant discovery in the $\nu$=5/2 state, characterized by an electrical Hall conductance, $\sigma_{xy}^{2D}=(5/2)(e^2/h)$.
This state represents an unconventional even-denominator FQH state. Crucially, concurrent measurements revealed a fractionalized thermal Hall conductance:
\begin{equation}
\kappa_{xy}^{2D}=\frac{5}{2}\mathcal{K}_0.
\end{equation}
\red{The half-integer quantization of $\kappa_{xy}^{2D}$ serves as a distinctive signature of non-Abelian anyonic excitations in this topological phase. This connection to FQH physics arises because the $\nu$=5/2 FQH state is uniquely described by the Moore-Read (Pfaffian) wave function, which differs fundamentally from conventional Laughlin-type FQH states at fillings such as $\nu$=1/3 and 1/5. While conventional FQH states support Abelian anyons that contribute integer multiples to thermal Hall conductivity, the Moore-Read state hosts non-Abelian anyons that contribute half-integer multiples. Our experimental observation of half-integer quantization therefore directly confirms that the $\nu$=5/2 FQH state realizes the Moore-Read wave function, providing evidence for non-Abelian anyons in this $p$-wave paired condensate of composite fermions. \cite{moore1991nonabelions}}

The experimental detection of half-integer quantized thermal Hall conductance in Kitaev QSL candidate materials will represent a significant phenomenon in condensed matter physics, particularly when contextualized alongside the well-established IQH and FQH effects observed in 2D electron systems.

\subsubsection{Half-integer quantized thermal Hall effect}

 \begin{figure}[b]
\includegraphics[clip,width=\linewidth]{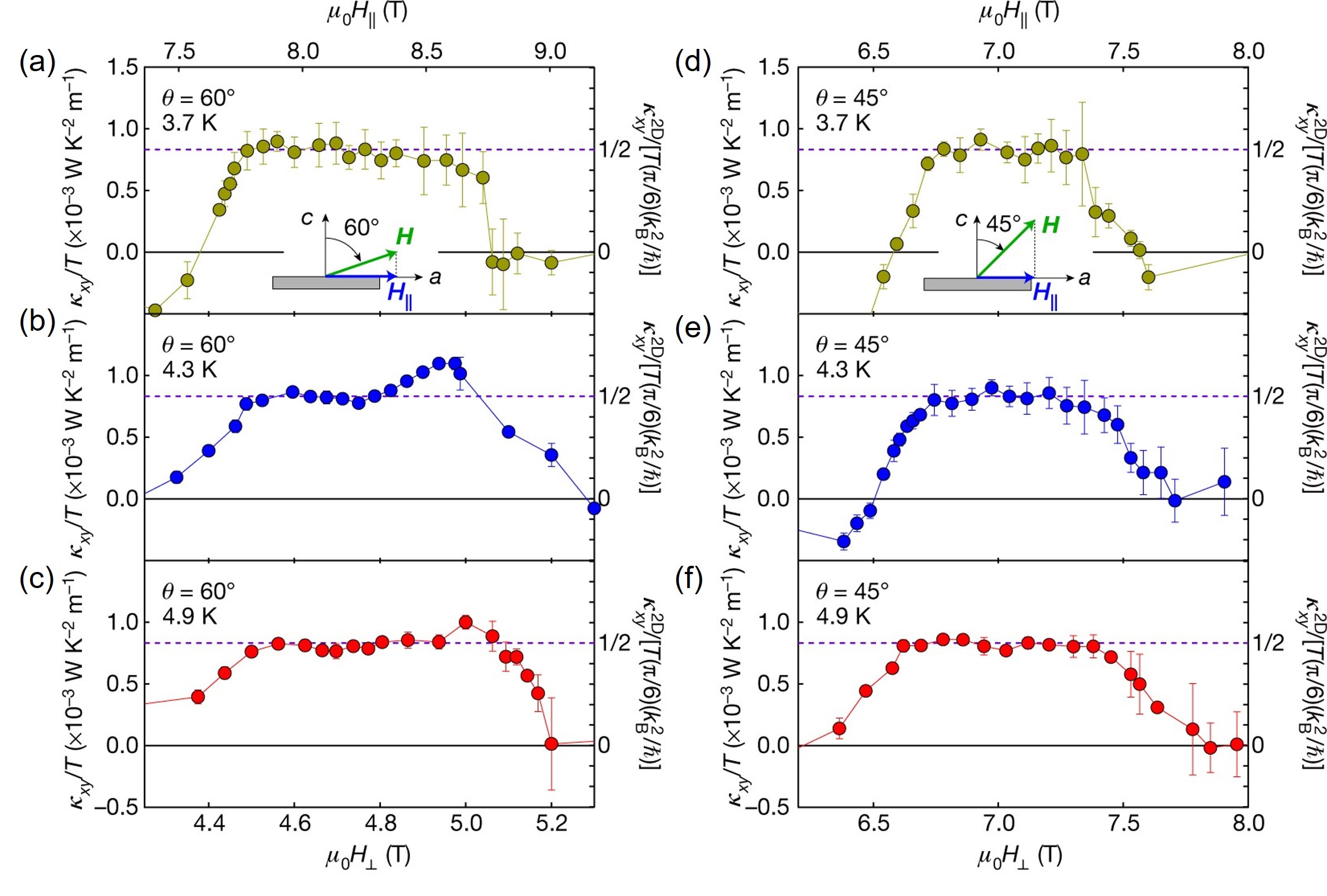}
	\caption{Half-integer thermal Hall conductance plateau observed above critical fields with magnetic fields tilted from the $c$-axis. The thermal Hall conductivity ($\kappa_{xy}/T$) was measured under magnetic fields tilted at angles $\theta=60^{\circ}$ (panels (a)-(c)) and 45$^{\circ}$ (panels (d)-(f)), plotted against the field component perpendicular to the $ab$ plane ($H_{\perp}$). The top axes display the parallel field component ($H_{||}$), while the right scales show the 2D thermal Hall conductance ($\kappa_{xy}^{2D}/T$) in units of $(\pi/6)(k_B^2/\hbar)$.  Violet dashed lines represent the half-integer thermal Hall conductance, $\kappa_{xy}^{2D}/[T(\pi/6)(k_B^2/\hbar)]=1/2$. All error bars indicate one standard deviation. \cite{kasahara2018majorana}
	}
    \label{fig:ThermalHall1}
\end{figure}

Recent studies have documented comprehensive field-dependent analyses of thermal Hall conductivity in the FIQD state under magnetic fields with components parallel to the $ab$ plane in $\alpha$-RuCl$_3$ \cite{kasahara2018majorana,yamashita2020sample,yokoi2021half,bruin2022robustness,czajka2023planar,kasahara2022quantized, zhang2024stacking,Kee2023Thermal}. Kasahara {\it et al.}\ studied the field dependence of the thermal Hall conductivity $\kappa_{xy}(H)$ under magnetic fields applied at angles of $\theta=-45^{\circ}$ and $-60^{\circ}$ relative to the $c$-axis, tilted toward the $-a$-axis direction $[-1,0,0]$  \cite{kasahara2018majorana}.  The negative angles indicate field rotation toward the direction of $-a$-axis.  In their experimental setup, the thermal current was applied parallel to the $a$-axis. 

Due to the extremely small magnitude of the thermal Hall conductivity $\kappa_{xy}$ compared to the longitudinal thermal conductivity $\kappa_{xx}$ in $\alpha$-RuCl$_3$, precise experimental protocols were essential to detect the intrinsic thermal Hall signal. All reported field dependence measurements employed a systematic approach where magnetic field values were held constant while the temperature was stabilized with high precision at each field point before data acquisition \cite{kasahara2018majorana,yamashita2020sample,yokoi2021half,bruin2022robustness,czajka2023planar,kasahara2022quantized, zhang2024stacking}.

\begin{figure}[t]
\includegraphics[clip,width=0.7\linewidth]{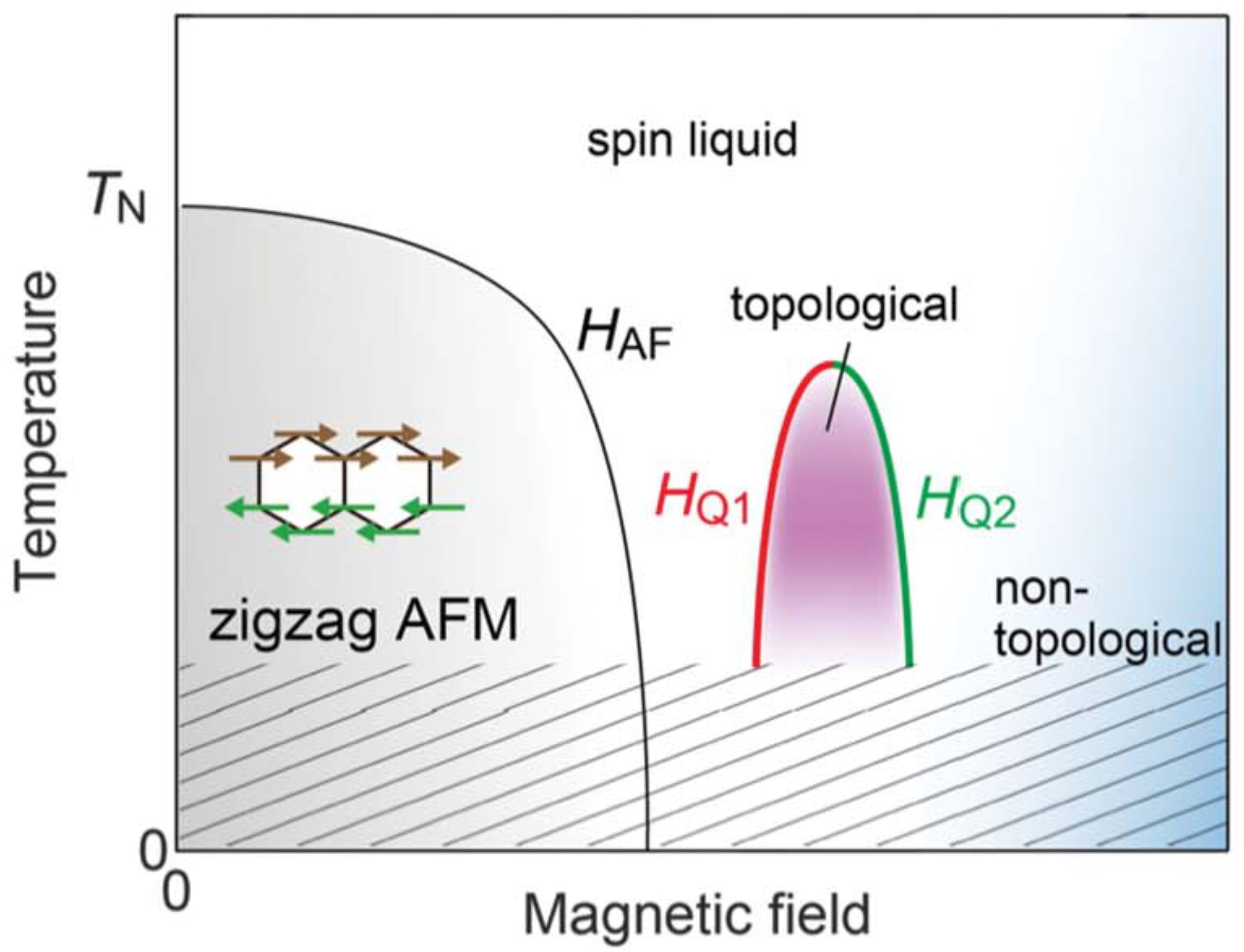}
	\caption{Schematic $H$-$T$ phase diagram of $\alpha$-RuCl$_3$ in magnetic field applied parallel to the $a$ axis \cite{yokoi2021half}.
Gray shaded area represents the region of zigzag AFM order.  The zz1-zz2 phase boundary within the magnetically ordered region shown in Fig.\,\ref{fig:Magneticphase}(a) is omitted.  The pink area represents the regime where half-integer quantized thermal Hall effect is observed; $\kappa_{xy}^{2D}$ exhibits the quantized plateau with ($\kappa_{xy}^{2D}=\mathcal{K}_0/2$) between $H_{\rm Q1}$ and $H_{\rm Q2}$. The Blue shaded
area represents the regime where $\kappa_{xy}^{2D}$ vanishes. The hatched region denotes experimental parameter spaces where thermal Hall conductivity measurements were inaccessible due to technical limitations in Yokoi {\it et al.}'s experimental apparatus. The critical field $H_{\rm Q2}$ potentially represents a phase boundary separating two distinct quantum states: one being a non-trivial QSL phase exhibiting topologically protected chiral Majorana edge modes, and another phase whose ground state properties remain unidentified. The nature of this high-field phase---whether it manifests as a conventional forced-ferromagnetic state or hosts an exotic magnetic order state---requires further experimental characterization and theoretical analysis to resolve. 
	}
    \label{fig:Phasediagram3}
\end{figure}

Figures\,\ref{fig:ThermalHall1}(a)-(c) and (d)-(f) depict the field dependence of $\kappa_{xy}/T$ at $\theta=-60^{\circ}$ and $\theta=-45^{\circ}$, respectively \cite{kasahara2018majorana}. $\kappa_{xy}/T$ exhibits negligible magnitude in the zigzag AFM ordered phase. A sharp increase in $\kappa_{xy}/T$ is observed upon the system's transition to the FIQD state. What is most remarkable is the emergence of a well-defined plateau in $\kappa_{xy}/T$ within the magnetic field range of 7.6\,T $< \mu_0H_{\parallel} <$ 8.4-8.7\,T and 6.8\,T $< \mu_0H_{\parallel} <$ 7.2-7.4\,T for $\theta=-60^{\circ}$ and $\theta=-45^{\circ}$, respectively, where $\mu_0H_{\parallel}$ denotes the applied magnetic field parallel to the $a$-axis. The right axes represent $\kappa_{xy}^{2D}/T$ in units of $\mathcal{K}_0/T$.  Notably, the plateau value corresponds to $\mathcal{K}_0/2$ within an experimental uncertainty of $\pm$3\%, which is half of the quantum thermal Hall conductance observed in integer quantum Hall systems \cite{banerjee2017observed}. At higher fields, $\kappa_{xy}^{2D}/T$ exhibits a rapid decrease, asymptotically approaching zero. 

The half-integer thermal Hall plateau has been independently reproduced by Yamashita {\it et al.}\ at $\theta=60^{\circ}$ \cite{yamashita2020sample} and Bruin {\it et al.}\  at $\theta=70^{\circ}$ \cite{bruin2022robustness}. These results confirm the presence of the half-integer quantized thermal Hall effect at various tilted field angles. Yamashita {\it et al.}\ and Bruin {\it et al.}\ reported that the sign of the half-integer thermal Hall effect is negative at $\theta=60^{\circ}$ and positive at $\theta=70^{\circ}$, respectively. Additionally, Yokoi {\it et al.}\  reproduced the half-integer quantized thermal Hall effect at $\theta=\pm 45^{\circ}$ and $\pm 60^{\circ}$, reporting positive values at $\theta=-45^{\circ}$ and $-60^{\circ}$, and negative values at $\theta=45^{\circ}$ and $60^{\circ}$ \cite{yokoi2021half}. Notably, the sign structure of the thermal Hall conductance at different $\theta$ is consistent with the sign structure of the Chern number predicted by the pure Kitaev model, as shown in Fig.\,\ref{fig:Chern}(c). We note that the sign structure is also expected from the generic model Eq.\,(\ref{eq:Hamiltonian2}), where the Kitaev model is a special limit, i.e., $J_{ab} = J_c = A = B$. Thus the predicted sign structure is a necessary condition for the Kitaev QSL, not sufficient. The smoking gun signature of the Kitaev QSL is the {\it quantized} 1/2-integer thermal Hall conductivity.

\subsubsection{Planar thermal Hall effect}

The most distinctive feature of the Kitaev QSL is the persistence of a non-zero Chern number even in the absence of an out-of-plane magnetic field component \cite{yokoi2021half}. This remarkable property manifests as a planar thermal Hall effect, wherein a transverse thermal gradient emerges in response to an in-plane magnetic field. Specifically, $C_h=-1$ when {\boldmath $H$} is parallel to the $a$ axis, and $C_h=+1$ when {\boldmath $H$} is antiparallel to the $a$ axis. While one might expect the net chirality of edge states to be indeterminate when the applied magnetic field lacks an out-of-plane component, this expectation holds true only when the field is aligned with a high-symmetry axis possessing two-fold rotational symmetry. Such symmetry would invert the chirality of edge states, thereby precluding their existence. Consequently, the thermal Hall conductance vanishes for a field oriented along the $b$ axis (Fig.\,\ref{fig:Chern}(d), dashed purple arrow), which exhibits two-fold rotational symmetry \cite{yokoi2021half,chern2021sign}. In contrast, a finite response is observed for a field along the $a$ axis, which lacks this symmetry. 
This symmetry-dependent behavior underscores the intricate relationship between lattice geometry, field orientation, and topological properties in Kitaev QSL systems, as discussed in Sec.\ IV.

The robustness of the sign change is due to the symmetry of the Hamiltonian (Eq.\,(\ref{eq:Hamiltonian1})) beyond the perturbative approach. 
Assuming that the field has induced the Kitaev QSL, where the ground state is protected by the gap $\Delta_K$, and we treat the non-Kitaev interactions perturbative if they become smaller than $\Delta_K$, the Chern number is given by, 
\begin{equation}
C_h = \text{sgn}{\{c_1(h_x + h_y + h_z) + c_3h_xh_yh_z + \cdots\}}, 
\label{eq:Ch}
\end{equation}
where $c_1$ contains non-Kitaev contributions. For in-plane fields, the linear term $(h_x + h_y + h_z)$ vanishes, leaving the third-order Kitaev contribution as the dominant term. This suggests that additional interactions should minimally affect the system's behavior. In contrast, for out-of-plane fields, together with non-Kitaev interactions, the persistent linear contribution can potentially modify both the Chern number's sign structure and the thermal Hall conductance.

Therefore, the detailed investigation of the planar thermal Hall effect within the FIQD phase promises to provide crucial insights into the Ising topological order, establishing a definitive experimental signature for the emergence of the Kitaev QSL state in $\alpha$-RuCl$_3$. The observation of this distinctive in-plane thermal transport behavior serves as a decisive test for detecting the fractionalized excitations and topological order predicted by the Kitaev model, bridging theoretical predictions with experimental evidence in candidate quantum magnets.

\begin{figure}[t]
\includegraphics[clip,width=1.0\linewidth]{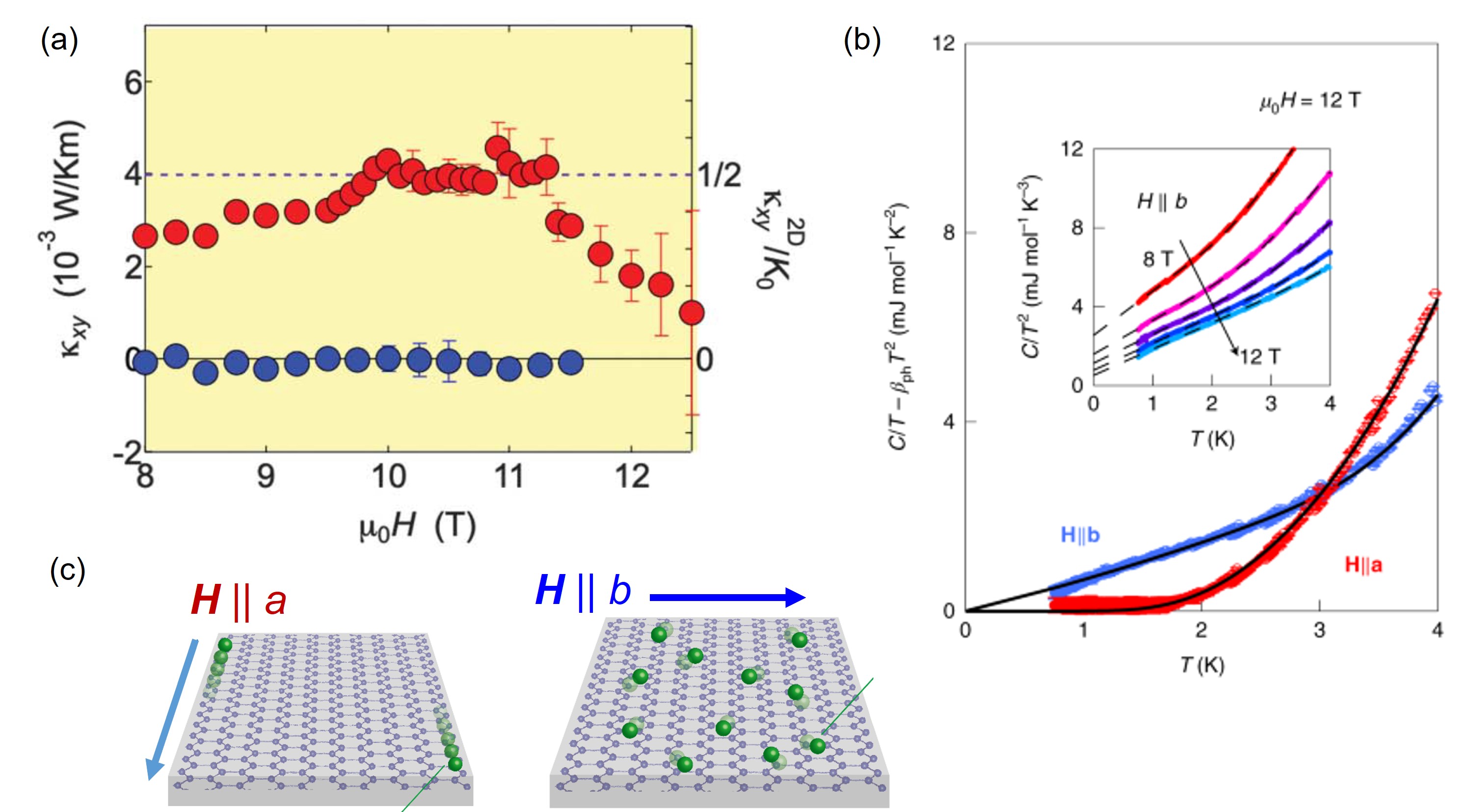}
	\caption{(a) Planar thermal Hall effect:
When {\boldmath $H$} is applied parallel to the $-a$ axis, the Chern number $C_h=+1$ corresponds to a half-integer quantized thermal Hall conductance ($\kappa_{xy}^{2D}=\mathcal{K}_0/2$) with a positive sign, as shown by red circles. In contrast, when {\boldmath $H$} is aligned parallel to the $b$ axis, the Chern number $C_h=0$ results in no observable thermal Hall effect, indicated by blue circles, consistent with the symmetry of Hamiltonian, Eq.\ref{eq:Hamiltonian2}.
(b) Specific heat:
The specific heat data, with subtracted phonon contribution ($C/T - \beta_{\rm ph}T^2$), is analyzed in the FIQD state above the critical magnetic field for both $\bm{H}\parallel a$ (red) and $\bm{H}\parallel b$ (blue) orientations. The inset displays $C/T^2$ versus $T$ for $\bm{H}\parallel b$, where a finite intercept as $T\rightarrow 0$ indicates the presence of a $T^2$ term in the specific heat ($C\propto T^2$). This behavior provides thermodynamic evidence for $E$-linear dispersion in the quasiparticle excitations, which is characteristic of a Dirac cone dispersion. (c) The transport behavior exhibits distinct characteristics depending on the magnetic field orientation. For $\bm{H}\parallel a$, a thermal Hall current is observed despite the presence of a bulk energy gap. This configuration demonstrates that thermal transport is confined to the edge states of the crystal, as the gapped bulk state prohibits bulk conduction. In contrast, for $\bm{H}\parallel b$, thermal transport occurs in the bulk state due to the gapless nature of the system. However, the system exhibits no thermal Hall response because the $C_2$ symmetry is preserved.
}
\label{fig:BulkEdge}
\end{figure}

The red circles in Fig.\,\ref{fig:BulkEdge}(a) show the magnetic field dependence of $\kappa_{xy}$ at 4.8 K in the FIQD regime \cite{yokoi2021half} when the magnetic field was applied parallel to the $-a$ axis. In the zigzag AFM phase, $\kappa_{xy}$ is negligible below approximately 5\,T. Upon entering the FIQD phase, $\kappa_{xy}$ with a positive sign appears and increases, reaching a maximum with a plateau-like behavior between approximately 10 and 11\,T. The right axis of Fig.\,\ref{fig:BulkEdge}(a) presents $\kappa_{xy}^{2D}$ in units of the quantized thermal Hall conductance $\mathcal{K}_0$. Within an experimental uncertainty of $\pm 10$\%, $\kappa_{xy}^{2D}$ exhibits a quantized plateau at half of $\mathcal{K}_0$ from 9.7 to 11.3\,T. At higher magnetic fields, $\kappa_{xy}^{2D}$ deviates significantly from this half-quantized value, showing strong suppression. This behavior suggests the occurrence of a phase transition to a distinct FIQD phase.

The observation of half-integer quantized thermal Hall conductance with a positive sign in the planar geometry with ${\bm H}\parallel -a$ axis supports the presence of a Chern number $C_h=+1$. (By convention, $C_h=-1$ is defined for the configuration ${\bm H}\parallel a$ axis.) In contrast, when the magnetic field is applied along the $b$ axis, $\kappa_{xy}$ exhibits null values across the entire field range, as demonstrated by the blue circles in Fig.\,\ref{fig:BulkEdge}(a). This behavior aligns with the theoretical expectation of $C_h=0$, which emerges from the system's symmetry constraints---specifically, the two-fold rotational symmetry or mirror symmetry with respect to the $b$ axis \cite{yokoi2021half,chern2021sign}. Thus the observed pattern of signs in $\kappa_{xy}$ measurements confirms the microscopic Hamiltonian (Eq.\,(\ref{eq:Hamiltonian2})), where $B$ term should be finite to get a finite $\kappa_{xy}$ because $C_{2a}$ symmetry is broken due to $B$ term \cite{chern2021sign}. The realization of the chiral QSL is supported by 1/2-quantized $\kappa_{xy}^{2D}$.

The experimental measurements reveal distinctive angular dependencies of $\kappa_{xy}$. Specifically, positive values are observed at $-45^{\circ}$ and $-60^{\circ}$, with sign reversals occurring in the angular intervals between $\pm 60^{\circ}$ and $\pm 45^{\circ}$. These characteristic features align closely with theoretical predictions derived from the pure Kitaev model. Through detailed analysis, Yokoi {\it et al.}\ demonstrated that these observations place stringent constraints on the ratio of coupling parameters $c_1/c_3$ in Eq.\,(\ref{eq:Ch}), establishing an upper bound of $c_1/c_3 \leq 0.1K^2$ \cite{yokoi2021half}.

\begin{figure}[t]
\includegraphics[clip,width=0.9\linewidth]{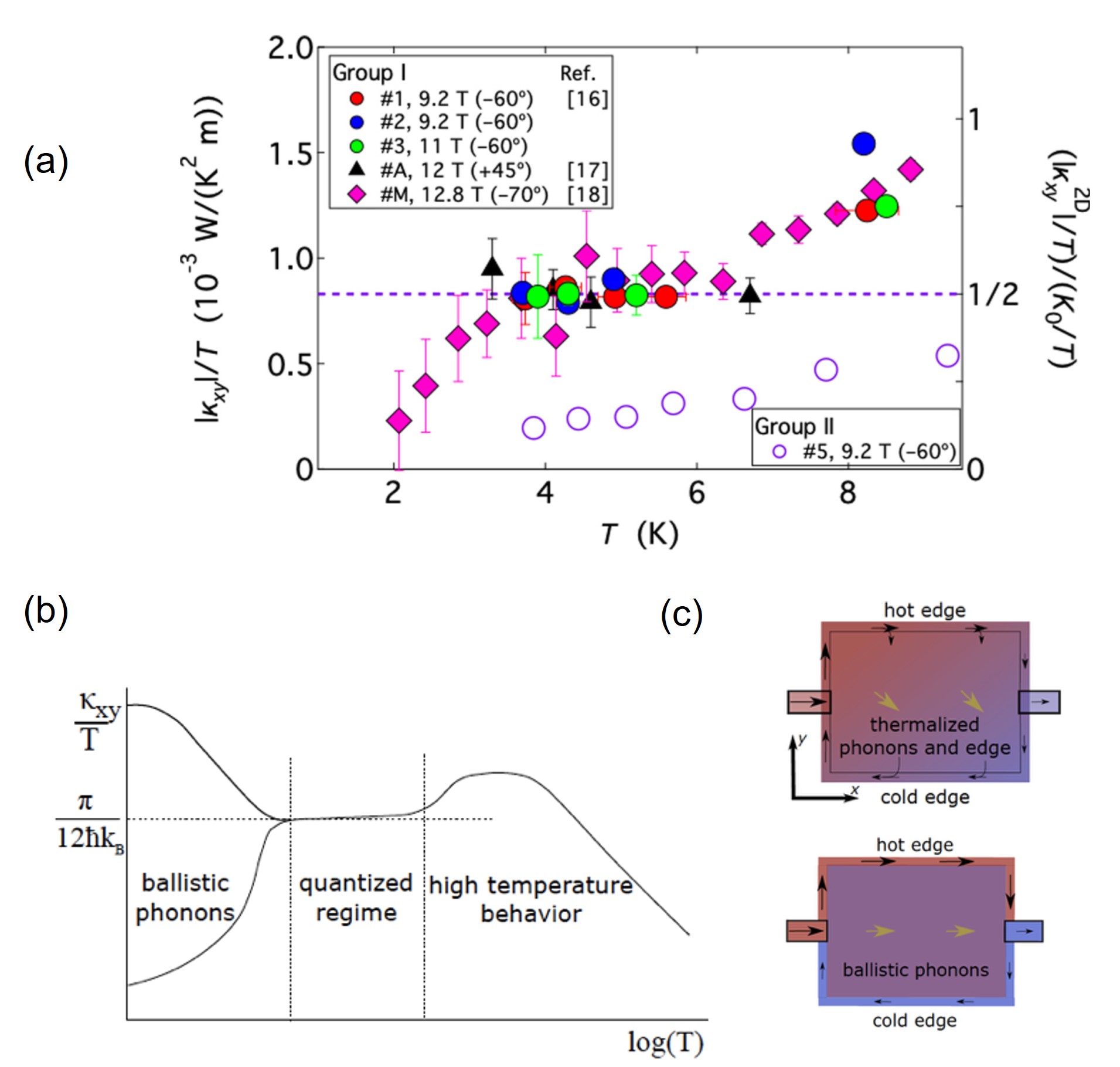}
	\caption{(a) The temperature dependence of $|\kappa_{xy}|/T$ was systematically investigated across six distinct $\alpha$-RuCl$_3$ single crystals \cite{kasahara2022quantized}. The experimental data for samples (\#1, \#2, \#3, \#A, and \#M) demonstrate half-integer quantization of the thermal Hall conductance, characterized by $\kappa_{xy}^{2D}/T=(\mathcal{K}_0/T)/2$ (denoted by dashed violet lines), within a temperature window of approximately 3.5-6.5\,K. Sample \#5, however, does not exhibit this quantization behavior. According to Kasahara {\it et al.}, the manifestation of thermal Hall quantization is strongly correlated with sample quality. Notably, a deviation from the half-quantized value is observed at lower temperatures. (b) The theoretical temperature dependence of the thermal Hall effect in the Kitaev QSL that considers the interaction between chiral Majorana edge modes and phonons \cite{vinkler2018approximately}. This coupling mechanism proves fundamental for the observation of quantized thermal Hall plateaus in Kitaev QSLs. The half-integer quantized thermal Hall conductance emerges exclusively when thermal equilibrium is established between the edge current and bulk phonons, a condition achieved only within an intermediate temperature range (upper panel of (c)). At low temperatures, significant deviations from quantization occur as the phonon mean free path ($\ell_{\rm ph}$) approaches the crystal dimensions (lower panel of (c)). In this ballistic phonon regime, the phonon-edge current coupling efficiency substantially decreases, resulting in departures from the quantized thermal Hall conductance. This theoretical framework aligns well with experimental observations.
    }
    \label{fig:Decoupling}
\end{figure}

Figure\,\ref{fig:Decoupling}(a) presents the temperature evolution of the normalized thermal Hall conductivity, $|\kappa_{xy}|/T$ and $|\kappa_{xy}^{2D}|/T$ (normalized by $\mathcal{K}_0/T$), measured at magnetic field values corresponding to the half-integer quantum thermal Hall effect. The data encompass multiple crystals, orientations, and experimental groups \cite{kasahara2022quantized}. 

At temperatures above the previously described regime, $|\kappa_{xy}|/T$ exhibits a significant enhancement, exceeding the half-integer quantization values. The temperature dependence of $|\kappa_{xy}|/T$ shows remarkable similarities to the behavior observed in $\bm{H}\parallel c$ configurations: it reaches a maximum at approximately 15\,K, followed by a monotonic decrease at higher temperatures. The thermal Hall conductance eventually becomes negligible at $\sim$ 70\,K, a temperature scale comparable to the Kitaev interaction strength.

Two competing theoretical frameworks have been proposed to explain the observed enhancement and peak behavior of $|\kappa_{xy}|/T$.  Go {\it et al}.\ attribute these phenomena to remnants of topological phase transitions within the Kitaev QSL state, suggesting the emergence of distinct Majorana excitation modes as a consequence of these transitions \cite{go2019vestiges}.  Joy and Rosch, and Chen {\it et al.} proposed alternative mechanisms for this phenomenon. Their theoretical explanations are based on the dynamics of $Z_2$ gauge fluxes \cite{joy2022dynamics} and the Berry phase of \red{visons} \cite{chen2022berry}, respectively.  In their framework, non-Kitaev interactions induce mobility in the otherwise static fluxes of the pure Kitaev model, generating a finite thermal Hall effect at elevated temperatures. This mechanism potentially accounts for both the enhancement and peak structure in the thermal response.

Experimental measurements of sample \#M (Fig.\,\ref{fig:Decoupling}(a)) demonstrate that the absolute value of $\kappa_{xy}^{2D}/T$ falls below the expected half-integer quantization at temperatures under 3.5\,K \cite{bruin2022robustness}. This low-temperature deviation from quantization offers vital insights into phonon-mediated processes in the quantized thermal Hall effect, as highlighted by Vinkler-Aviv and Rosch, and Ye {\it et al}.\ \cite{vinkler2018approximately,ye2018quantization}.

While phonon coupling to thermal edge currents is essential for observing the quantized thermal Hall effect, the presence of gapless acoustic phonons interacting with gap excitations in the Kitaev QSL leads to energy dissipation into the bulk. This dissipation disrupts the ballistic thermal transport along the edge modes, fundamentally altering the conventional edge picture of the thermal Hall effect. The interaction between chiral Majorana edge modes and phonons thus emerges as a critical factor in the observation of quantized thermal Hall plateaus in the Kitaev chiral state. The deviation from quantization becomes particularly pronounced at low temperatures, where the phonon mean free path ($\ell_{ph}$) approaches the crystal dimensions. In this ballistic phonon regime, the effectiveness of phonon-edge current coupling significantly diminishes. 

Figures\,\ref{fig:Decoupling}(b) and (c) illustrate the thermal Hall effect with these mechanisms taken into account, revealing that near-quantized thermal Hall conductivity emerges only within a specific intermediate temperature window. At low temperatures, a distinctive phenomenon manifests: as $\ell_{ph}$ approaches the sample dimensions, the thermal edge current decouples from the phonon system, causing the thermal Hall conductivity to deviate from its quantized value. This behavior stands in marked contrast to the electrical Hall effect, where robust quantization persists in real materials---highlighting a fundamental difference between thermal and electrical transport in general.

To test this hypothesis, Kasahara {\it et al.}\ estimated  $\ell_{ph}$ at the temperature where deviation from quantization occurs. They found $\ell_{ph}$ to be approximately 40-100\,$\mu$m \cite{bruin2022robustness,kasahara2022quantized}, a value approaching or comparable to the crystal dimensions of $\sim$100-300\,$\mu$m. Vinkler-Aviv and Rosch's theoretical work further predicted that enhanced phonon mean free paths in cleaner crystals would reduce phonon-edge coupling, resulting in a narrower plateau field range for samples with lower disorder \cite{vinkler2018approximately}. Initially, the half-integer quantized thermal Hall effect was observed exclusively in crystals grown by the Bridgman method. A recent breakthrough came when Xing {\it et al}.\ demonstrated this effect in ultraclean crystals grown by a two-step crystal growth method \cite{namba2024two} characterized by higher thermal conductivity and larger $\ell_{ph}$ \cite{xing2024magnetothermal}. Notably, the field range supporting quantization in Bridgman-grown samples is approximately 2.5\,times wider than in ultraclean samples, aligning with Vinkler-Aviv and Rosch's predictions.
The observed low-temperature deviations from quantization thus show semi-quantitative agreement with theoretical frameworks. These findings underscore the critical role of phonon-edge current coupling in achieving quantized thermal Hall conductance.
However, it is important to note that a more systematic study involving variations in crystal size is necessary to validate and refine the theoretical models fully.

{A direct experimental approach to investigate the hypothesized decoupling between bulk phonon and edge thermal currents is through systematic measurement of sample size-dependent thermal transport properties. This will be discussed quantitatively in Sec. IX D3.}

 \begin{table*}
        \caption{A comparative study of the thermal Hall effect from various growth methods. The symbol $\checkmark$ indicates compatibility with the Kitaev QSL, while X indicates incompatibility.   For Bridgman I, $\kappa_{xy}<0$ at $\theta=+45^{\circ}$ and $+60^{\circ}$, $\kappa_{xy}>0$ at $\theta=-60^{\circ}$ and $-45^{\circ}$, and $\kappa_{xy}>0$ at $\bm{H}\parallel -a$. For Bridgman II, $\kappa_{xy}<0$ at $\theta=+60^{\circ}$.  For Bridgman III, $\kappa_{xy}>0$ at $\theta=-70^{\circ}$. For CVT II, $\kappa_{xy}>0$ at $\bm{H}\parallel -a$. For two step growth, $\kappa_{xy}>0$ at $\bm{H}\parallel -a$. [1]\cite{kasahara2018majorana}. [2]\cite{yokoi2021half}. [3]\cite{yamashita2020sample}. [4]\cite{bruin2022robustness}. [5]\cite{czajka2023planar}. [6]\cite{zhang2024stacking}. [7]\cite{xing2024magnetothermal}.
        }
   \begin{center}
  		\begin{tabular}{c|c|c|c|c}
  			\hline \hline
        	Method of sample growth & Plateau in {$\kappa_{xy}(H)$} & \begin{tabular}{c} {size of $\kappa_{xy}$ near} \\ {$(\kappa_{xy}^{2D}=\mathcal{K}_0/{2})$} \end{tabular} & sign of $\kappa_{xy}$ & Planar thermal Hall effect\\
  			\hline \hline
Bridgman I~[1][2]& \checkmark & \checkmark & \checkmark & \checkmark 
  			\\
            \hline
           Bridgman II~[3] &  \checkmark &  \checkmark & \checkmark  & 
  			\\
            \hline
            Bridgman III~[4] & \checkmark  & \checkmark  & \checkmark  & 
  			\\
            \hline
           CVT I~[5]&  X  & X &  & \checkmark
  			\\
            \hline
             CVT  II~[6] & \checkmark  & $\sim 75$\% of $\mathcal{K}_0/2$ & \checkmark  & \checkmark 
  			\\
             \hline
             Two-step growth~[7] & \checkmark  & \checkmark & \checkmark  & \checkmark \\
  			\hline \hline
  		\end{tabular}
        \end{center}
  \end{table*}

\subsubsection{Reproducibility}

\begin{figure}[b]
	\includegraphics[clip,width=1.0\linewidth]{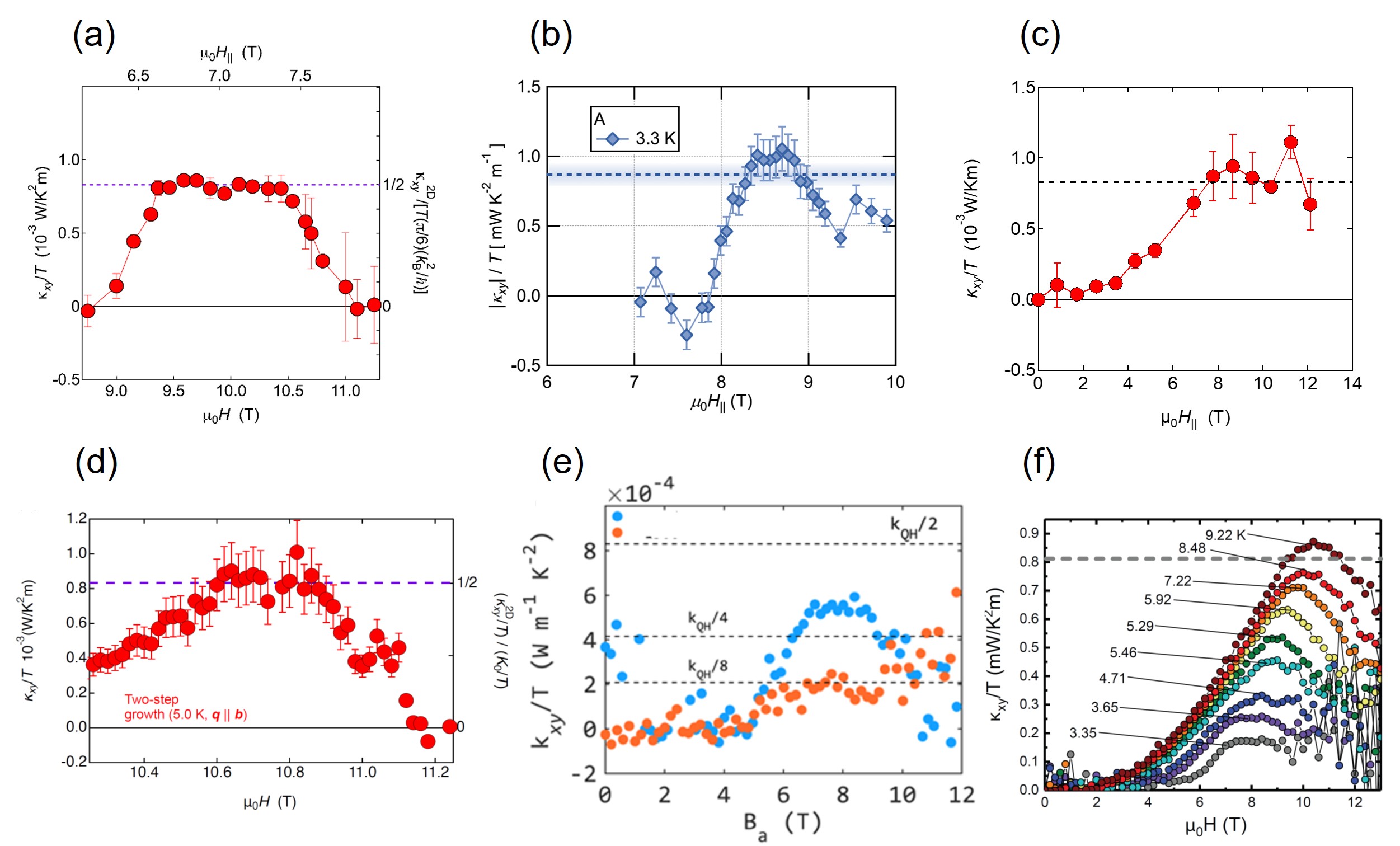}
	\caption{Field dependence of the thermal Hall effect investigated across multiple crystal samples synthesized using different growth methods, as reported by various research groups.  Bridgman-grown samples (a) \cite{kasahara2018majorana}, (b) \cite{yamashita2020sample}, and (c) \cite{bruin2022robustness} and two-step growth sample (d) \cite{xing2024magnetothermal} well-defined half-integer thermal Hall conductance plateaus. Samples (e) \cite{zhang2024stacking} and (f) \cite{czajka2023planar} are prepared using CVT method.  Sample (e), while showing a plateau feature, displays a reduced magnitude of approximately 75\% of the expected half-integer thermal Hall value. Sample (e) shows neither the plateau feature nor the half-quantized thermal Hall conductance. For more detailed quantitative results from these measurements, refer to Table II. 
	}
    \label{fig:Reproducibility}
\end{figure}

Figures\,\ref{fig:Reproducibility}(a)-(f) display the field dependence of the thermal Hall effect investigated across multiple crystal samples synthesized using different growth methods, as reported by various research groups.  The crystals grown by the Bridgman method (a), (b), and (c) \cite{kasahara2018majorana,yamashita2020sample,yokoi2021half,bruin2022robustness,kasahara2022quantized} and two-step growth method (d) \cite{xing2024magnetothermal} exhibit well-defined half-integer thermal Hall conductance plateaus. 

Samples (e) and (f) are prepared using the chemical vapor transport (CVT) method.  Sample (e),  while showing a plateau feature, displays a reduced magnitude of approximately 75\% of the expected half-integer thermal Hall value \cite{zhang2024stacking}. Although their crystals exhibit relatively high thermal conductivity, diffuse X-ray scattering observed at low temperatures suggests the presence of twin domains associated with stacking faults \cite{zhang2024stacking,kim2024structural}. It has been pointed out that such twin domains lead to partial cancellation of the thermal Hall signals \cite{kurumaji2023symmetry}, which can be a source of deviations from the half-integer quantization. Further details of the results of these measurements are provided in Table II. 

In contrast to the above findings, Czajka {\it et al}.\ reported divergent observations in the FIQD phase of $\alpha$-RuCl$_3$ \cite{czajka2023planar}, as shown in Fig.\,\ref{fig:Reproducibility}(f). Their measurements of $\kappa_{xy}$ were conducted with $\bm{H}\parallel a$. While confirming the presence of a planar thermal Hall effect, their results revealed that $\kappa_{xy}$ exhibits a non-monotonic field dependence: $\kappa_{xy}$ initially increases with the applied field, reaches a broad maximum, and subsequently decreases without manifesting the plateau behavior.

\subsubsection{Thermal Hall effect of topological boson}

Czajka {\it et al.}\ demonstrated that $\kappa_{xy}$ is strongly temperature-dependent in the range of 0.5-10\,K, with a temperature profile consistent with the distinct form expected for topological bosonic modes in a Chern-insulator-like model \cite{matsumoto2011rotational}, as described by Eq.\,(\ref{eq:thermalHall}). Their analysis suggests magnon band energies with an energy gap, and that spin excitations evolve into magnon-like modes with a Chern number of approximately 1 at high fields.
The spin excitation energy ($\omega_1 \approx 1$\,meV), derived from fits of $\kappa_{xy}/T$, closely aligns with the energy of the dominant sharp mode observed in ESR \cite{ponomaryov2017unconventional,ponomaryov2020nature}, microwave absorption \cite{wellm2018signatures} and INS experiments \cite{balz2019finite}.

Based on these findings, Czajka {\it et al.}\ argue that the observed topological boson character contradicts the previously proposed Majorana fermionic particle interpretation \cite{czajka2023planar}. Regarding the sign behavior of $\kappa_{xy}$, Zhang {\it et al}.\ and Chern {\it et al}.\ developed an alternative theoretical framework based on topological magnons, demonstrating that the sign of $\kappa_{xy}$ can be explained by the topological magnon within the specific parameter ranges \cite{zhang2021topological,chern2021sign}.

The identification of the quasiparticle nature---fermionic or bosonic---underlying the thermal Hall effect is crucial for understanding the quantum spin state of $\alpha$-RuCl$_3$. Two critical distinctions differentiate these scenarios. First, the presence of a half-integer quantized thermal Hall effect serves as a key discriminator between the two frameworks. Second, the quasiparticle excitation spectrum in the FIQD phase exhibits markedly different characteristics in each scenario. This will be discussed in Sec.\ IX.D in detail.

\subsubsection{Phonon thermal Hall effect}

\begin{figure}[b]
\includegraphics[clip,width=\linewidth]{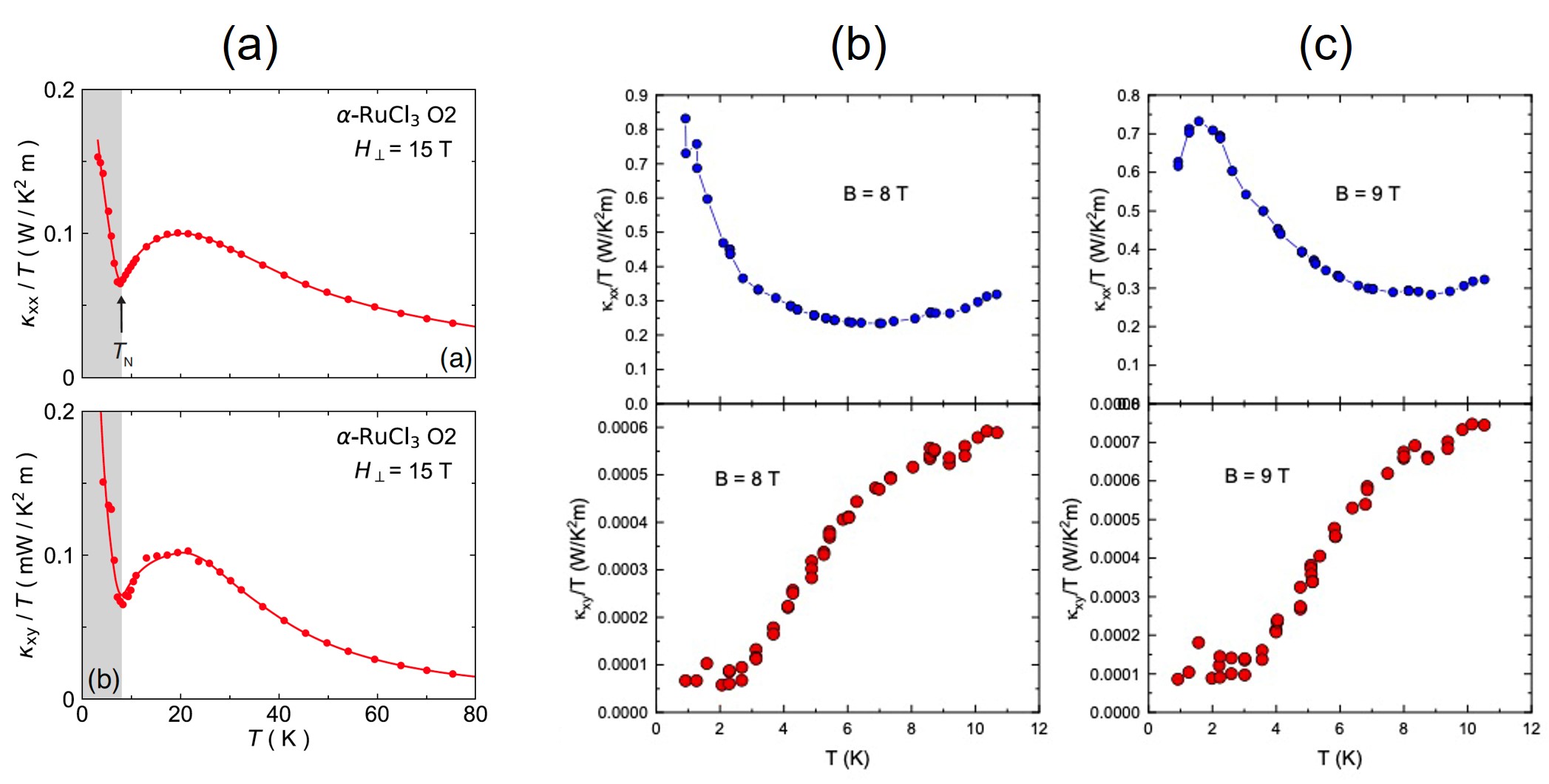}
	\caption{ (a) Temperature evolution of the longitudinal ($\kappa_{xx}/T$) and transverse ($\kappa_{xy}/T$) thermal conductivity coefficients, measured at $\mu_0H = 15\text{ T}$ for $\bm{H} \parallel c$-axis \cite{lefranccois2022evidence}.  Lefran{\c{c}}ois {\it et al.}\ interpret the correlative $T$-dependence between these components as evidence for phonon-mediated thermal Hall effect. Note a clear difference in $\kappa_{xy}$ with $\bm{H} \parallel c$-axis shown in Fig.\,\ref{fig:Thermalconductivity}(b) \cite{hentrich2019large}.  (b)(c) $T$-dependence of $\kappa_{xx}/T$ and $\kappa_{xy}/T$ under different magnetic fields with $\bm{H} \parallel a$-axis configuration in the FIQD state \cite{czajka2023planar}. Czajka {\it et al.}\ observed the anticorrelation between $\kappa_{xx}/T$ and $\kappa_{xy}/T$]. As temperature rises, $\kappa_{xx}/T$ decreases, while $\kappa_{xy}/T$ increases across most of the observed temperature range.}
    \label{fig:phonon}
\end{figure}

Traditionally, phonons were not considered capable of generating a thermal Hall signal since phonons, as neutral quasiparticles, cannot directly couple to the magnetic field.  However, {it was noted that a correction to the Born-Oppenheimer approximation gives rise to the Berry curvature of a phonon band under a magnetic field, leading to a finite phonon thermal Hall effect in nonmagnetic band insulators.\cite{saito2019berry,Behnia2025AB} Furthermore,} recent experimental findings have reported finite thermal Hall effects in several nonmagnetic insulating materials.  For instance, phonon thermal Hall effects have been proposed in such materials, including Tb$_3$Ga$_5$O$_{12}$ \cite{strohm2005phenomenological}, SrTiO$_3$ \cite{sim2021sizable}, and black phosphorus \cite{li2023phonon}. The planar thermal Hall conductivity has been measured in Kitaev candidate material Na$_2$Co$_2$TeO$_6$ \cite{takeda2022planar,chen2024planar} for {\boldmath $H$}$\parallel a^*$.  In this configuration, the two-fold rotation symmetry around the $a^*$ axis of the honeycomb lattice prohibits the thermal Hall effect.  Takeda {\it et al.}\ observed small but finite $\kappa_{xy}$ and discuss that it mainly stems from the topological magnons in the magnetically ordered states \cite{takeda2022planar}. On the other hand, the phonon contribution to $\kappa_{xy}$ has also been suggested \cite{gillig2023phononic,chen2024planar}. The phonon planar thermal Hall effect in this material with inversion and mirror symmetries presents an intriguing paradox \cite{chen2024planar}, as symmetry constraints conventionally forbid this phenomenon.

Theoretically, it has been shown that a finite angular momentum of the phonon occurs in the presence of spin-phonon interaction\cite{Zhang2014angular}, and a surge of interest in the chiral phonon has recently emerged. Three primary scenarios have been proposed to explain the phonon thermal Hall effects: the intrinsic scenario, which arises from Berry curvature effects \cite{qin2012berry,saito2019berry}, the extrinsic skew scenario \cite{chen2020enhanced,guo2021extrinsic,flebus2022charged,mangeolle2022phonon}, and the extrinsic side-jump scattering scenario \cite{guo2022resonant}. It should be noted that the magnitude of $\kappa_{xy}$ predicted by all of these scenarios is much smaller than experimentally observed values. \red{Recently it has been suggested that the phonon thermal Hall effect can potentially reach levels comparable to those observed in $\alpha$-RuCl$_3$ \cite{dhakal2024theory}.}

Lefran{\c{c}}ois {\it et al.}\  reported that the thermal Hall conductivity of $\alpha$-RuCl$_3$ may incorporate significant phonon thermal Hall contributions, based on observed similarities in the temperature dependence between $\kappa_{xx}$ and $\kappa_{xy}$ for {\boldmath $H$}$\parallel c$ shown in Fig.\,\ref{fig:phonon}(a) \cite{lefranccois2022evidence}. However, these findings diverge notably from earlier reports by Kasahara {\it et al.}\ \cite{kasahara2018unusual} and Hentrich {\it et al.}\ \cite{hentrich2019large}.   Of particular significance, data presented by Czajka {\it et al.}\ \cite{czajka2023planar}, Yokoi {\it et al.}\ \cite{yokoi2021half}, and Bruin {\it et al.}\ \cite{bruin2022robustness} demonstrate a clear anticorrelation between $\kappa_{xx}/T$ and $\kappa_{xy}/T$, as demonstrated in Figs.\,\ref{fig:phonon}(b) and (c); in temperature or field regimes where $\kappa_{xx}/T$ exhibits an increase, $\kappa_{xy}/T$ correspondingly shows a decrease as the temperature lowers. This inverse relationship in the behavior of $\kappa_{xx}$ and $\kappa_{xy}$ presents a stark contrast to the results of Lefran{\c{c}}ois {\it et al.}, demonstrating significant differences in the thermal transport data. The discrepancies are not limited to quantitative variations but extend to qualitative differences in the thermal Hall behavior. \red{Furthermore, the magnitude of the thermal Hall effect exhibits a significant sample size dependence at low temperatures in $\alpha$-RuCl$_3$ \cite{zhang2025evidence}. This finding challenges the interpretation of a phononic thermal Hall effect (see Section IX D3). Given these conflicting observations and lack of theory regarding phonon thermal Hall effects, it remains to resolve the}
phonon contributions 
to the thermal Hall conductivity in $\alpha$-RuCl$_3$.

\subsection{Anomalous thermal conductivity}

Parallel to investigations of the planar thermal Hall effect, Czajka {\it et al.}\ reveal that longitudinal thermal conductivity exhibits oscillatory behavior as a function of the strength of applied in-plane fields \cite{czajka2021oscillations}. These oscillations manifest along both $a$-axis and $b$-axis field directions and become particularly pronounced below 0.5\,K, while remaining undetectable above 1\,K. Czajka {\it et al.}\ proposed that these oscillations are analogous to quantum oscillations in metals under magnetic fields, where they arise from Landau quantization of electrons \cite{czajka2021oscillations}. They demonstrated that the oscillations show periodicity in $1/H$ and persist even within the AFM-ordered state. Given the insulating nature of the system, Czajka {\it et al}.\ attributed this phenomenon to hypothetical charge-neutral fermions.  In fact, Sodemann, Chowdhury, and Senthil had previously proposed quantum oscillations in insulators with neutral Fermi surfaces \cite{sodemann2018quantum,villadiego2021pseudoscalar}. The presence of such emergent neutral fermions would support the existence of a QSL state, albeit one distinct from that predicted by the Kitaev model.
Multiple research groups have confirmed the oscillations in $\kappa_{xx}$ as a function of magnetic field \cite{bruin2022origin,suetsugu2022evidence,lefranccois2023oscillations,zhang2024stacking,zhang2024anisotropic,xing2024magnetothermal}. 

\begin{figure}[t]
	\includegraphics[clip,width=1.0\linewidth]{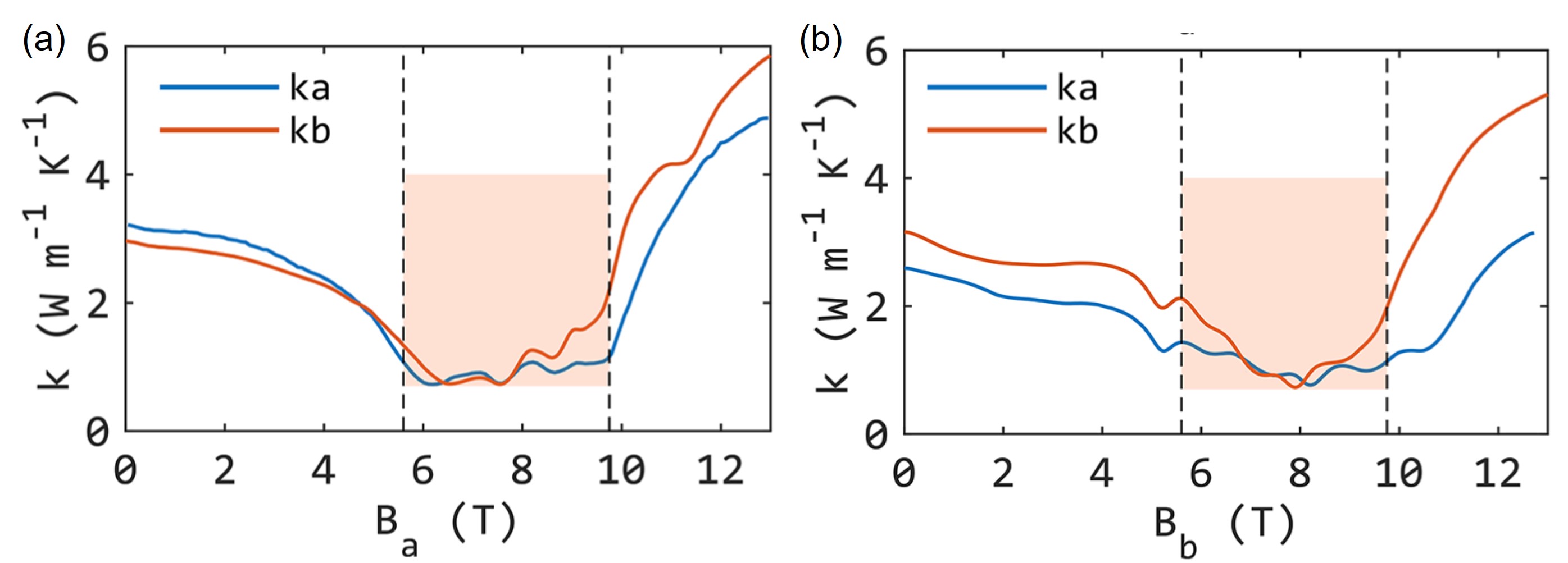}
	\caption{Field dependence of longitudinal thermal conductivity ($\kappa$) with the thermal current ({\boldmath $j$}$_q$) along the $a$ ($\kappa_a$) and $b$ ($\kappa_b$) axes and for (a) $\bm{H} \parallel a$ and (b) $\bm{H} \parallel b$ at 2\,K  \cite{zhang2024anisotropic}.  Pronounced oscillations in $\kappa$ are observed across all four measurement configurations ({\boldmath $j$}$_q \parallel a$, {\boldmath $H$}$||a$; {\boldmath $j$}$_q ||b$, {\boldmath $H$}$||a$; {\boldmath $j$}$_q ||a$, {\boldmath $H$}$||b$;  {\boldmath $j$}$_q ||b$, {\boldmath $H$}$||b$). The underlying mechanism of these oscillations remains a subject of active investigation, with two primary questions under debate: (1) whether the oscillatory behavior originates from intrinsic quantum mechanical effects or extrinsic factors such as defects or stacking faults, and (2) the fundamental physical mechanism driving these thermal transport oscillations.}
    \label{fig:Oscillation}
\end{figure}

However, their fundamental origin remains debated: while some researchers interpret them as evidence of emergent quasiparticles, others suggest they result from a sequence of magnetic phase transitions. Zhang {\it et al.}\ claim that these thermal conductivity oscillations are an intrinsic property of $\alpha$-RuCl$_3$. Their investigation of the oscillatory features utilized high-quality single crystals displaying sharp specific heat jumps at $T_N$, with no magnetic susceptibility or specific heat anomalies caused by stacking faults above $T_N$. Through comprehensive measurements of thermal conductivity as functions of both thermal current direction and magnetic field vector orientation (Figs.\,\ref{fig:Oscillation}(a) and (b)), they revealed distinct behavioral patterns. When measuring thermal conductivity along both the $a$- and $b$-axes, the oscillations exhibited synchronized behavior for $\bm{H} \parallel a$, yet became notably phase-shifted for {\boldmath $H$}$\parallel b$. Furthermore, systematic variation of the angle between thermal current and magnetic field vectors revealed higher-order angular oscillations in the thermal resistivity,  reminiscent of theoretical predictions for the Kitaev QSL. Based on this extensive experimental evidence, Zhang {\it et al.}\ proposed that the oscillatory thermal conductivity in $\alpha$-RuCl$_3$ shares a fundamental connection with the putative Kitaev QSL state and its associated Majorana fermion excitations in the ground state. 

On the other hand, several research groups, including Bruin {\it et al.}\ \cite{bruin2022origin}, Lefrançois {\it et al.}\ \cite{lefranccois2023oscillations}, and Xing {\it et al.}\ \cite{xing2024magnetothermal}, argue that these oscillations have an extrinsic origin. Bruin {\it et al.}\ conducted comprehensive measurements of $\kappa_{xx}(H)$ at temperatures down to 100\,mK, complemented by magnetization ($M$) studies, using single crystals synthesized via two distinct methods. Their results revealed characteristic dips in $\kappa_{xx}(H)$ that coincide precisely with peaks in the field derivative of $M$ at identical magnetic field values, with this correlation remaining consistent across different crystal growth techniques. This consistency supports their interpretation that these features stem from field-induced phase transitions rather than quantum oscillations. Furthermore, their observations of the magnitude of these features in $\kappa_{xx}(H)$ contradict quantum oscillation expectations in two key aspects: the features become diminished in crystals with higher thermal conductivity and show a marked decrease upon cooling below 1\,K. Adding further weight to this interpretation, Xing {\it et al.}\ demonstrated that these oscillations are significantly suppressed in high-quality single crystals grown using a sophisticated two-step method. They propose a mechanical mechanism for the observed phenomena: sample bending during thermal conductivity measurements may introduce stacking faults, which could subsequently trigger magnetic ordering and introduce additional scattering mechanisms.

\red{However, Zhang {\it et al.} reported that the various anomalies at fields above 7.3\,T are well-defined and reproducible in samples with a dominant magnetic transition around 7\,K but are suppressed in samples with defects showing additional 14-K magnetic transition \cite{zhang2023sample}. In the temperature and field dependence of thermal conductivity, no anomaly was observed to be associated with the magnetic order around 14\,K or 10\,K.}

\red{Thus, there is no consensus on the origin of the anomalies at high fields.}
To resolve these conflicting observations of $\kappa_{xx}(H)$, future investigations 
are essential to definitively establish whether this oscillatory behavior is intrinsic or extrinsic.

\subsection{Bulk-edge correspondence}

An important aspect of the topological properties is the bulk-edge correspondence. While the edge modes of Majorana fermions are determined by the Chern number $C_h$, the bulk excitations are determined by the Majorana gap $\Delta_M$ given by Eq.\,\ref{eq:Majoranagap} in the Kitaev QSL. These two quantities $C_h$ and $\Delta_M$ are closely related through the product $h_xh_yh_z$ in the Kitaev model. As discussed in the previous section, the Majorana edge modes can be detected by the quantized thermal Hall effect, whose sign is driven by the Chern number (see also Fig.\,\ref{fig:Dirac3}). The Chern number has characteristic field-angle dependence through the product $h_xh_yh_z$ as depicted in Fig.\,\ref{eq:Chern}. Correspondingly, bulk excitations are also expected to depend strongly on the field directions, which specific heat measurements can test \cite{Hwang2022}. 

\begin{figure*}[t]
	\includegraphics[clip,width=0.8\linewidth]{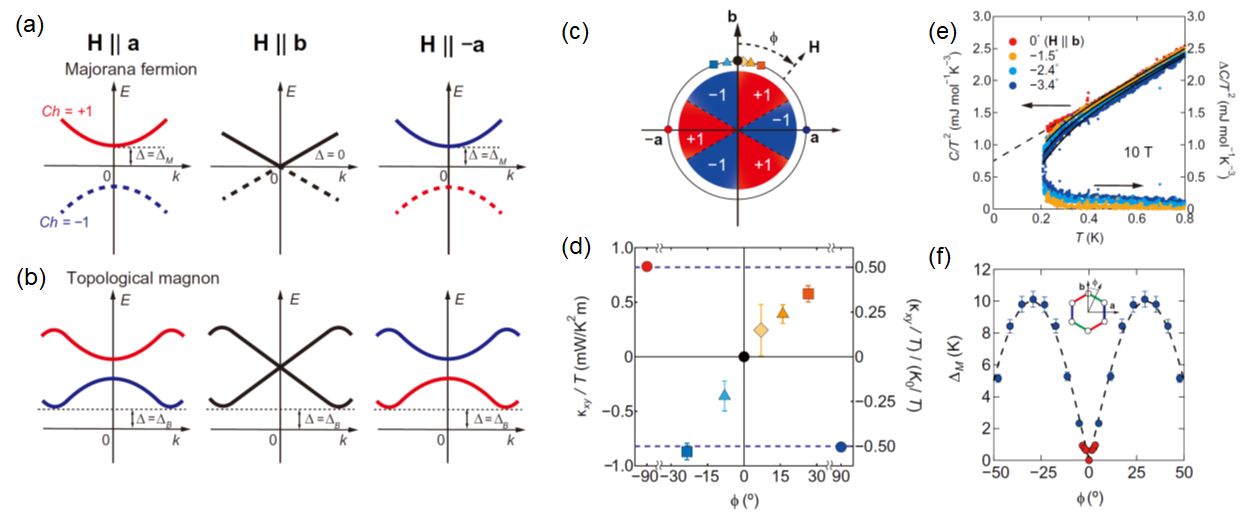}
	\caption{(a) Schematic energy dispersion $E(k)$ of the itinerant Majorana fermions in a Kitaev QSL. The red and blue curves have different topological Chern numbers $+1$ and $-1$, respectively. The $E < 0$ part is redundant (dashed lines), but its Chern number determines the topological properties. For field $\bm{H}$ parallel to the $a$ (left) or $-a$ axis (right), the Majorana excitations have a finite energy gap described by $\Delta_M$, which closes for $\bm{H}\parallel b$ (middle). (b) Similar plots for topological magnons. In the spin-polarized phase, the lowest energy band, whose Chern number determines the topological properties, always has a finite gap $\Delta_B$. (c) In-plane field-angle dependence of the Chern number $C_h$. (d) Field-angle dependence of $\kappa_{xy}/T$ at a fixed field of 11\,T at $T = 4.8$\,K. (e) Temperature dependence of $C/T^2$ at 10\,T for tilt angles $\phi$ from the bond direction ($b$ axis). Dashed lines represent the linear fitting for $\phi = 0^\circ$, and solid lines are fitting curves with the excitation gap $\Delta_M$ for finite angles. The data for the change in $C/T^2$ from $\phi = 0^\circ$ are also shown (right axis). These are the raw data without phonon contributions subtracted. (f) Angle dependence of Majorana gap $\Delta_M$ at 10\,T. The data at low angles (red circles, deduced from (e)) are obtained in dilution-fridge measurements, and the high-angle data (blue circles) are from $^3$He measurements. $\Delta_M(\phi)$ shows a $|\sin 3\phi|$ angle dependence (dashed lines). \cite{imamura2024majorana} 
	}
    \label{fig:MajoranavsMagnon}
\end{figure*}

\subsubsection{Specific heat at high fields}

Specific heat is one of the most fundamental physical quantities and thus there have been several studies to discuss spin fractionalization in $\alpha$-RuCl$_3$ \cite{wolter2017field,sears2017phase,widmann2019thermodynamic,bachus2020thermodynamic,Bachus2021,tanaka2022thermodynamic}. Unlike the experimental spin probes that may not be able to access fractionalized excitations directly, the thermodynamic probes can get the information on the Majorana and vison excitations that both have entropy. Early reports found gapped excitations in the specific heat, and the gap estimated from the temperature dependence of specific heat $C(T)$ tends to increase as a function of the magnetic field in the FQID phase \cite{wolter2017field,sears2017phase}. However, the field direction was not the focus of these studies. 

Tanaka {\it et al.}\ have reported high-resolution specific heat measurements in a single crystal of $\alpha$-RuCl$_3$ under field rotation within the honeycomb plane down to $\sim0.7$\,K \cite{tanaka2022thermodynamic}. They found that the field-angle dependence of low-temperature specific heat $C$ shows distinctly different behaviors between the low-field AFM and high-field FIQD states. From the analysis of the temperature dependence of $C$ at different field angles, they found that the extracted excitation gap shows strong angle dependence.
For $\bm{H}\parallel a$, they use the analysis involving two different gaps, a high-energy gap corresponding to the vison gap $\Delta_K$ that is related to the peak in magnetic $C(T)$, and the other low-energy gap corresponding to the Majorana gap $\Delta_M$. From this, it is found that the field dependence of the low-energy gap indeed shows an increasing trend with the field. This field dependence of the excitation gap has been discussed in comparison with Eq.\,(\ref{eq:Majoranagap}), by using the field dependence of the peak temperature $T_{\rm max}$ of magnetic $C(T)$ as a measure of the $Z_2$ flux gap $\Delta_K$ through the relation $T_{\rm max}\propto\Delta_K$ as found in a calculation of the pure Kitaev model \cite{motome2020hunting}. The $H^3$ dependence expected from the product $h_xh_yh_z$ for $\bm{H}\parallel a$ can then be reproduced when they plot $T_{\rm max}^2\Delta_M$ as a function of $H^3$. The presence of the low-energy gap in the bulk excitations implies that the edge mode discussed in the thermal Hall effect is protected by topology through the bulk-edge correspondence.

It should be noted that the analysis was done by subtracting the phonon contribution, and as mentioned previously, the spin-phonon coupling is significant in magnetic materials with strong SOC. Thus, it is important to see the zero-temperature limit behavior where the phonon contributions vanish. 

\subsubsection{Quasiparticles; fermions or bosons?}

An important difference in the excitation spectra between Majorana fermions and bosonic magnons has been pointed out by Imamura {\it et al.}\ \cite{imamura2024majorana}. As shown in Fig.\,\ref{fig:MajoranavsMagnon}(a), the Majorana excitation gap is closed for $\bm{H}\parallel b$ (bond direction), which comes from the $h_xh_yh_z$ factor in Eq.\,(\ref{eq:Majoranagap}). In sharp contrast, the bosonic magnons are always gapped in the spin-disordered phase above the critical field of AFM long-range order. 
In other words, when the field angle is rotated, topological band crossing occurs at {\em zero} energy in the Majorana fermion case, whereas the band crossing occurs at a {\em finite} energy in the magnon case. 

Based on this idea, the bulk-edge correspondence of the Kitaev QSL has been further tested through the combination of the thermal Hall conductivity and the specific heat measurements under field rotation near the Ru-Ru bond axis (the $b$ axis). It is found that the sign of the thermal Hall effect changes when the field direction is rotated through the $b$ axis as shown in Figs.\,\ref{fig:MajoranavsMagnon}(c) and (d) \cite{imamura2024majorana}. For $\bm{H}\parallel b$, $\kappa_{xy}$ becomes zero, and concurrently the specific heat down to $\sim200$\,mK shows a residual $C/T^2$ component in the low-temperature limit, which was interpreted as an indication of the gapless Dirac-cone-line excitations (Fig.\,\ref{fig:MajoranavsMagnon}(e)). 
Tilting of field direction from the $b$ axis results in the vanishing of the residual $C/T^2$, which can be consistently described by the gap opening in the $|\sin(3\phi)|$ form, as demonstrated in Fig.\,\ref{fig:MajoranavsMagnon}(f). (Here $\phi$ is defined as the in-plane field angle from the $b$ axis.)
The authors emphasize that the data in Fig.\,\ref{fig:MajoranavsMagnon}(e) is raw data without subtracting the phonon contribution, and the low-temperature change from gapless to gapped behaviors by only a few degree rotations is a remarkable result. This sharp transition suggests a distinct phase change in the material's fundamental properties.
From the general discussion of the topological bands introduced above, it is claimed that this gap-closing behavior with the sign-changing thermal Hall effect is the hallmark of an angle rotation-induced topological transition of fermions, ruling out the bosonic scenarios of planer thermal Hall effect in the FIQD state of $\alpha$-RuCl$_3$.

In $\alpha$-RuCl$_3$, non-Kitaev interactions are present, leading to the AFM order at low fields.  Assuming that the field induces the Kitave spin liquid, and one may apply Eq.\,(\ref{eq:Ch}) in the perturbative approach, where these non-Kitaev interactions are now smaller than the gap $\Delta_K$, which lead to an additional term  $\propto(h_x+h_y+h_z)$ in the Majonara excitation gap given by Eq.\,(\ref{eq:Majoranagap}). This additional term is zero for the field within the honeycomb plane ($[111]$ plane), in which $h_xh_yh_z$ dependence is expected \cite{huang2021heat,tanaka2022thermodynamic,imamura2024majorana}. Thus the above results are consistent with the Kitaev QSL even in the presence of perturbative non-Kitaev interactions. 
The challenge in this approximation lies in understanding why and how the non-Kitaev interactions are smaller than the gap $\Delta_K$ due to the field, despite being significant enough to induce the AFM order in the absence of the field. This issue remains unresolved.

{
\subsubsection{ 1D thermal edge currents and their decoupling from bulk phonons}

Zhang {\it et al.} investigated edge thermal currents by measuring thermal transport properties in samples of different sizes using focused ion beam nanofabrication to create samples with precise micrometer-scale geometry \cite{zhang2025evidence}.
The thermal Hall resistivity $\lambda_{yx}$ shows negligible size dependence between 1\,mm and $\sim 100\mu$m wide samples at 4-6\,K. However, at lower temperatures around 2\,K, $\lambda_{yx}$ vanishes in large samples (1\,mm) but remains finite and enhanced in small samples ($\sim 100\mu$m). When sample dimensions approach the phonon mean free path ($\sim100\mu$m), edge thermal currents become thermally decoupled from the bulk phonon reservoir due to the ballistic transport regime.
The results provide experimental evidence that geometric constraints enforce ballistic phonon transport, preserving edge contributions to thermal Hall transport. Zhang {\it et al}. conclude that chiral fermion edge modes are responsible for the observed thermal Hall effect\cite{zhang2025evidence,halasz2025geometry}.
Nonetheless, further theoretical studies on phonon contributions taking into account boundary scattering in confined geometries and anharmonic effects are desired to better distinguish them from edge contributions.}

\subsubsection{Possible field-induced topological transition}

An important feature of the thermal Hall effect is that the half-integer quantization is observed in a limited field window, and above the plateau regime $\kappa_{xy}$ becomes close to zero \cite{kasahara2018majorana,yokoi2021half}. This high-\red{field} fate of the quantization implies that the topological properties governed by the Chern number vanish at higher fields, suggesting a kind of topological phase transition. Near this field range, the field-angle dependence of specific heat also shows a change from 6-fold to 2-fold rotational symmetry \cite{tanaka2022thermodynamic}. This rotational symmetry breaking appears with a small anomaly in the specific heat at low temperatures, suggesting a field-induced phase transition. Such a field-induced transition has also been discussed from anomalies in the field dependence of specific heat and longitudinal thermal conductivity by Suetsugu {\it et al.}\ \cite{suetsugu2022evidence}. However, the measurements of magnetic Gr\"uneisen parameter did not provide clear evidence for a phase transition \cite{bachus2020thermodynamic}, \red{and the angular dependence of the magnetization below $\sim7$\,T also reported small two-fold anisotropy even above $T_N$ \cite{balz2021field}.}

To explain the observed change in the rotational symmetry, Takahashi {\it et al.}\ proposed a theory of topological nematic transition which can be induced by a magnetic field \cite{Takahashi2021}. They consider four-body interactions among Majorana fermions, which can trigger a nematic phase transition of Majorana bonds. This corresponds to a field-induced change from the isotropic non-Abelian phase (B phase in the Kitaev model) with $C_h=\pm1$ to an anisotropic toric-code phase (A phase) with $C_h=0$, which can account for the vanishing thermal Hall plateau at high fields. The toric-code phase is characterized by the nematic order parameter, and thus this transition between topologically different phases can be called a field-induced topological nematic transition. So far, only the measurements of specific heat $C(\phi)$ have detected such field-induced rotational symmetry breaking, and further experiments are required to clarify this point.

\section{Effect of disorder}

\subsection{Unquantized thermal Hall conductance}

The observation of quantized thermal conductance remains controversial, with different research groups reporting conflicting results (Fig.\,\ref{fig:Reproducibility}). For instance, sample \#5 in Fig.\,\ref{fig:Decoupling} exhibits a monotonic decrease in $|\kappa_{xy}|/T$ with decreasing temperature, with values consistently below the half-quantized threshold throughout the measured temperature range. Kasahara {\it et al}.\ identified crucial correlations between the half-integer quantum thermal Hall effect and key sample-dependent parameters \cite{kasahara2022quantized}. Their analysis revealed that the magnitude of longitudinal thermal conductivity and the N\'{e}el temperature at zero field play decisive roles in this phenomenon. Specifically, samples displaying the half-integer quantum thermal Hall effect possess zero-field thermal conductivity above a critical threshold, indicating that extended heat carrier mean free paths are essential for its manifestation.
In $\alpha$-RuCl$_3$, the N\'{e}el temperature shows subtle sample-dependent variations \cite{kasahara2022quantized,zhang2024stacking}. Kasahara {\it et al.}\ reported that specimens with higher N\'{e}el temperatures require stronger magnetic fields to achieve quantization. These findings strongly indicate that the quantization phenomenon is intimately tied to both impurity scattering and non-Kitaev interactions, revealing the intricate relationship between material properties and quantum thermal transport in these systems.

Zhang {\it et al}.\ revealed the critical impact of stacking disorder on thermal transport in $\alpha$-RuCl$_3$ \cite{zhang2024stacking}. Their research demonstrates that even subtle stacking irregularities can dramatically influence the N\'{e}el temperature while suppressing both structural transitions and thermal conductivity. Crystals exhibiting minimal stacking disorder achieve N\'{e}el temperatures above 7.4\,K, while those with more pronounced stacking faults show N\'{e}el temperatures below 7\,K.  Their investigation revealed that $\alpha$-RuCl$_3$ crystals with minimal stacking disorder display enhanced thermal conductivity, which drives the thermal Hall conductivity closer to the half-integer quantized value. Within a specific magnetic field range, Zhang {\it et al}.\ observed a planar thermal Hall effect accompanied by a distinct thermal Hall plateau, with the thermal Hall conductance reaching approximately 70\% of the expected half-quantized value.

These results align with the earlier findings of Kasahara {\it et al.}\ \cite{kasahara2022quantized}, emphasizing the fundamental importance of sample quality and structural integrity in quantum thermal phenomena. The observed correlation between reduced stacking disorder, elevated N\'{e}el temperatures, and enhanced thermal conductivity strengthens the premise that crystalline perfection plays a decisive role in the emergence of quantized thermal transport in these materials.

\subsection{Atomic substitution and electron irradiation}

It has been discussed that site vacancies and bond disorder inevitably exist in real materials can give rise to intriguing phenomena in Kitaev magnets \cite{willans2010disorder,willans2011site,Knolle2019,Nasu2020,Sreenath2012,Petrova2013,Udagawa2018,Dantas2022,Singhania2023,takahashi2023nonlocal,Yamada2020,Kao2021,Banerjee2023emergent}. The effect of a relatively large amount of site vacancies has been experimentally studied by partial substitution of Ru site with Ir and Rh, which reveals that the magnetic order in $\alpha$-RuCl$_3$ can be suppressed by disorder \cite{Lampen-Kelley2017,Do2018,Do2020,bastien2022dilution,morgan2024structure}. 

To investigate how dilute disorder changes the Kitaev QSL, electron irradiation provides an ideal method for introducing uniformly distributed point defects. Using this technique, Imamura {\it et al.}\ conducted a specific heat study that examined the effects of impurities on low-energy excitations in the FIQD state of $\alpha$-RuCl$_3$ \cite{imamura2024defect}. Their findings revealed that the disorder induces low-energy excitations in specific heat, consistent with theoretical calculations on bond disorder and site dilution effects \cite{Yamada2020,Kao2021}. From the scaling behavior proposed for disordered quantum spin systems \cite{kimchi2018scaling}, it is reported that disorder-induced low-energy excitations are very sensitive to the field. From these results, the possible weak localization effect of Majorana fermions has been discussed \cite{imamura2024defect}. 

In relation to this, it has been theoretically discussed that a chiral spin liquid can emerge in amorphous lattices that can be induced by ion irradiation, implying the importance of structural disorder in Kitaev materials \cite{Grushin2023,Cassella2023}.

\subsection{Ultraclean crystals}

Another direction for studying the effect of disorder is to make the samples cleaner. Recent efforts on single crystal growth have made available high-quality crystals with superior cleanness compared with those in the conventional CVT or Bridgman crystals. These new generation crystals have been synthesized by self-selective vapor phase growth (SSVG) \cite{Yan2023} and more recently by the two-step growth consisting of a purification process by CVT followed by the main crystal growth process by sublimation \cite{namba2024two}. These crystals exhibit large longitudinal thermal conductivity $\kappa_{xx}$ in the zero-field AFM state, indicating long mean free paths of the heat carrier. From comparisons of physical properties in different quality crystals, it has been shown that higher quality crystals tend to have higher $T_N$ (up to 7.7\,K) \cite{namba2024two}. Note that even in these ultraclean crystals, stacking faults can be easily introduced by external strain or cutting, which can be checked by additional higher temperature AFM transitions at $\sim10$ and 14\,K. The intrinsic $T_N\sim7$\,K is reduced by electron irradiation \cite{imamura2024defect}, so there seems to be a systematic trend that disorder suppresses $T_N$, and a higher $T_N$ can be used as a measure of cleanliness. \red{Indeed, a clear correlation between $T_N$ and the phonon mean free path estimated from $\kappa_{xx}$ in the zero-field AFM state has been recently demonstrated \cite{Imamura2025}.}

As discussed in Fig.\,\ref{fig:Reproducibility}(d), the ultraclean crystals show half-integer quantization \cite{xing2024magnetothermal}, which supports the fermionic edge current. \red{Further confirmations on the bulk excitations have been recently reported by field-angle dependent specific heat measurements in these ultraclean crystals \cite{Imamura2025}, which show gapped excitations for $\bm{H}\parallel a$ and gapless excitations for $\bm{H}\parallel b$. It should be noted that the residual values $\alpha$ of $C/T^2$ for $\bm{H}\parallel b$ in these ultraclean crsytals are found to be in close correspondence with the theoretical estimates of $\alpha$ (which is inversely proportional to the Fermi velocity squared in the linear Majorana dispersions) with the reported Kitaev interactions $K= 5$-10\,meV.} 

\section{Other candidate Kitaev materials}

Intense efforts continue to explore other candidate materials. 
As several nice reviews of Kitaev materials have already been available \cite{takagi2019concept,trebst2022kitaev,garlea2024review,hermanns2018physics,motome2020hunting,rousochatzakis2024beyond}, here we will briefly discuss recent highlights related to these materials.

\subsection{Ir-based materials}

Together with the $4d$ electron system $\alpha$-RuCl$_3$, Ir-oxide honeycomb magnets with $5d$ electrons have been extensively studied \cite{takagi2019concept,motome2020hunting,trebst2022kitaev}. Due to the Jackeli-Khaliullin mechanism \cite{jackeli2009mott,chaloupka2010kitaev}, honeycomb structures based on IrO$_6$ octahedra with Ir$^{4+}$ ($d^5$) configuration can lead to the Kitaev interactions between localized $j_{\rm eff}=1/2$ spins. The Ir-based candidate materials include $\alpha$-$A_2$IrO$_3$ ($A =$ Li, Na) \cite{singh2010antiferromagnetic,singh2012relevance} and $A'_3$LiIr$_2$O$_6$ ($A' =$ H, Cu, Ag) \cite{kitagawa2018spin,takagi2019concept} with layered honeycomb planes similar to $\alpha$-RuCl$_3$, as well as $\beta$-Li$_2$IrO$_3$ and $\gamma$-Li$_2$IrO$_3$ with 3D hyperhoneycomb network \cite{Takayama2015,Modic2014}. 

\red{We note that 3D Kitaev models have been developed, demonstrating that fractionalized Majorana fermions generically form exotic metallic phases \cite{o2016classification}. These phases exhibit diverse topological characteristics, including Majorana Fermi surfaces, nodal line semimetals, and topologically protected Weyl nodes. The critical breakthrough lies in demonstrating that the nature of these emergent Majorana metals can be systematically predicted through projective symmetry analysis of time-reversal and inversion symmetries, providing a unified theoretical framework for understanding quantum fractionalization phenomena in 3D frustrated magnetic systems.}

Among them, H$_3$LiIr$_2$O$_6$ is unique in that no antiferromagnetic order has been observed down to $\sim0.05$\,K \cite{kitagawa2018spin}, which is quite important to study possible QSL ground state at zero field. Unfortunately, however, the effect of strong disorder is evident in the available polycrystalline samples \cite{Kao2021}, which prevents the nature of the phase, i.e., disorder-driven disordered phase vs.\ QSL, despite the presence of disorder. Further studies are desired to enhance the sample quality to study this important candidate material.

\subsection{Co-based materials}

Another candidate system is Co$^{2+}$-based honeycomb magnets with $3d$ electrons. It has been theoretically proposed that the electron hopping for the $e_g$ channel, which exists in the high-spin configuration of $3d^7$, suppresses the Heisenberg interaction and strengthens the Kitaev interaction \cite{Liu2018,Sano2018,Motome2020}. Candidate materials include Na$_2$Co$_2$TeO$_6$ \cite{Lin2021,Hong2021} and K$_2$Co$_2$TeO$_6$ \cite{Xu2023} with different stacking types, as well as CaCo$_2$TeO$_6$ \cite{Haraguchi2024}, Na$_3$Co$_2$SbO$_6$ \cite{Yan2019,Li2022} and BaCo$_2$(AsO$_4$)$_2$ \cite{Zhong2020,Zhang2023,Halloran2023}.

In these materials, the zero-field ground state is antiferromagnetic, and in-plane magnetic fields can suppress the magnetic order similar to the case of $\alpha$-RuCl$_3$. However, these are relatively new materials and, at the present stage, no solid evidence for a Kitaev QSL with Majorana excitations has been obtained in the field-induced disordered phase above the critical field, which requires further investigation. In particular, BaCo$_2$(AsO$_4$)$_2$ exhibits low AFM critical field of only $\sim0.5$\,T, which is promising for wider studies \cite{Zhong2020}.

\begin{figure}[b]
	\includegraphics[clip,width=\linewidth]{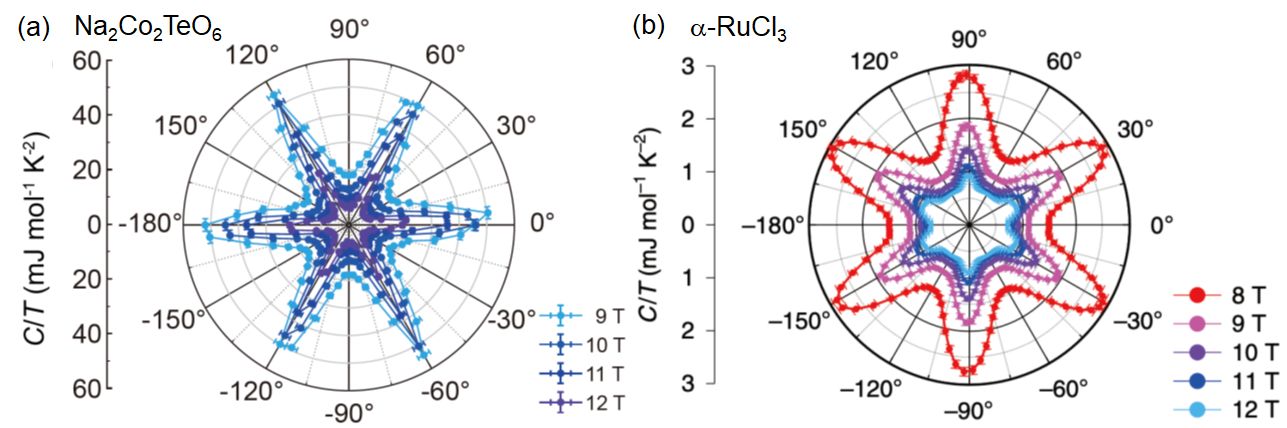}
	\caption{Comparison of the field-angle dependence of specific heat for two Kitaev candidate materials at several fields above the AFM critical field at constant temperatures. (a) For Na$_2$Co$_2$TeO$_6$ at 2.5\,K \cite{fang2024}.(b) For $\alpha$-RuCl$_3$ at 0.7\,K \cite{tanaka2022thermodynamic}. The phonon contribution is not subtracted.
	}
\label{fig:NCTO}
\end{figure}

In the most studied Na$_2$Co$_2$TeO$_6$, the AFM order sets in below $T_N=27$\,K at zero field \cite{Lin2021,Hong2021}. At low fields, a collinear single-$q$ zigzag state \cite{Yao2020,Lefrancois2016,Bera2017} or non-collinear triple-$q$ state \cite{Chen2021,Lee2021,Yao2023,Kruger2023} have been discussed, and at higher fields above $\sim6$\,T a canted
AFM state appears up to $\sim10$\,T above which the field-induced disordered phase is considered as a candidate of QSL. Neutron scattering experiments suggest that the Kitaev interaction is significant as theoretically predicted, but conflicting ferromagnetic ($-9.0$\,meV \cite{Songvilay2020}) and antiferromagnetic ($+3.6$\,meV \cite{Kim2021}) estimates have been reported. Takeda {\it et al.}\ have found that planar thermal Hall conductivity emerges above $\sim6$\,T for both fields along the bond and zigzag direction in Na$_2$Co$_2$TeO$_6$, which persists above the AFM critical field of $\sim10$\,T. They do not observe the quantization of the thermal Hall effect and conclude that topological magnons are responsible for the planar thermal Hall effect. 

This conclusion is supported by the specific heat measurements under in-plane field rotation \cite{fang2024}, which revealed fully gapped excitations at low energies without having the gap closure expected for the bond direction in the Kitaev QSL. This is completely different from the $\alpha$-RuCl$_3$ case, which can be readily seen in Fig.\,\ref{fig:NCTO} comparing the field-angle dependence of specific heat in the field-induced disordered states in the two materials. In the bond direction ($90^\circ$ angle in the notation of Fig.\,\ref{fig:NCTO}), where the gapless excitations are expected in the Kitaev QSL, the specific heat shows a broad maximum in $\alpha$-RuCl$_3$, whereas it shows a minimum in Na$_2$Co$_2$TeO$_6$. These results evidence that the low-energy physics cannot be described by the Kitaev QSL, although the signatures of Kitaev interaction may be witnessed by higher-energy probes such as neutron scattering. In cobaltate, a recent theory pointed out that the relationship between the Kitaev and Heisenberg interactions is strongly influenced by the inter- and intra-orbital exchange paths \cite{Liu2023}, indicating that reducing the direct intra-orbital path enhances the Kitaev interaction and that small variations in crystal structures can lead to different ground states.

\subsection{$f$-electron systems}

In addition to the above $d$-electron systems, $f$-electron systems, where strong spin-orbit coupling is expected, have been considered as candidate materials \cite{Li2017,Rau2018}. For instance, Yb$^{3+}$-based compounds with $4f^{13}$ electron configurations were nominated \cite{Rau2018,Luo2020}, and YbCl$_3$ isostructural to $\alpha$-RuCl$_3$ have been studied \cite{Xing2020,Sala2023,Matsumoto2024}. The  AFM transition sets in at a relatively low temperature $T_N=0.6$\,K, which is completely suppressed by an in-plane critical field of 6\,T \cite{Matsumoto2024}. Although the proximity to Kitaev physics was discussed \cite{Xing2020}, it has been recently pointed out that at the critical field, a transition to a fully polarized state occurs, which is described by a quantum critical 2D Bose gas \cite{Matsumoto2024}.

Based on {\it ab initio} calculations, Jang {\it et al.}\ proposed that another Pr-based candidate material $A_2$PrO$_3$ ($A =$ alkali metals) can be realized in the forms of layered honeycomb and three-dimensional hyperhoneycomb structures of PrO$_6$ octahedra \cite{Jiang2019,Jiang2020}. In these materials, it has been shown that the $4f^1$ states under an octahedral crystal field and strong SOC are well approximated by the $\Gamma_7$ Kramers doublet with the effective angular momentum $j_{\rm eff} = 1/2$ for all the $A$-site substitutions. Indeed, it has been experimentally demonstrated that Na$_2$PrO$_3$ crystallizes in two polymorphs ($\alpha$ and $\beta$) comprising honeycomb and hyperhoneycomb lattices of octahedrally coordinated Pr$^{4+}$ ($4f^1$) \cite{Okuma2024IC,Okuma2024NC}. The 3D hyper-honeycomb $\beta$-NaPrO$_3$ exhibits an AFM transition at $T_N=5.2$\,K. From neutron scattering experiments, Okuma {\it et al.}\ found a strongly noncollinear magnetic order with highly dispersive gapped excitations, implying the off-diagonal $\Gamma$-type interactions \cite{Okuma2024NC}. These results suggest that the rare-earth $4f$ honeycomb systems may offer a platform for the exploration of quantum compass spin models in the extreme spin-orbit regime.

In addition, it has been theoretically suggested that other Kitaev materials with rare-earth ions such as Nd$^{3+}$ ($4f^{3}$) and Er$^{3+}$ ($4f^{11}$) exhibit strong AFM Kitaev interactions \cite{Jang2024}. It was also reported that $4f$ honeycomb magnet SmI$_3$ shows no sign of long-range magnetic order down to 0.1\,K \cite{Ishikawa2022}. These results suggest that the material design of $f$-electron systems shows potential for future Kitaev QSL studies.



\subsection{Atomic layer thick $\alpha$-RuCl$_3$}

Recent STM studies of  1T-TaSe$_2$, a candidate for hosting a QSL state \cite{law20171t,ribak2017gapless,klanjvsek2017high,murayama2020effect}, with monolayer thickness have revealed signatures of fractionalized quantum spin excitations  \cite{ruan2021evidence,chen2022evidence}. These results suggest that low-energy magnetic excitations may be accessible above the Mott gap via tunneling electrons.

\begin{figure}[b]
	\includegraphics[clip,width=0.95\linewidth]{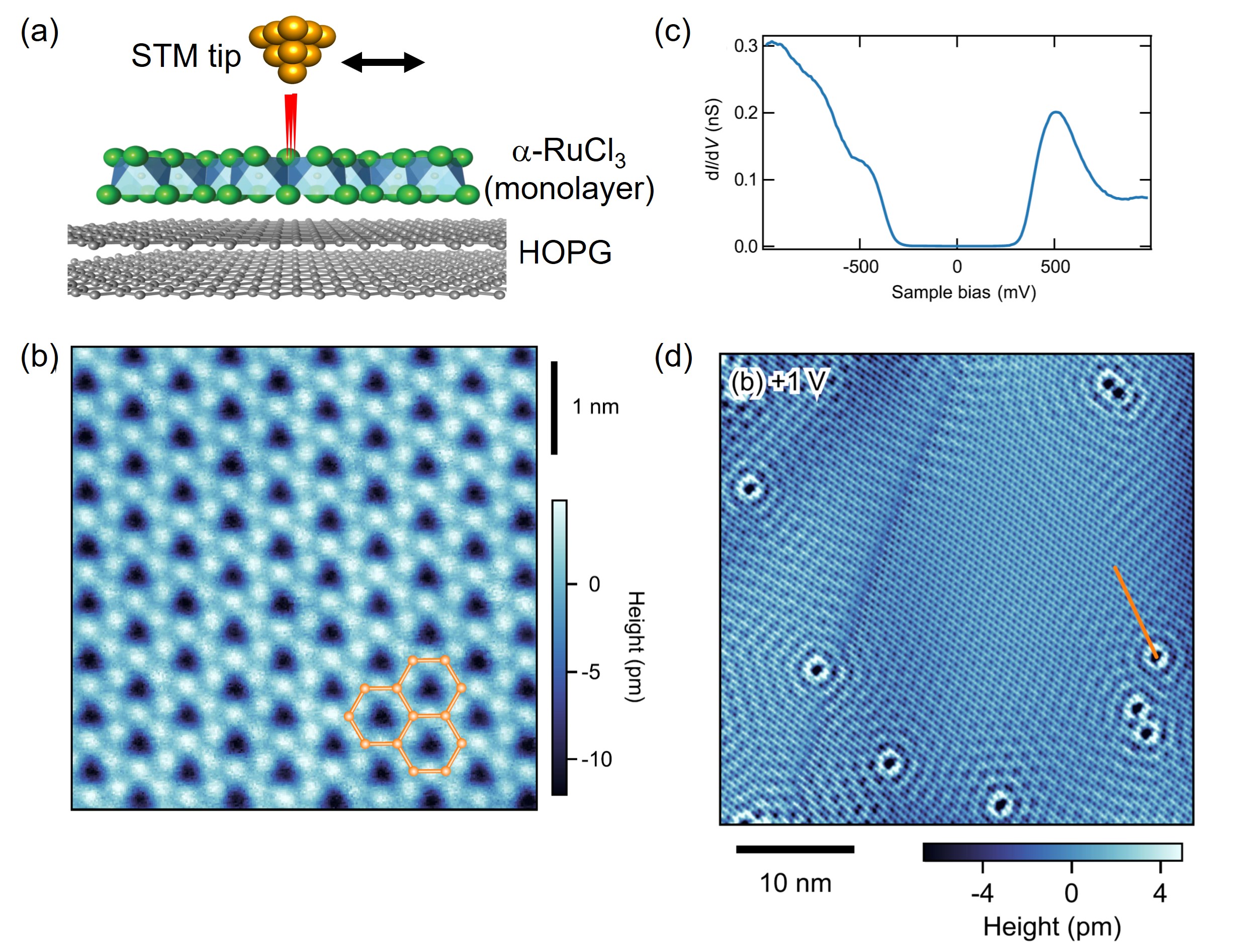}
	\caption{(a) Schematic diagram of STM measurement setup for monolayer $\alpha$-RuCl$_3$ deposited on HOPG. (b) Atomic-resolution topographic image of monolayer $\alpha$-RuCl$_3$, which reveals a kagome-like lattice of regularly arranged circular protrusions. The overlaid illustration depicts the position of the Ru honeycomb. (c) The differential conductance ($dI/dV$) of STM spectroscopy. A Mott gap of approximately 0.6\,eV between the lower and upper Hubbard bands near the Fermi level within the $t_{2g}$-$p_{\pi}$ manifold is observed. (d) Long-wavelength spatial modulations in the LDOS decaying
away from defects reported by Kohsaka {\it et al.}\ \cite{kohsaka2024imaging}.  Similar oscillations have also been reported by Qiu {\it et al.}\ \cite{qiu2024evidence} and  Zheng {\it et al.}\ \cite{zheng2024incommensurate}.
	}
    \label{fig:STM}
\end{figure}

Recent theoretical work suggests that STM may provide direct experimental access to properties of Kitaev QSLs \cite{trousselet2011effects,nasu2021spin,udagawa2021scanning,willans2010disorder,willans2011site,sreenath2012localized,halasz2014doping,Udagawa2018,Nasu2020,Kao2021,dantas2022disorder,takahashi2023nonlocal,jang2021vortex,pereira2020electrical,feldmeier2020local,konig2020tunneling}. Of particular interest is the potential detection of non-Abelian anyons through local STM measurements. Site vacancies have emerged as a promising approach to overcome experimental challenges in these systems. Extensive theoretical studies have explored the effects of vacancies in Kitaev quantum magnets, revealing that they can stabilize Majorana zero modes in the ground state.

Recent advances in sample preparation methodologies have enabled multiple research groups to successfully perform STM measurements on pristine monolayer and atomic-thickness $\alpha$-RuCl$_3$ samples (Fig.\,\ref{fig:STM}(a)). 
Wang {\it et al.}\ successfully synthesized monolayer $\alpha$-RuCl$_3$ on highly oriented pyrolytic graphite (HOPG) substrates using molecular beam epitaxy (MBE) \cite{wang2022direct}. Their atomic-resolution topographic imaging revealed a kagome-like lattice of regularly arranged circular protrusions (Fig.\,\ref{fig:STM}(b)), confirming the successful synthesis. The measurements demonstrated the Mott insulating nature of these monolayer films. Kohsaka {\it et al.}\ later corroborated these findings using samples grown by pulsed laser deposition (PLD) on HOPG substrates (Fig.\,\ref{fig:STM}(c)) \cite{kohsaka2024imaging}.

However, studies of mechanically exfoliated monolayer $\alpha$-RuCl$_3$ by Zheng {\it et al.}\ and Yang {\it et al.}\ \cite{zheng2023tunneling,zheng2024insulator,yang2023magnetic} reported metallic states within the Mott-Hubbard gap, markedly different from MBE- and PLD-grown films.
Further investigations by Qiu {\it et al.}\ using graphene/few-layer $\alpha$-RuCl$_3$ heterostructures on graphite or hexagonal boron nitride substrates showed energy gaps smaller than those in MBE- and PLD-grown monolayers. These results contrast with observations of substantial electron transfer from HOPG substrates in certain experiments \cite{leeb2021anomalous,zheng2023tunneling,zheng2024insulator,yang2023magnetic}. The mechanism behind this enhanced electron transfer in exfoliated systems compared to epitaxially grown films remains open.

Massicotte {\it et al.} fabricated magnetic tunnel junctions by sandwiching atomically thin $\alpha$-RuCl$_3$ flakes between two graphite layers. The $\alpha$-RuCl$_3$ flakes were obtained through mechanical exfoliation. They claim that AFM ordering persists at 14\,K even in few-layer $\alpha$-RuCl$_3$, and this ordering can be suppressed by applying an in-plane magnetic field \cite{massicotte2024giant}.  However, whether AFM ordering occurs in monolayer $\alpha$-RuCl$_3$ remains an open question.

Notably, STM measurements have revealed remarkable long-wavelength spatial modulations in the LDOS decaying away from defects, as independently documented by  Qiu {\it et al.}\ \cite{qiu2024evidence}, Zheng {\it et al.}\ \cite{zheng2024incommensurate}, and Kohsaka {\it et al.}\ \cite{kohsaka2024imaging}, as displayed in Fig.\,\ref{fig:STM}(d). While these oscillatory patterns share superficial phenomenological similarities, they exhibit fundamental differences, particularly in the spatial characteristics of the LDOS.  Both Qiu {\it et al.}\ and Kohsaka {\it et al.}\ report asymmetry between spectra in positive and negative bias regions, with different wave numbers of oscillations at opposite bias voltages \cite{qiu2024evidence,kohsaka2024imaging}. This particle-hole asymmetry suggests the emergence of unusual electronic states around defects. However, a significant dichotomy exists between their findings. Qui {\it et al.}, studying few-layer specimens, observed electronic ordering phenomena at energies corresponding to both the upper and lower Hubbard bands, with electronic superstructures maintaining commensurability with the underlying $\alpha$-RuCl$_3$ lattice periodicity. They interpret these findings within the framework of electron-hole crystallization in a doped Mott insulating state. In contrast, Kohsaka {\it et al.}\ reported that spatial modulations in the LDOS exhibit incommensurability with lattice periodicity. Through comprehensive numerical simulations, they showed that these modulations can be quantitatively reproduced by considering the scattering of itinerant Majorana fermions across a Majorana Fermi surface within a Kitaev QSL state. They suggested an analogy between the observed quantum interference patterns and Friedel oscillations in metallic systems \cite{kohsaka2024imaging,jahin2024theory}, though with Majorana fermions as the underlying quasiparticles rather than conventional electrons.

The microscopic mechanism behind these spatial oscillation patterns remains an open challenge. 
These characteristic responses to local symmetry-breaking perturbations provide unprecedented microscopic access to fundamental excitations and correlations that remain inaccessible to traditional bulk measurements.

\section{Perspective}

Over the last decade, $\alpha$-RuCl$_3$ has emerged as a prime candidate for the Kitaev QSL, prompting significant efforts from both theoretical and experimental communities to understand the underlying physics of this fascinating material.  Despite exhibiting a particular zigzag order, $\alpha$-RuCl$_3$ displays a dominant Kitaev interaction. The generic model developed for materials with strong SOC includes not only the Kitaev interaction but also the off-diagonal symmetric interaction, known as the $\Gamma$ interaction, as well as the standard Heisenberg interaction. These non-Kitaev interactions, especially their role in destabilizing the Kitaev spin liquid, have been extensively investigated. 

Even in the presence of non-Kitaev interactions that lead to the zigzag order, experimental data---such as neutron scattering, Raman response, and THz spectroscopy above the Néel temperature---have revealed spin-liquid-like features, such as continuum excitations. These findings suggest that $\alpha$-RuCl$_3$ is close to realizing the Kitaev QSL. The ordered state is proximate to the Kitaev spin liquid, which governs the high-temperature physics above the zigzag-ordered state.

While various experimental discoveries indicate that $\alpha$-RuCl$_3$ is close to a Kitaev spin liquid, the definitive signature of such a state---namely, the half-integer thermal Hall conductivity in the presence of a magnetic field along a direction that breaks two-fold rotational symmetry---has yet to be conclusively observed. Recent experiments have reported thermal Hall conductivity near the half-integer value, suggesting that the Kitaev spin liquid could be induced under a magnetic field starting from the zigzag ordered state. Despite this exciting discovery, debates regarding the thermal transport data persist, and several open questions remain for future investigation.
We outline a few challenges below.

From a theoretical perspective, the spin exchange parameters are constrained by experimental findings. The most well-established behavior is the zigzag ordering at low temperatures in the absence of a field. Additionally, the magnetic moment is pinned in the $ac$-plane, making the angle of $\sim32^\circ$ within the plane, which limits the possible parameter sets.  The exchange parameters that fit these conditions are the FM Kitaev and AFM $\Gamma$ interactions with a small FM $\Gamma'$ and/or a third n.n.\ Heisenberg interaction. While the precise values of these interactions are yet to be determined, a generic model has been established.

On the experimental side, most data have been compared with the pure Kitaev model (or the Kitaev contribution has been subtracted to isolate the non-Kitaev contributions), as the pure Kitaev model at finite temperatures is more accessible via numerical methods. However, it remains unclear how non-Kitaev interactions intertwine with the Kitaev interaction, and a simple reduction approach requires further theoretical justification.

The Kitaev QSL starting from the zigzag order under a field along the $a$-axis (or $ac$-plane direction) has not been found in theoretical studies. To induce the Kitaev spin liquid under the magnetic field, the non-Kitaev interactions responsible for the zigzag order must become insignificant under the magnetic field. Understanding this is a difficult task. On the other hand, a $c$-axis field could potentially induce a quantum spin liquid due to the AFM $\Gamma$ interaction, though this remains an open question.

The effects of the interlayer coupling in $\alpha$-RuCl$_3$ have been less explored theoretically, as it is a van der Waals material with weak interlayer coupling and is prone to stacking faults. Stacking faults in the crystal structure have been extensively studied, and further research is required to precisely determine the values of microscopic spin interactions. It is worth noting that the spin interactions, particularly the Kitaev interaction, are sensitive to the Cl$^-$ positions, as the interorbital hopping between Ru ions via $p$-orbitals is the main exchange process leading to the Kitaev interaction. It is possible that a series of 3D magnetic orders could emerge under the influence of a magnetic field, with just a few Tesla potentially altering the interlayer magnetic structure. Further studies are needed to clarify this possibility.

Spin-phonon interactions at finite temperatures represent another important avenue for future research. It is natural to expect significant coupling between spin and phonon interactions in magnetic materials with strong SOC. Since orbitals are responsible for the bond-dependent pseudospin interaction due to their shape via SOC, even a small change in the lattice could significantly impact the dynamics of the pseudospin. The magnetic field may also alter the pseudospin dynamics as the orbital angular momentum is not quenched in magnetic materials with SOC, including $\alpha$-RuCl$_3$.

From a materials perspective, diversification of material platforms is essential for advancing the field. Particularly promising are systems exhibiting AFM Kitaev interactions, which theoretical studies predict could stabilize the Kitaev QSL phase under applied magnetic fields. Engineering a $j_{\rm eff}=1/2$ single layer with reduced direct exchange process, such as ensuring well-distance $j_{\rm eff}=1/2$ ions, is desirable as it enhances the Kitaev interaction relative to non-Kitaev interactions.

Crystallographic stacking faults significantly influence the magnetic, thermal, and transport properties of samples, leading to notable sample-to-sample variations. Future work should focus on two key areas: (1) systematic characterization of stacking faults using high-resolution TEM and XRD techniques, combined with synthesis of high-purity samples, and (2) precise measurement of in-plane magnetic responses using vector magnetometry.

The existence of a quantum spin liquid phase beyond the critical magnetic field ($H_c$) of the AFM ground state remains contentious at low temperatures ($T\ll T_N$). Verification of such a phase requires identification of quantum critical points or thermodynamic phase transitions at magnetic fields $H\gg H_c$ through heat capacity measurements, neutron scattering studies of magnetic correlations, and nuclear magnetic resonance investigations of local spin dynamics. These experimental probes must be conducted under precisely controlled field and temperature conditions to map the phase diagram and characterize the nature of potential quantum phase transitions.

Theoretical studies indicate that STM measurements could provide direct experimental access to signatures of the Kitaev QSL state. Of particular significance is the predicted non-Abelian anyon excitation, which manifests as a $Z_2$ vortex flux binding a Majorana zero mode. While atomic-resolution STM measurements on monolayer $\alpha$-RuCl$_3$ could potentially detect these exotic quasiparticle states, two fundamental obstacles impede their experimental realization and manipulation: the thermal proliferation of $Z_2$ flux excitations above the flux gap energy scale, and perturbative non-Kitaev exchange interactions that can destabilize the topological ground state. These factors present significant challenges for achieving the controlled nucleation and braiding of non-Abelian anyons required for unambiguous experimental verification of their non-trivial exchange statistics and potential applications in topological quantum computation. A promising strategy to overcome these challenges involves a systematic investigation of crystallographic vacancies, atomic-scale defects, and chemical substitution. Critical to this approach is elucidating the microscopic origin of long-wavelength spatial oscillations in the LDOS observed via STM measurements in monolayer $\alpha$-RuCl$_3$ near defect sites. 

To conclude, this review complements existing works while offering a unique focus on the latest studies of $\alpha$-RuCl$_3$, a material that has only been intensively studied in the context of Kitaev physics for approximately 10 years. We hope that this review provides a comprehensive summary of the exciting experimental findings and theoretical studies, while also highlighting open questions that will inspire further research, not only within $\alpha$-RuCl$_3$ but also in other candidate materials.

\section{Acknowledgements}
We thank T. Asaba, K. Behnia, K. S. Burch, J. Bruin, K-Y. Choi, P. Czajka, S. Fujimoto, A. Go, K. Hashimoto, C. Hickey, K. Hwang, K. Imamura, N. Kawakami,  G. Khaliullin, S. Kasahara, Y. Kasahara, I. Kimchi, J. Knolle, Y. Kohsaka, N. Kurita, T. Kurumaji, P. A. Lee, S. Lin, S. Ma, Y. Mizukami, T. Mizushima, E-G. Moon, Y. Motome, S. E. Nagler, J. Nasu, P. Noh, N.P. Ong, N. Perkins, A. Scheie, S. Suetsugu, H. Takagi, H. Tanaka, T. Terashima, K. Totsuka, S. Trebst, R. Valent\'{i}, D. Wulferding, Y. Xing, M. G. Yamada, M. Yamashita, T. Yokoi for insightful discussions. This work is supported by Grants-in-Aid for Scientific Research (KAKENHI) (No.\ JP23H00089), on Innovative Areas ``Quantum Liquid Crystals'' (No.\ JP19H05824), and on Transformative Research Areas (A) ``Correlation Design Science'' (No.\ JP25H01248) from the Japan Society for the Promotion of Science, by JST CREST (JPMJCR19T5), and by the NSERC Discovery Grant No. 2022-04601. H.Y.K acknowledges support from the Canada Research Chairs Program.
\bibliography{KitaevRef}
\end{document}